\documentclass[reprint,notitlepage,aps,prd,superscriptaddress,nofootinbib,nobibnotes,longbibliography]{revtex4-2}

\usepackage{hyperref}
\usepackage[nolist]{acronym}
\usepackage{amssymb,amsmath,amsfonts}
\usepackage[per-mode=symbol]{siunitx}
\usepackage[dvipsnames]{xcolor}
\usepackage{graphicx}
\usepackage{placeins}

\usepackage{tabularx}
\newcolumntype{Y}{>{\centering\arraybackslash}X}
\newcolumntype{K}{>{\centering\arraybackslash}m{1.85cm}}
\newcolumntype{L}{>{\centering\arraybackslash}m{2.5cm}}

\usepackage{setspace}
\begin{document}

\title{Design of Dedicated Tilt-to-Length Calibration Maneuvers for LISA}

\author{Henry Wegener}
\email[]{henry.wegener@aei.mpg.de}
\author{Sarah Paczkowski}
\author{Marie-Sophie Hartig}
\author{Martin Hewitson}
\author{Gerhard Heinzel}
\author{Gudrun Wanner}
\affiliation{Max Planck Institute for Gravitational Physics (Albert Einstein Institute)}
\affiliation{Leibniz Universität Hannover}

\date{\today}

% ABSTRACT
\begin{abstract}
    Tilts of certain elements within a laser interferometer can undesirably couple into measurements as a form of noise, known as \ac{TTL} coupling. This \ac{TTL} coupling is anticipated to be one of the primary noise sources in the \ac{LISA} mission, after \ac{TDI} is applied. Despite the careful interferometer design and calibration on the ground, \ac{TTL} is likely to require in-flight mitigation through post-processing subtraction to achieve the necessary sensitivity. Past research has demonstrated \ac{TTL} subtraction in simulations through the estimation of 24 linear coupling coefficients using a noise minimization approach. This paper investigates an approach based on performing rotation maneuvers for estimating coupling coefficients with low uncertainties. In this study, we evaluate the feasibility and optimal configurations of such maneuvers to identify the most efficient solutions. We assess the efficacy of \ac{TTL} calibration maneuvers by modulating either the \ac{SC} attitude or the \ac{MOSA} yaw angle. We found that sinusoidal signals with amplitudes of around \SI{30}{\nano\radian} and frequencies near \SI{43}{\milli\Hz} are practical and nearly optimal choices for such modulations. Employing different frequencies generates uncorrelated signals, allowing for multiple maneuvers to be executed simultaneously. Our simulations enable us to estimate the \ac{TTL} coefficients with precision below \SI{15}{\micro\m\per\radian} (1-$\sigma$, in free space) after a total maneuver time of $20$~minutes. The results are compared to the estimation uncertainties that can be achieved without using maneuvers.
\end{abstract}

\maketitle

% BODY

\section{Introduction\label{sec:intro}}

\begin{figure}[htp]
   \centering
   \includegraphics[width=.45\textwidth]{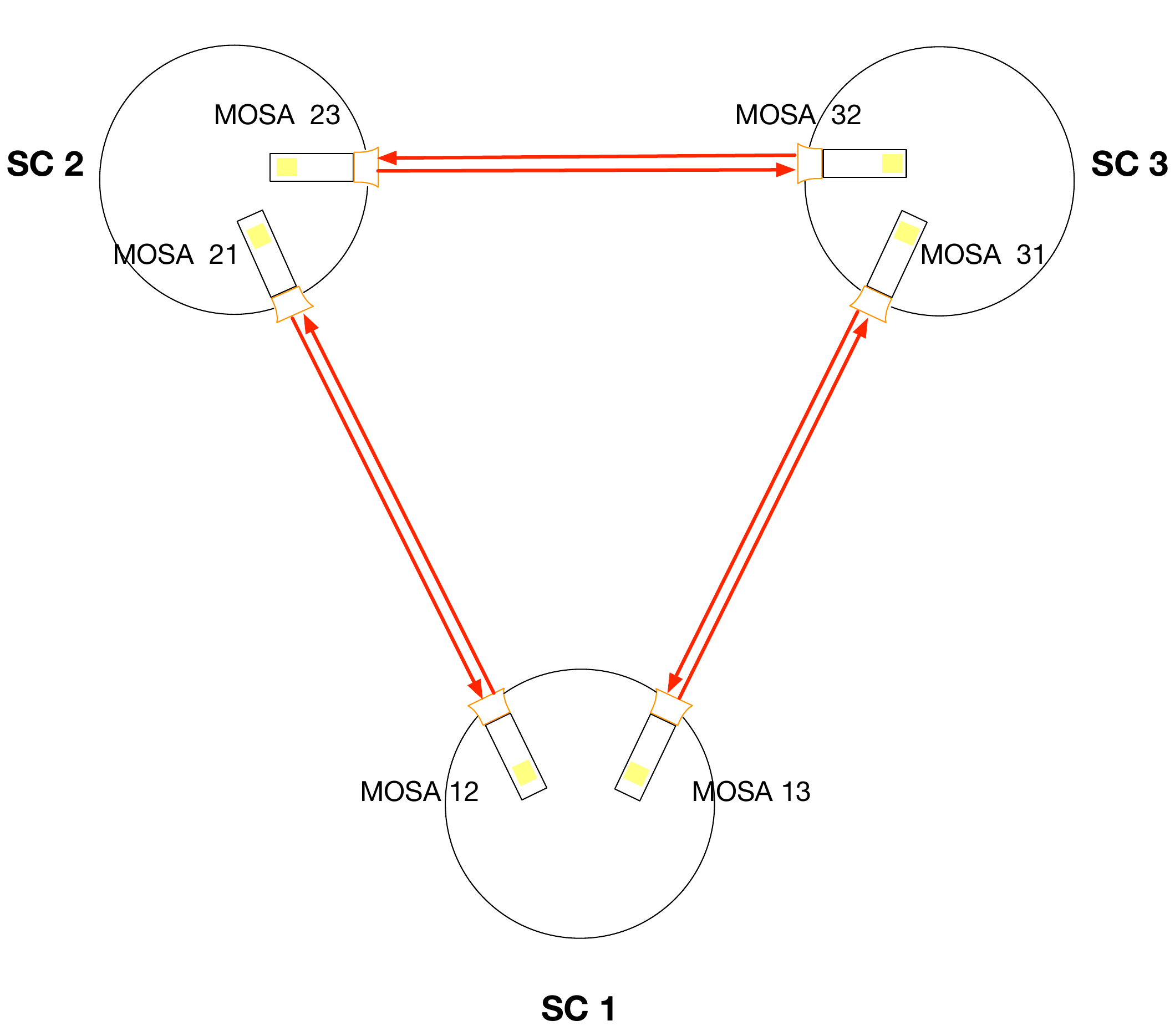}
	\caption{\ac{LISA} \ac{SC} constellation, \ac{MOSA} index notation. Image credit: \cite{paczkowski_2022}.}
\label{fig:sc_constellation}
\end{figure}

\ac{LISA} is a space-based \ac{GW} detector that will operate within a measurement band ranging from approximately \SI{0.1}{\milli\Hz} to \SI{1}{\Hz} \cite{lisa_redbook}. The detector is composed of three \ac{SC}, arranged in an almost equilateral triangle with $2.5$~million~\si{\km} arm length, that will trail behind the Earth in a heliocentric orbit. The \ac{LISA} mission is led by the \ac{ESA} and was recently adopted for an expected launch in the 2030s. Each \ac{SC} will host two \acp{MOSA}, each comprising a telescope, an \ac{OB} and a \ac{TM}. Each \ac{MOSA} will be pointed towards one of the two remote \ac{SC}. The \acp{MOSA} will be designated using the notation shown in Fig.~\ref{fig:sc_constellation}, where \ac{MOSA}\,$ij$ refers to the assembly on \ac{SC}\,$i$ facing \ac{SC}\,$j$. The lengths of the three arms of the \ac{LISA} constellation will vary with time, unlike ground-based \ac{GW} detectors. This leads to the interferometric measurements being highly affected by laser frequency noise. To mitigate this noise, the \ac{TDI} \cite{tinto_2020} algorithm will be applied to generate \ac{TDI} output variables, which mimic three virtual equal-arm interferometers. In this study we examine the second generation Michelson $X,Y,Z$ combinations. These variables will be affected by \ac{TTL} coupling, i.e. they contain error terms which depend on the \ac{MOSA} tilt angles \ac{w.r.t.} the incident beam. This \ac{TTL} should be estimated and subtracted from the measurements in post-processing.

In this paper we investigate the possibility of using rotation maneuvers to estimate the \ac{TTL} coefficients, i.e. the parameters of the \ac{TTL} model. By injecting a modulation signal into the angles that cause \ac{TTL}, the signal for the fit is enhanced. This can reduce the uncertainty in the \ac{TTL} coefficient estimation. This option could be used if \ac{TTL} is less well separable from other noise terms or \ac{GW} signals than anticipated. In such a case, the maneuvers could serve as a beneficial backup plan. It may also be decided to perform such maneuvers once within the commissioning phase. \ac{TTL} maneuvers have already been performed in the \ac{LPF} mission, cf. \cite{armano_2023,audley_2020}. A comparable approach has successfully been used in the \ac{GFO} mission and is considered for future geodesy missions as well \cite{wegener_phd,wegener_2020}.

Other sources addressing \ac{TTL} in \ac{LISA} include \cite{wanner_2024,paczkowski_2022,paczkowski_2025,george_2022,houba_2022a,houba_2022b}. Wanner et al. \cite{wanner_2024} provide a comprehensive analytical description of \ac{TTL} in the individual interferometers of \ac{LISA} as well as in the \ac{TDI} Michelson variables. In \cite{paczkowski_2022} and \cite{paczkowski_2025}, it is described how the \ac{TTL} error can be estimated through noise minimization and subtracted from the \ac{TDI} variables, utilizing pointing angles measured by \ac{DWS} \cite{morrison_1994}. George et al. \cite{george_2022} apply a Fisher information matrix analysis to derive lower bounds for the uncertainty with which the \ac{TTL} coefficients can be estimated and use these to analyze the residual \ac{TTL} noise after post-processing subtraction.

In \cite{houba_2022a}, the observability of \ac{TTL} in the \ac{TDI} Michelson variables is shown by propagating the \ac{TTL} contributions through the \ac{TDI} algorithm. The two options of estimating the \ac{TTL} coefficients with or without rotation maneuvers are discussed. Periodic maneuvers at frequencies outside the \ac{LISA} measurement band are considered, in order not to degrade the science measurements. Thus, large amplitudes are required, however, the feasibility of such maneuvers is not discussed. This study is extended in \cite{houba_2022b} by additionally considering \ac{GW} signals in the measurements and introducing a separation of \ac{TDI} variables that allows performing \ac{TTL} maneuvers without disturbing science operations. The maneuvers discussed in \cite{houba_2022b} are stochastically generated, instead of periodic stimuli. A quantitative analysis of the estimation error is performed, however, a rather long integration time of $15$ hours was assumed.

This paper follows a different approach of designing dedicated \ac{TTL} maneuvers, focussing on sinusoidal stimuli at frequencies within the \ac{LISA} measurement band. We investigate what angular amplitudes are achievable when implemented via \ac{SC} or \ac{MOSA} rotations. A detailed analysis of the estimation uncertainty shows a strong dependency on the maneuver frequency, which can be optimized subsequently. In order to maximize the efficiency, we develop a plan to perform several maneuvers simultaneously. This facilitates very good estimation of the \ac{TTL} coefficients after an integration time of merely $20$ minutes. With simulations we quantify the improvement that such maneuvers provide over the noise minimization approach.

The notation and the \ac{TTL} model are defined in Sec.~\ref{sec:notation}. In Sec.~\ref{sec:simulation}, the simulator settings are specified. The parameter estimation method is briefly described in Sec.~\ref{sec:pe}. Section~\ref{sec:design} on the maneuver design is the main part of this paper. In particular, we discuss the optimal maneuver frequency, and how multiple maneuvers can be performed simultaneously. The simulation results are reported in Sec.~\ref{sec:results}.

\section{Notation\label{sec:notation}}

We analyze the \ac{TTL} in the \ac{TDI} 2.0 variables $X,Y,Z$, which is the sum of the \ac{TTL} noise terms of the \ac{ISI} and the \ac{TMI}. Both are mainly dependent on the $\eta$ (pitch) and $\phi$ (yaw) angles of the \acp{MOSA} \ac{w.r.t.} the incident beam. See \cite{wanner_2024} for more details. Consequently, our model does not distinguish between the two effects. Each angle couples with the local measurement, referred to as \textbf{Rx} \ac{TTL} coupling, as well as through the measurement on the opposing \ac{MOSA} located on the distant \ac{SC}, called \textbf{Tx} \ac{TTL} coupling. Considering that there are six \acp{MOSA}, each of which has an $\eta$ and a $\phi$ angle, we evaluate a total of $24$ \ac{TTL} contributions that transfer to the \ac{TDI} variables. Owing to the small magnitude of angles, we can linearize the contributions resulting in $24$ linear coupling coefficients. As the coefficients are unknown, it will be necessary to determine them in-flight. The coefficients will also be measured on ground and minimized by aligning the interferometers as precisely as possible. However, they are expected to change during launch, rendering any a priori knowledge of them unreliable.

\subsection{MOSA Angles\label{sec:angles}}

Let us denote by $\eta_{ij}$ and $\phi_{ij}$ the pitch and yaw angles of \ac{MOSA}\,$ij$, $i,j=1,2,3$, $i \neq j$, \ac{w.r.t.} the incident beam received from \ac{SC}\,$j$. These angles describe the combined rotation of \ac{MOSA}\,$ij$ \ac{w.r.t.} \ac{SC}\,$i$ and \ac{SC}\,$i$ \ac{w.r.t.} inertial space. The former is represented by the angles $\eta_{ij}^\mathrm{MOSA}$ and $\phi_{ij}^\mathrm{MOSA}$, respectively. The latter is represented by the roll, pitch, and yaw angles of \ac{SC}\,$i$, denoted by $\theta_{i}^\mathrm{SC}$, $\eta_{i}^\mathrm{SC}$, and $\phi_{i}^\mathrm{SC}$. We used the same frame definitions as given for example in \cite{wanner_2024} or \cite{houba_2022a}, as well as the following small-angle approximations. For $\phi_{ij}$ we have
\begin{equation}
	\phi_{ij} \approx \phi_{i}^\mathrm{SC} + \phi_{ij}^\mathrm{MOSA}. \label{eq:phi_ij}
\end{equation}
For $\eta_{ij}$ we have
\begin{align}
	\eta_{ij} &\approx \cos(\pi/6) \cdot \eta_{i}^\mathrm{SC} \pm \sin(\pi/6) \cdot \theta_{i}^\mathrm{SC} + \eta_{ij}^\mathrm{MOSA} \label{eq:eta_ij_1} \\
	&= \frac{\sqrt{3}}{2} \cdot \eta_{i}^\mathrm{SC} \pm \frac{1}{2} \cdot \theta_{i}^\mathrm{SC} + \eta_{ij}^\mathrm{MOSA}, \label{eq:eta_ij_2}
\end{align}
where the $\pm$ is a plus for $\eta_{13}$, $\eta_{21}$, $\eta_{32}$, and a minus for $\eta_{12}$, $\eta_{23}$, $\eta_{31}$.

We call the \ac{DWS} measurements of these angles $\eta_{ij}^\mathrm{DWS}$ and $\phi_{ij}^\mathrm{DWS}$, respectively. These \ac{DWS} angles contain measurement noise, denoted by $n_{\eta_{ij}}^\mathrm{DWS}$ and $n_{\phi_{ij}}^\mathrm{DWS}$, respectively. We write this as:
\begin{align}
	\eta_{ij}^\mathrm{DWS} &= \eta_{ij} + n_{\eta_{ij}}^\mathrm{DWS}, \label{eq:eta_ij_dws} \\
	\phi_{ij}^\mathrm{DWS} &= \phi_{ij} + n_{\phi_{ij}}^\mathrm{DWS}. \label{eq:phi_ij_dws}
\end{align}
For our analyses, the individual \ac{DWS} sensing noise terms $n_{\eta_{ij}}^\mathrm{DWS}$ and $n_{\phi_{ij}}^\mathrm{DWS}$ are assumed to be uncorrelated. Further, we assume that there is no cross-talk between $\eta_{ij}^\mathrm{DWS}$ and $\phi_{ij}^\mathrm{DWS}$, i.e. real angular motion in a pitch angle does not leak into the yaw angle measured by the \ac{DWS}, nor does real yaw motion leak into \ac{DWS} pitch measurements.

\subsection{TTL in the TDI Variables\label{sec:tdi}}

Building the \ac{TDI} variables requires delay operators such as $\mathcal{D}_{ij}$, which accounts for the light travel time from \ac{SC}\,$j$ to \ac{SC}\,$i$. A cascaded delay operator such as $\mathcal{D}_{ij} \mathcal{D}_{jk}$ applies the delay from \ac{SC}\,$k$ to \ac{SC} $j$ first, followed by the delay from \ac{SC}\,$j$ to \ac{SC}\,$i$. We use index contraction, where the last index of the left operator is merged with the first index of the right operator. For instance, $\mathcal{D}_{ij} \mathcal{D}_{jk} = \mathcal{D}_{ijk}$, and iteratively $\mathcal{D}_{i \cdots k} \mathcal{D}_{kl} = \mathcal{D}_{i \cdots kl}$, and so on.

\begin{figure*}[ht!]
	\centering
	\includegraphics[width=\textwidth]{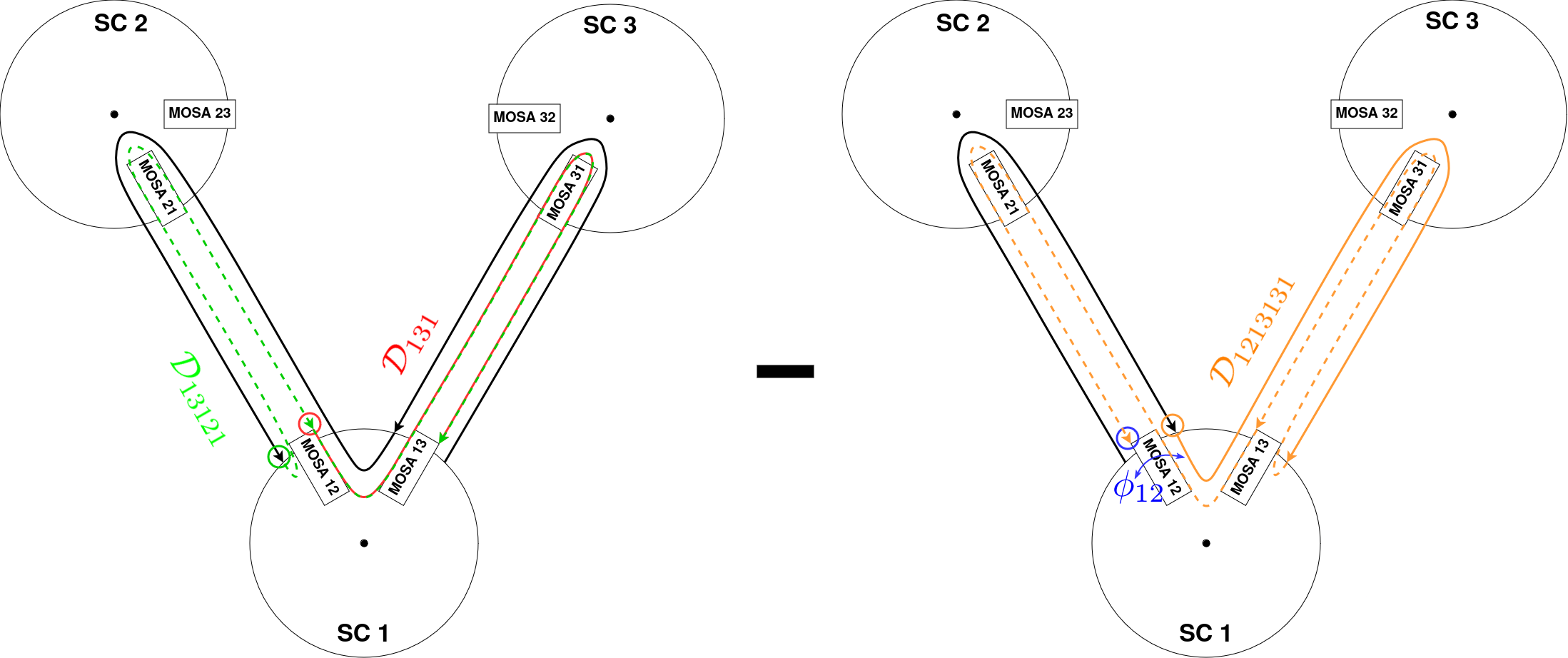}
	 \caption{Schematic of the construction of the \ac{TDI} 2.0 $X$ variable (left minus right combination), cf. \cite{bayle_phd}. The colors illustrate how to build the \ac{TDI} angle $\mathcal{X}_{12\phi Rx}$, which is used to define the \ac{TTL} contribution $X_{12 \phi Rx}^\mathrm{TTL}$ given in Eq.~\eqref{eq:X_12phirx_ttl}.}
\label{fig:tdi_x_phi12rx}
\end{figure*}

The \ac{TDI} 2.0 variables $X,Y,Z$ are defined for instance in \cite{bayle_phd} and \cite{otto_phd}. Due to the combination of measurements in \ac{TDI}, each \ac{TTL} contribution appears with multiple echoes in one \ac{TDI} variable. Figure \ref{fig:tdi_x_phi12rx} illustrates how to construct the \ac{TDI} 2.0 $X$ variable. We take this as an example to trace the Rx-coupling of the angle $\phi_{12}$ into the \ac{TDI} $X$ variable. By following the arrows from the points of incidence at \ac{MOSA}\,$12$ (marked by colored circles), one can infer the correct delays for the four terms. The terms from the left-hand side (with a ``+'') are combined with those from the right-hand side (with a ``-''). This way we construct the term
\begin{equation}
	\mathcal{X}_{12 \phi Rx} = (-\textcolor{blue}{1} + \textcolor{red}{\mathcal{D}_{131}} + \textcolor{green}{\mathcal{D}_{13121}} - \textcolor{orange}{\mathcal{D}_{1213131}}) \phi_{12}, \label{eq:X_12phirx_angle}
\end{equation}
which we refer to as the \ac{TDI} angle for the Rx \ac{TTL} contribution of $\phi_{12}$ in $X$. By multiplying this \ac{TDI} angle with the \ac{TTL} coupling coefficient $C_{12 \phi Rx}$, one obtains the corresponding \ac{TTL} contribution, called $X_{12 \phi Rx}^\mathrm{TTL}$:
\begin{equation}
	X_{12 \phi Rx}^\mathrm{TTL} = C_{12 \phi Rx} \cdot \mathcal{X}_{12 \phi Rx}. \label{eq:X_12phirx_ttl}
\end{equation}
Note that we consider $X,Y,Z$ in time domain to be in the unit of \si{\m}, so the unit of the \ac{TTL} is \si{\m} as well, the unit of the coefficients is \si{\m\per\radian}, and the unit of the \ac{TDI} angles is \si{\radian}.

For each of the \ac{TTL} contributions in $X$, $Y$, and $Z$, we build \ac{TDI} angles denoted by $\mathcal{X}_{ij\alpha\beta}$, $\mathcal{Y}_{ij\alpha\beta}$, and $\mathcal{Z}_{ij\alpha\beta}$, respectively. Then the individual \ac{TTL} contributions are given by
\begin{align}
	X_{ij\alpha\beta}^\mathrm{TTL} &= C_{ij\alpha\beta} \cdot \mathcal{X}_{ij\alpha\beta} \label{eq:X_ttl_contr} \\
	Y_{ij\alpha\beta}^\mathrm{TTL} &= C_{ij\alpha\beta} \cdot \mathcal{Y}_{ij\alpha\beta} \label{eq:Y_ttl_contr} \\
	Z_{ij\alpha\beta}^\mathrm{TTL} &= C_{ij\alpha\beta} \cdot \mathcal{Z}_{ij\alpha\beta}, \label{eq:Z_ttl_contr}
\end{align}
where $i,j \in \{1,2,3\}$, $i \neq j$, $\alpha \in \{\eta,\phi\}$, and $\beta \in \{\mathrm{Rx,Tx}\}$. In total we have $24$ factors $C_{ij\alpha\beta}$, which are called the \ac{TTL} coupling coefficients and which we assume to be constant in our analysis. The equations for all \ac{TTL} contributions are provided in Eqs.~\eqref{eq:x_eta12_rx}~to~\eqref{eq:z_eta21_tx} in App.~\ref{a:ttl_in_tdi}. We write the sum of all \ac{TTL} contributions in $X$ as
\begin{equation}
	X^\mathrm{TTL} = \sum\limits_{i,j,\alpha,\beta} X_{ij\alpha\beta}^\mathrm{TTL}, \label{eq:X_ttl}
\end{equation}
where $8$ of $24$ summands are zero, cf. App.~\ref{a:ttl_in_tdi}. Likewise, $Y^\mathrm{TTL}$ and $Z^\mathrm{TTL}$ denote the total \ac{TTL} in $Y$ and $Z$.

\subsection{Data Notation\label{sec:data}}

In this paper we work with simulated sampled data. For the handling of these data and for the parameter estimation we introduce the following compact notation. Let $N$ be the number of considered data points (samples). Throughout this paper, data \ac{TS} are represented by vectors and matrices in which the rows are associated with time stamps. For instance, let
\begin{equation}
	\eta_{ij} = \begin{pmatrix} \eta_{ij}(t_1) \\ \eta_{ij}(t_2) \\ \vdots \\ \eta_{ij}(t_N) \end{pmatrix} \in \mathbb{R}^{N \times 1}, \label{eq:eta_ij_vector}
\end{equation}
where the $n^\mathrm{th}$ component is $\eta_{ij}(t_n)$, i.e. the angle at time $t_n, n=1,\ldots,N$. Other quantities can be written in the same way, e.g., we denote by $X,Y,Z \in \mathbb{R}^{N \times 1}$ the \ac{TS} of the \ac{TDI} 2.0 variables, each with $N$ samples.

\begin{table}[!ht]
	\caption{Numbering of \ac{TTL} contributions, for all index combinations $ij\alpha\beta$ with $i,j=1,2,3$, $i~\neq~j$, $\alpha~\in~\{\eta,\phi\}$, and $\beta~\in~\{\mathrm{Rx},\mathrm{Tx}\}$.}
	
	\begin{tabularx}{\linewidth}{ c | c | c | c | c | c | c | c | c | c | c | c | c }

	\# & 1 & 2 & 3 & 4 & 5 & 6 & 7 & 8 & 9 & 10 & 11 & 12  \\ \hline
	\textbf{Rx} & $\eta_{12}$ & $\eta_{23}$ & $\eta_{31}$ & $\eta_{13}$ & $\eta_{32}$ & $\eta_{21}$ & $\phi_{12}$ & $\phi_{23}$ & $\phi_{31}$ & $\phi_{13}$ & $\phi_{32}$ & $\phi_{21}$ \\ \hline \hline
	\# & 13 & 14 & 15 & 16 & 17 & 18 & 19 & 20 & 21 & 22 & 23 & 24 \\ \hline
	\textbf{Tx} & $\eta_{12}$ & $\eta_{23}$ & $\eta_{31}$ & $\eta_{13}$ & $\eta_{32}$ & $\eta_{21}$ & $\phi_{12}$ & $\phi_{23}$ & $\phi_{31}$ & $\phi_{13}$ & $\phi_{32}$ & $\phi_{21}$ \\
	
	\end{tabularx}
	
	\label{tab:C_order}
\end{table}

In Tab.~\ref{tab:C_order} we assign a number between $1$ and $24$ to each \ac{TTL} contribution with indices $ij\alpha\beta$. According to this numbering we define $\mathcal{X}$ to be the $N \times 24$ matrix in which each column is one \ac{TDI} angle \ac{TS} $\mathcal{X}_{ij\alpha\beta}~\in~\mathbb{R}^{N \times 1}$:
\begin{equation}
	\mathcal{X} = \begin{pmatrix} \mathcal{X}_{12 \eta Rx}, & \ldots, & \mathcal{X}_{21 \phi Tx} \end{pmatrix} \in \mathbb{R}^{N \times 24}, \label{eq:X_contr_matrix}
\end{equation}
and accordingly for $\mathcal{Y} \in \mathbb{R}^{N \times 24}$ and $\mathcal{Z} \in \mathbb{R}^{N \times 24}$. Let us denote by $C \in \mathbb{R}^{24 \times 1}$ the column vector containing the $24$ true \ac{TTL} coefficients in the order according to Tab.~\ref{tab:C_order}, and by $\hat C \in \mathbb{R}^{24 \times 1}$ the vector of estimated \ac{TTL} coefficients. With this notation the entire \ac{TTL} in $X$ can be written as
\begin{align}
	X^\mathrm{TTL} &= \sum\limits_{i,j,\alpha,\beta} X_{ij\alpha\beta}^\mathrm{TTL} = \sum\limits_{i,j,\alpha,\beta} C_{ij\alpha\beta} \cdot \mathcal{X}_{ij\alpha\beta} \nonumber \\
	&= \mathcal{X} \cdot C. \label{eq:X_ttl_whole}
\end{align}

Moreover, we define the following concatenated quantities. We denote by $\mathcal{A}$ the concatenation of \ac{TDI} angle matrices,
\begin{equation}
	\mathcal{A} = \begin{pmatrix} \mathcal{X} \\ \mathcal{Y} \\ \mathcal{Z} \end{pmatrix} \in \mathbb{R}^{3N \times 24}, \label{eq:alpha}
\end{equation}
and by $\mathcal{A}^\mathrm{DWS} = \mathcal{A} + n_\mathcal{A}^\mathrm{DWS}$ the measured quantity, derived from \ac{DWS} angles containing noise. For the \ac{TDI} variables we define
\begin{equation}
	V = \begin{pmatrix} X \\ Y \\ Z \end{pmatrix} \in \mathbb{R}^{3N \times 1}. \label{eq:V_concat}
\end{equation}
Now, the total \ac{TTL} in the concatenated \ac{TDI} vector $V$ can be expressed as
\begin{align}
	V^\mathrm{TTL} &= \begin{pmatrix} X^\mathrm{TTL} \\ Y^\mathrm{TTL} \\ Z^\mathrm{TTL} \end{pmatrix} \in \mathbb{R}^{3N \times 1} \\
	&= \mathcal{A} \cdot C, \label{eq:V_ttl}
\end{align}
where the right-hand side is a matrix multiplication with $\mathcal{A} \in \mathbb{R}^{3N \times 24}$ and $C \in \mathbb{R}^{24 \times 1}$. Since we do not consider any \ac{GW} signals within this study, we assume that $V$ is the sum of $V^\mathrm{TTL}$ and other noise:
\begin{equation}
	V = V^\mathrm{TTL} + n_V, \label{eq:V_noise}
\end{equation}
i.e. $n_V$ denotes the concatenation of the noise terms in $X,Y,Z$. Assumptions on the noise will be made in Sec.~\ref{sec:pe} on parameter estimation. The most frequently used notation in this paper is summarized in Tab.~\ref{tab:notation}.

\begin{table*}[!ht]
	\caption{Summary of the notation used in this paper.}
\begin{tabularx}{\linewidth}{ c | Y | c | c }

	variable & description & dimension & unit \\ \hline \hline
	
	$N$ & number of samples & $1$ & $1$  \\ \hline
	$\eta_{ij}$, $\phi_{ij}$ & \ac{MOSA} angles \ac{w.r.t.} incident beam (\textit{\ac{TS}}) $\rightarrow$~Eqs.~\eqref{eq:phi_ij}-\eqref{eq:eta_ij_2} and \eqref{eq:eta_ij_vector} & ${N \times 1}$ & \si{\radian} \\ \hline
	$\eta_{ij}^\mathrm{DWS}$, $\phi_{ij}^\mathrm{DWS}$ & \ac{DWS} measurements of $\eta_{ij}$, $\phi_{ij}$ (\textit{\ac{TS}}) $\rightarrow$~Eqs.~\eqref{eq:eta_ij_dws},\eqref{eq:phi_ij_dws} & ${N \times 1}$ & \si{\radian} \\ \hline
	$\mathcal{D}_{ij}$ & delay operators accounting for the light travel time from \ac{SC}\,$j$ to \ac{SC}\,$i$ & $1$ & $1$ \\ \hline
	$X,Y,Z$ & \ac{TDI} 2.0 variables (\textit{\ac{TS}}) & ${N \times 1}$ & \si{\m} \\ \hline
	$\mathcal{X}_{ij\alpha\beta}, \mathcal{Y}_{ij\alpha\beta}, \mathcal{Z}_{ij\alpha\beta}$ & \ac{TDI} angles for indices $ij\alpha\beta$ in $X,Y,Z$ (\textit{\ac{TS}}) $\rightarrow$~Eqs.~\eqref{eq:X_ttl_contr}-\eqref{eq:Z_ttl_contr} & $N \times 1$ & \si{\radian} \\ \hline
	$C_{ij\alpha\beta}$ & true TTL coefficient for indices $ij\alpha\beta$ $\rightarrow$~Eqs.~\eqref{eq:X_ttl_contr}-\eqref{eq:Z_ttl_contr} & $1$ & \si{\m\per\radian} \\ \hline
	$X_{ij\alpha\beta}^\mathrm{TTL}, Y_{ij\alpha\beta}^\mathrm{TTL}, Z_{ij\alpha\beta}^\mathrm{TTL}$ & \ac{TTL} contributions for indices $ij\alpha\beta$ in $X,Y,Z$ (\textit{\ac{TS}}) $\rightarrow$~Eqs.~\eqref{eq:X_ttl_contr}-\eqref{eq:Z_ttl_contr} & $N \times 1$ & \si{\m} \\ \hline
	$X^\mathrm{TTL}, Y^\mathrm{TTL}, Z^\mathrm{TTL}$ & entire \ac{TTL} in $X,Y,Z$ (\textit{\ac{TS}}) $\rightarrow$~Eqs.~\eqref{eq:X_ttl} and \eqref{eq:X_ttl_whole} & $N \times 1$ & \si{\m} \\ \hline
	$\mathcal{X}, \mathcal{Y}, \mathcal{Z}$ & \ac{TDI} angle matrices (\textit{each column one \ac{TDI} angle \ac{TS}}) $\rightarrow$~e.g. Eq.~\eqref{eq:X_contr_matrix} & ${N \times 24}$ & \si{\radian} \\ \hline
	$C$ & true TTL coefficients (\textit{vector}) $\rightarrow$~Eq.~\eqref{eq:X_ttl_whole} & ${24 \times 1}$ & \si{\m\per\radian} \\ \hline
	$\hat C$ & estimated TTL coefficients (\textit{vector}) & ${24 \times 1}$ & \si{\m\per\radian} \\ \hline
	$\mathcal{A}$ & $= (\mathcal{X}^T, \mathcal{Y}^T, \mathcal{Z}^T)^T$ (\textit{concatenated matrices}) $\rightarrow$~Eq.~\eqref{eq:alpha} & ${3N \times 24}$ & \si{\radian} \\ \hline
	$V$ & $=(X^T,Y^T,Z^T)^T$ (\textit{concatenated \ac{TS}}) $\rightarrow$~Eq.~\eqref{eq:V_concat} & ${3N \times 1}$ & \si{\m} \\ \hline
	$V^\mathrm{TTL}$ & $=((X^\mathrm{TTL})^T,(Y^\mathrm{TTL})^T,(Z^\mathrm{TTL})^T)^T$ (\textit{concatenated \ac{TS}}) $\rightarrow$~Eq.~\eqref{eq:V_ttl} & $3N \times 1$ & \si{\m} \\ \hline
	$n_V$ & noise in $V$ (\textit{concatenated \ac{TS}}) $\rightarrow$~Eq.~\eqref{eq:V_noise} & $3N \times 1$ & \si{\m}
	
\end{tabularx}
\label{tab:notation}
\end{table*}

\section{Simulation\label{sec:simulation}}

We utilized the Matlab-implemented simulator \textit{LISASim} \cite{lisasim} to produce simulated \ac{LISA} data. This includes \ac{TTL} coupling, \ac{DWS} measurements, and \ac{TDI} 2.0 \cite{tinto_2020} output variables $X,Y,Z$. The analysis of the simulated data was also performed in Matlab.

\subsection{Noise in the TDI Variables}

Within this study, we do not consider data glitches or \ac{GW} signals in the \ac{TDI} variables. The impact of \ac{GW} signals on the \ac{TTL} estimation accuracy using the noise minimization approach has been studied in \cite{paczkowski_2025} and \cite{LISA-LCST-INST-TN-017}, where no significant impediment has been found. When maneuvers are employed, we expect the estimation to be less vulnerable to disturbances than the noise minimization approach since it is the purpose of the maneuver to produce a \ac{TTL} signal which stands out of the \ac{TTL} noise. Large glitches are likely to be infrequent and of very short duration, so that it is expected that they can be removed from the affected data, even if they occur during the maneuver time.

\begin{figure}[ht!]
	\centering
    \includegraphics[width=.45\textwidth]{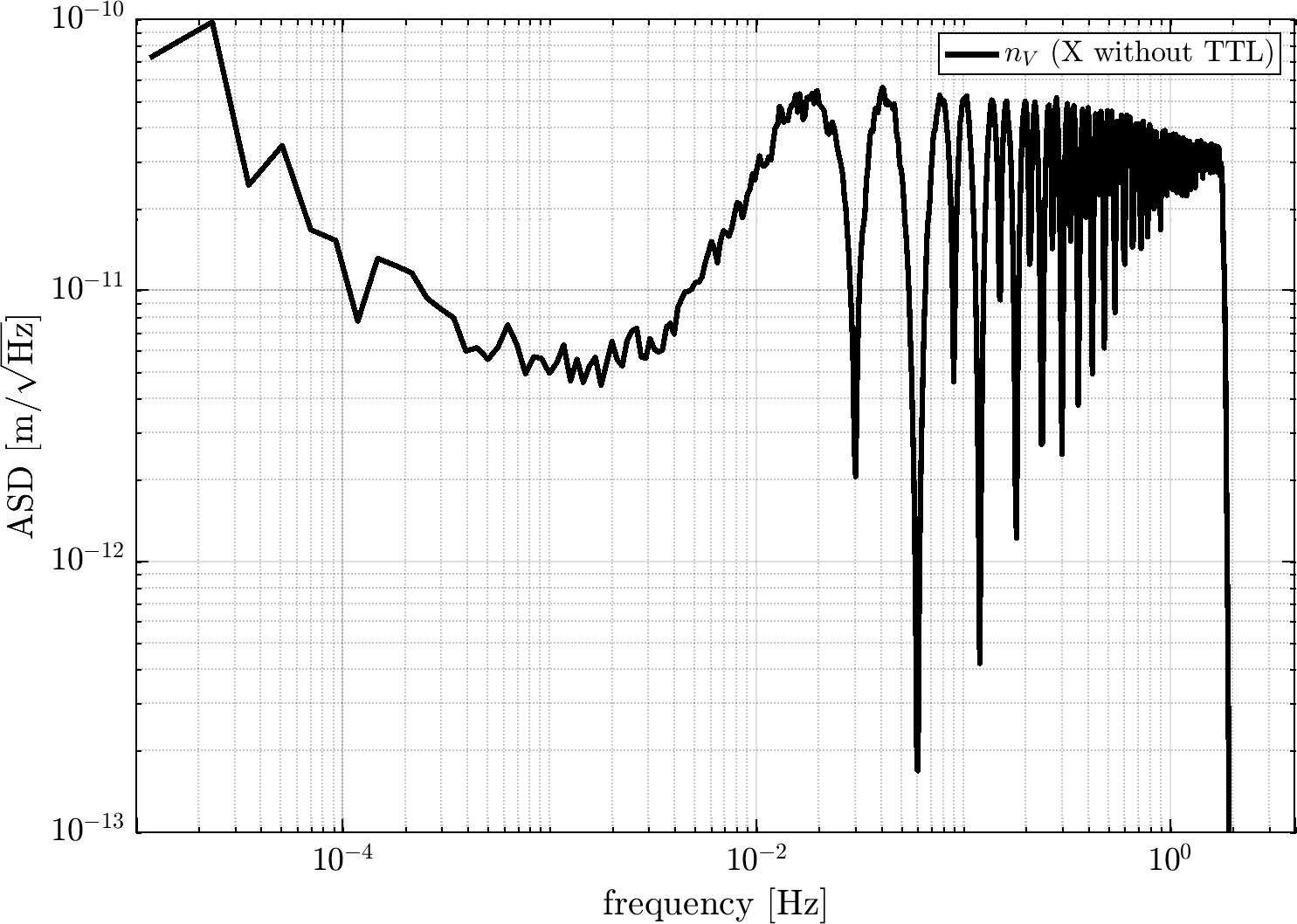}
\caption{Reference \ac{ASD} for $X$ without \ac{TTL}, i.e. $X~-~X^\mathrm{TTL}$. Assuming the spectrum to be identical for $Y$ and $Z$, it also indicates the level of the noise $n_V$ in the concatenated \ac{TS} $V$.}
\label{fig:X_noTTL}
\end{figure}

The magnitude of the noise $n_V$ in $V$, introduced in Eq.~\eqref{eq:V_noise}, will be required to compute the estimator uncertainties, see also Sec.~\ref{sec:pe}. We generated an \ac{ASD} of $n_V$ by simulating the \ac{TDI} output with all \ac{TTL} coefficients set to zero, which is displayed in Fig.~\ref{fig:X_noTTL}. For this reference, the standard deviation of $n_V$ at a data sampling rate of \SI{4}{Hz} is
\begin{equation}
	\sigma \left( n_V \right) \approx \SI{43.5}{\pico\m}. \label{eq:sigma_V}
\end{equation}

\begin{table*}[!ht]
\caption{The noise levels in LISASim used throughout this analysis.}

\begin{tabularx}{\linewidth}{ c | Y }

noise source & noise level in LISASim \\ \hline \hline

clock noise & not implemented \\ \hline

ranging noise & 0  \\ \hline

\acs{SC} angular jitter in $\theta^\mathrm{SC},\eta^\mathrm{SC}$ and $\phi^\mathrm{SC}$ & $\SI{5}{\nano\radian\per\sqrt{\Hz}}$ \\ \hline

\acs{MOSA} angular jitter in $\eta^\mathrm{MOSA}$ w.r.t \ac{SC} & $\SI{1}{\nano\radian\per\sqrt{\Hz}}$ \\ \hline

\acs{MOSA} angular jitter in $\phi^\mathrm{MOSA}$ w.r.t \ac{SC} & $\SI{5}{\nano\radian\per\sqrt{\Hz}}$ \\ \hline

\acs{DWS} sensing noise at detector level & $\SI{70}{\nano\radian\per\sqrt{\Hz}}$ \\ \hline

beam magnification factor & 335 \\ \hline

\acs{ISI} sensing noise & $\SI{6.35}{\pico\metre\per\sqrt{Hz}}$  \\ \hline

\acs{TMI} sensing noise &  $\SI{1.42}{\pico\metre\per\sqrt{Hz}}$ \\ \hline

\acs{RFI} sensing noise &  $\SI{3.32}{\pico\metre\per\sqrt{Hz}}$ \\ \hline

telescope pathlength noise & simulated via thermal expansion with $\mathrm{CTE} = 10^{-8}$/K, assuming a telescope optical path length of \SI{0.7}{\m} and thermal noise as given in \cite{lisasim}. We are aware that the true telescope optical path length is about \SI{1.8}{\m}, see text. \\ \hline

backlink/fibre noise & $\SI{3}{\pico\metre\per\sqrt{\Hz}}$ \\ \hline

\acs{TM} acceleration noise & $\SI{2.4}{\femto\metre\per\second^2\per\sqrt{\Hz}}$ (assuming a \ac{TM} weight of 1.96\,kg) \\ \hline

laser frequency noise  &  $0$ (set to zero because we used \ac{TDI} and assumed static arm lengths) \\ \hline

$x$ - \ac{SC} motion  & $\SI{10}{\nano\metre\per\sqrt{\Hz}}$

\end{tabularx}

\label{tab:LISASim_spec}
\end{table*}

The \textit{LISASim} settings utilized in this analysis are outlined in Tab.~\ref{tab:LISASim_spec}. Note that the laser frequency noise was set to zero. We used only post-\ac{TDI} data, where the laser frequency noise is cancelled due to the static and perfectly known arm lengths assumed by \textit{LISASim}. The noise term $n_V$ introduced above comprises realizations of the noise sources listed in Tab.~\ref{tab:LISASim_spec}, simulated by \textit{LISASim}, and propagated through \ac{TDI}. To ensure consistency, we applied the same settings for all simulations presented in this study. Note that we accidentally assumed the telescope optical path length to be \SI{0.7}{\m}, while the a more realistic value is \SI{1.8}{\m}, however, even with the larger value the effect is still negligible compared to other noise sources. Further, note that the $x$ - \ac{SC} motion also cancels in the \ac{TDI} variables.

Before the parameter estimation, a high-pass filter with a cutoff frequency of \SI{1}{\milli\Hz} was applied to all data, solely to ensure consistent frequency range consideration regardless of the length of the simulated data. The filter had no significant effect on the coefficient estimation. The chosen cutoff frequency is \SI{1}{\milli\Hz} because \ac{TTL} dominates the \ac{TDI} variables only for frequencies above this value.

\subsection{Angular Jitter and Noise}

\begin{figure}[ht!]
	   \centering
	   \includegraphics[width=.45\textwidth]{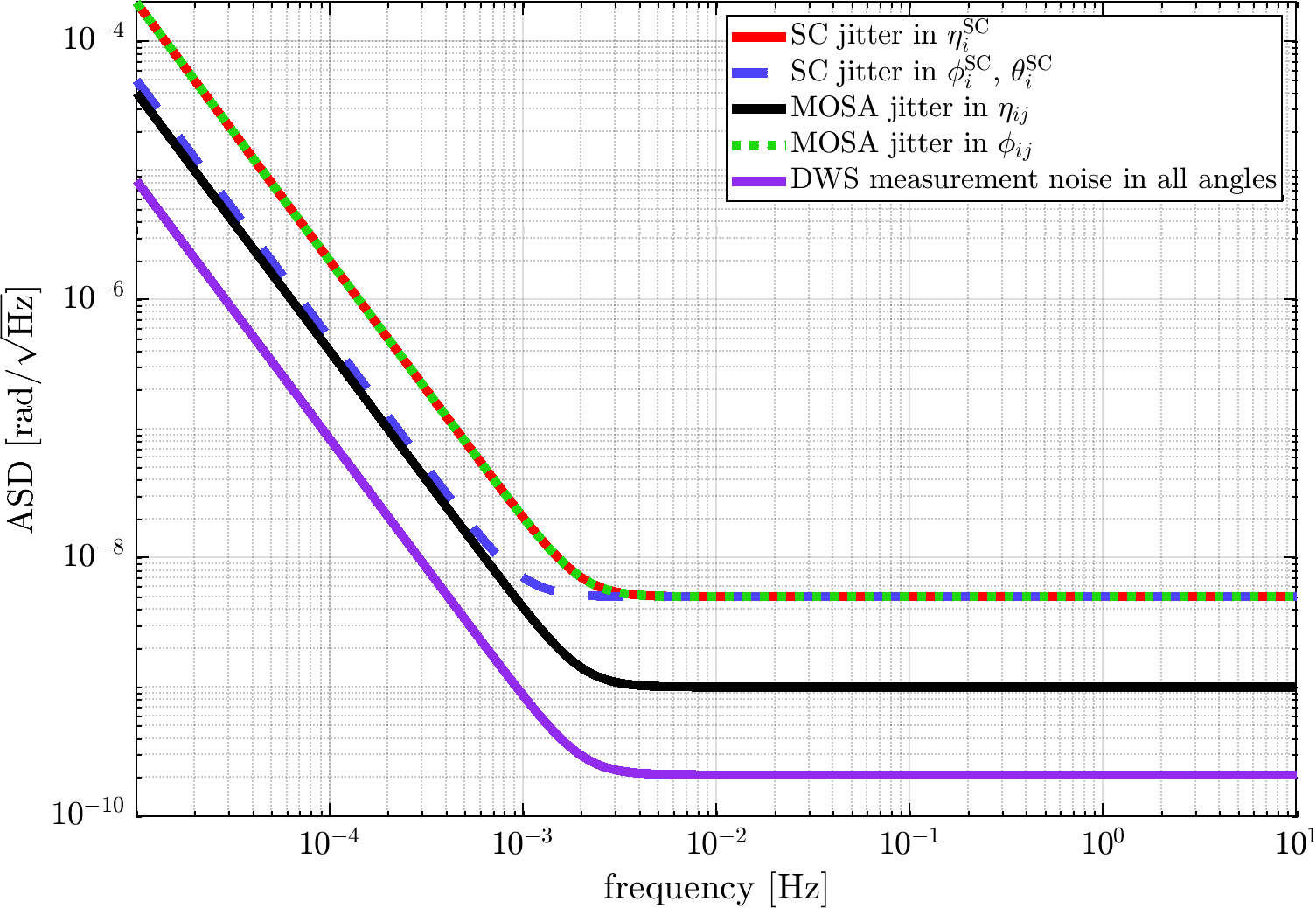}
\caption{\acp{ASD} of \ac{SC} jitter (black, red), \ac{MOSA} jitter (blue, green), and \ac{DWS} sensing noise (purple).}
\label{fig:lisasim_spec}
\end{figure}

For our analysis, it is important to consider the magnitude and spectral shapes of the \ac{DWS} sensing noise as well as of the jitter in both \ac{SC} and \ac{MOSA}. Unless explicitly stated otherwise, the following levels of noise and jitter were implemented for the simulations discussed here:
\begin{itemize}
	\item \ac{DWS} noise per angle: \SI{0.2}{\nano\radian\per\sqrt\Hz}
	\item \ac{SC} jitter in $\phi_i^\mathrm{SC}, ~ \eta_i^\mathrm{SC}, ~ \theta_i^\mathrm{SC}$: \SI{5}{\nano\radian\per\sqrt\Hz}
	\item \ac{MOSA} jitter in $\eta_{ij}^\mathrm{MOSA}$: \SI{1}{\nano\radian\per\sqrt\Hz}
	\item \ac{MOSA} jitter in $\phi_{ij}^\mathrm{MOSA}$: \SI{5}{\nano\radian\per\sqrt\Hz}
\end{itemize}
These levels refer to \ac{ASD} levels for frequencies above \SI{3}{\milli\Hz}. The corresponding spectra are presented in Fig.~\ref{fig:lisasim_spec}. The currently expected \ac{DWS} sensing noise level at the \ac{QPD} is \SI{70}{\nano\radian\per\sqrt\Hz}. This value must be divided by the beam magnification factor of the telescope \cite{chwalla_2020} of $335$. This results in $70/335$\,\si{\nano\radian\per\sqrt\Hz} $\approx$~\SI{0.2}{\nano\radian\per\sqrt\Hz} effective noise in the angles derived from \ac{DWS}.

The jitter levels were chosen according to the performance model \cite{lisa_perf_model} (version of 2021). However, tests with a different simulator, which incorporates a closed \ac{DFACS} control loop, have shown significantly lower \ac{MOSA} jitter levels. Thus we decided to consider an alternative scenario with reduced \ac{MOSA} jitter levels of \SI{0.1}{\nano\radian\per\sqrt\Hz} for $\eta_{ij}^\mathrm{MOSA}$ and \SI{0.5}{\nano\radian\per\sqrt\Hz} for $\phi_{ij}^\mathrm{MOSA}$, in addition to the scenario with the jitter settings listed above. Note that each \ac{MOSA} is hinged in the yaw axis, but rigid in the pitch axis, so the jitter of the \ac{MOSA} w.r.t. \ac{SC} in $\eta$ should be negligible. These reduced \ac{MOSA} jitter levels will be used for comparison in Sec.~\ref{sec:results}.

\section{Parameter Estimation\label{sec:pe}}

We selected a \ac{LSQ} fit in the time domain to estimate the \ac{TTL} coefficients due to its straightforward implementation and favorable performance. For the \ac{LSQ} estimation we assume that the terms $n_V(t_n)$ are independent and identically distributed normal random variables with zero mean and variance $\sigma^2(n_V)$, i.e. the three noise terms in $X,Y,Z$ are uncorrelated Gaussian white noise processes. This is a simplification as can be seen in Fig.~\ref{fig:X_noTTL}. Furthermore, the \ac{LSQ} estimator is optimal under the assumption that $n_\mathcal{A}^\mathrm{DWS}~=~0$, i.e. that there is no \ac{DWS} noise, which does not apply to our data. Note that there might be more accurate estimation methods, see for example \cite{hartig_2024}. However, due to its simplicity the \ac{LSQ} estimator is a beneficial option for the evaluation of \ac{TTL} maneuvers.

With the \ac{TDI} angles $\mathcal{A}^\mathrm{DWS} \in \mathbb{R}^{3N \times 24}$ derived from \ac{DWS} measurements and the concatenated \ac{TDI} variable $V \in \mathbb{R}^{3N \times 1}$, the approach of the \ac{LSQ} estimation is to minimize the residual $V - \mathcal{A}^\mathrm{DWS} C$ in the least squares sense \ac{w.r.t.} $C$. I.e., the goal is to minimize the cost function $R$:
\begin{align}
	R(C) = &\sum\limits_n \left( V(t_n) - \mathcal{A}^\mathrm{DWS}(t_n) C \right)^2 \label{eq:lsc_residual} \nonumber \\
	= &\left( V - \mathcal{A}^\mathrm{DWS} C \right)^T \left( V - \mathcal{A}^\mathrm{DWS} C \right) \nonumber \\
	= &~V^T V - 2 C^T \left( \mathcal{A}^\mathrm{DWS} \right)^T V \\
	&+ C^T \left( \mathcal{A}^\mathrm{DWS} \right)^T \mathcal{A}^\mathrm{DWS} C. \nonumber
\end{align}
This expression can be minimized by calculating the derivative \ac{w.r.t.} $C$,
\begin{equation}
	\frac{\partial}{\partial C} R(C) = -2 \left( \mathcal{A}^\mathrm{DWS} \right)^T V + 2 \left( \mathcal{A}^\mathrm{DWS} \right)^T \mathcal{A}^\mathrm{DWS} C,
\end{equation}
and equating it with the zero vector. Hence, the \ac{LSQ} estimated coefficients are computed as
\begin{equation} \label{eq:lsq}
	\hat C = \left( (\mathcal{A}^\mathrm{DWS})^T \mathcal{A}^\mathrm{DWS} \right)^{-1} \left( \mathcal{A}^\mathrm{DWS} \right)^T V,
\end{equation}
see chapter 1 in \cite{crassidis_2004} for some theoretical background.

When assuming zero \ac{DWS} measurement noise, i.e. if $\mathcal{A}^\mathrm{DWS} = \mathcal{A}$, the covariance matrix of $\hat C$ is given by
\begin{equation}
	\mathrm{cov}\left( \hat C \right) = \sigma^2(n_V) \left( \mathcal{A}^T \mathcal{A} \right)^{-1}.
\end{equation}
Hence, the standard deviations of the estimated coefficients are given by
\begin{equation}
	\sigma_\mathrm{LSQ}\left( \hat C \right) = \sigma(n_V) \cdot \mathrm{diag} \left( \sqrt{ \left( \mathcal{A}^T \mathcal{A} \right)^{-1} } \right),
	\label{eq:std_lsq}
\end{equation}
where $\sqrt{\cdot}$ is taken element-wise and diag denotes the column vector of diagonal matrix entries. Note that formula \eqref{eq:std_lsq}, which we also call formal error, may underestimate the true uncertainty since the assumptions of the estimator are not perfectly fulfilled. From a series of statistical tests we have performed, we have concluded that $\sigma_\mathrm{stat}$, defined by
\begin{equation}
	\sigma_\mathrm{stat} := 1.5 \cdot \sigma_\mathrm{LSQ},
\end{equation}
is a value which is very close to the true uncertainty in our simulation setting. These statistical tests are briefly described in App.~\ref{b:lsq}.

\section{Maneuver Design\label{sec:design}}

The aim of \ac{TTL} calibration maneuvers is to minimize the uncertainty of \ac{TTL} coefficient estimation by modulating the pitch and yaw angles of the \acp{MOSA} relative to incident beams. This induces a \ac{TTL} error, which can be interpreted as a signal instead of noise for the calibration purpose. With the estimated \ac{TTL} coefficients, the \ac{TTL} can be subtracted in post-processing. Additionally, the interferometers can be realigned using the \acp{BAM}. These \acp{BAM} will be used for minimizing the \ac{TTL} already on ground, however, they are planned to be remotely adjustable as well.

The injected signal shall be well distinguishable from the noise background, e.g. with the form of a sine wave or similar, in particular it shall be a repetitive signal with a well defined period $P$. In the frequency domain, such a signal then shows as a peak at $f_\mathrm{man} = 1/P$, which we call the maneuver frequency. The optimal choice of this frequency for the purpose of estimating \ac{TTL} coefficients is investigated in detail in Sec.~\ref{sec:f_man}.

For \ac{LISA}, one option would be to introduce such a modulation signal through the \ac{SC} attitude, which entails rotating the \ac{SC} using the cold gas thrusters. Alternatively, the signal can potentially be injected directly into the \ac{MOSA} angles, specifically rotating the \acp{MOSA}. Given that the \acp{MOSA} are designed for angular adjustments in $\phi$, it is expected that such injections via the \acp{MOSA} are only possible for the $\phi$ angle. These two options are analyzed in Sec.~\ref{sec:inj}. Afterwards, in Sec.~\ref{sec:simultaneous}, we fathom ways of performing simultaneous maneuvers. Section \ref{sec:problem_phi} is concerned with an issue that arises when injecting calibration signals into the \ac{SC} $\phi$ angles.

\subsection{Optimal Maneuver Frequency\label{sec:f_man}}

Since the purpose of \ac{TTL} maneuvers is to minimize the estimation uncertainty, our approach for optimizing the maneuver frequency is to investigate how it affects the \ac{LSQ} uncertainty $\sigma_\mathrm{LSQ}$ given in Eq.~\ref{eq:std_lsq}. To this end, we examine the formulas for the different \ac{TTL} contributions given in App.~\ref{a:ttl_in_tdi}. In order to reach a quantitative conclusion, we first make some general observations in the following.

We take the \ac{TTL} contributions of $\eta_{12}$ in $X$ as an example, but the conclusions will hold for all other angles as well. Firstly, note that for the estimation of $C_{12 \eta Rx}$ and $C_{12 \eta Tx}$, mainly the $X$ variable contains useful information since both contributions are zero in $Z$, and are perfectly correlated to each other in $Y$. I.e., Eqs.~\eqref{eq:y_eta12_rx} and \eqref{eq:y_eta12_tx} imply:
\begin{equation}
	Y_{12 \eta Tx}^\mathrm{TTL} = \frac{C_{12 \eta Tx}}{C_{12 \eta Rx}} \cdot Y_{12 \eta Rx}^\mathrm{TTL}. \label{eq:Y_12eta_corr}
\end{equation}
This circumstance will be addressed again in Secs.~\ref{sec:simultaneous} and \ref{sec:problem_phi} below. Thus, we focus on the $X$ variable in this example.

When assuming constant and equal arm lengths, the following approximations of Eqs.~\eqref{eq:x_eta12_rx} and \eqref{eq:x_eta12_tx} hold:
\begin{align}
	X_{12 \eta \mathrm{Rx}}^\mathrm{TTL} &\approx C_{12 \eta \mathrm{Rx}} \cdot (-1 + \mathcal{D}^2 + \mathcal{D}^4 - \mathcal{D}^6) \eta_{12} \label{eq:approx_12etarx} \\
	X_{12 \eta \mathrm{Tx}}^\mathrm{TTL} &\approx C_{12 \eta \mathrm{Tx}} \cdot (-1 + \mathcal{D}^2 + \mathcal{D}^4 - \mathcal{D}^6) \mathcal{D}^2 \eta_{12}, \label{eq:approx_12etatx}
\end{align}
where $\mathcal{D}^2$ corresponds to a fixed time delay, say $\delta_t$, where $\delta_t~\approx$~\SI{16.7}{\s}. Now, when a periodic signal is injected into $\eta_{12}$, Eq.~\eqref{eq:approx_12etarx} shows that 4 copies of this signal, all multiplied with the same factor $C_{12 \eta Rx}$, will appear in $X$. These echoes can interfere constructively or destructively, which depends on the period of the injected signal. It can be seen immediately that a period corresponding to $\mathcal{D}^2$ would result in the first two copies cancelling each other, as well as the last two copies, except for one cycle in the beginning and one cycle in the end. That is, if $P = \delta_t$, the \ac{TTL} contribution in $X$ would almost vanish, rendering the injection useless for \ac{TTL} calibration. Similarly, with a period of $2\delta_t$, term $1$ would cancel with term $3$, and term $2$ with term $4$. In fact, if the period is any fraction $2\delta_t/n$, $n \in \mathbb{N}$, the injections will nearly cancel. Note that a period of $2\delta_t~\approx~33.\bar 3$\,\si{\s} yields a frequency of about \SI{30}{\milli\Hz}. Consequently, all multiples of \SI{30}{\milli\Hz} are poor choices for the maneuver frequency.

\begin{figure*}[ht!]
	\begin{minipage}[c]{0.49\textwidth}
	   \centering
	   \includegraphics[width=\textwidth]{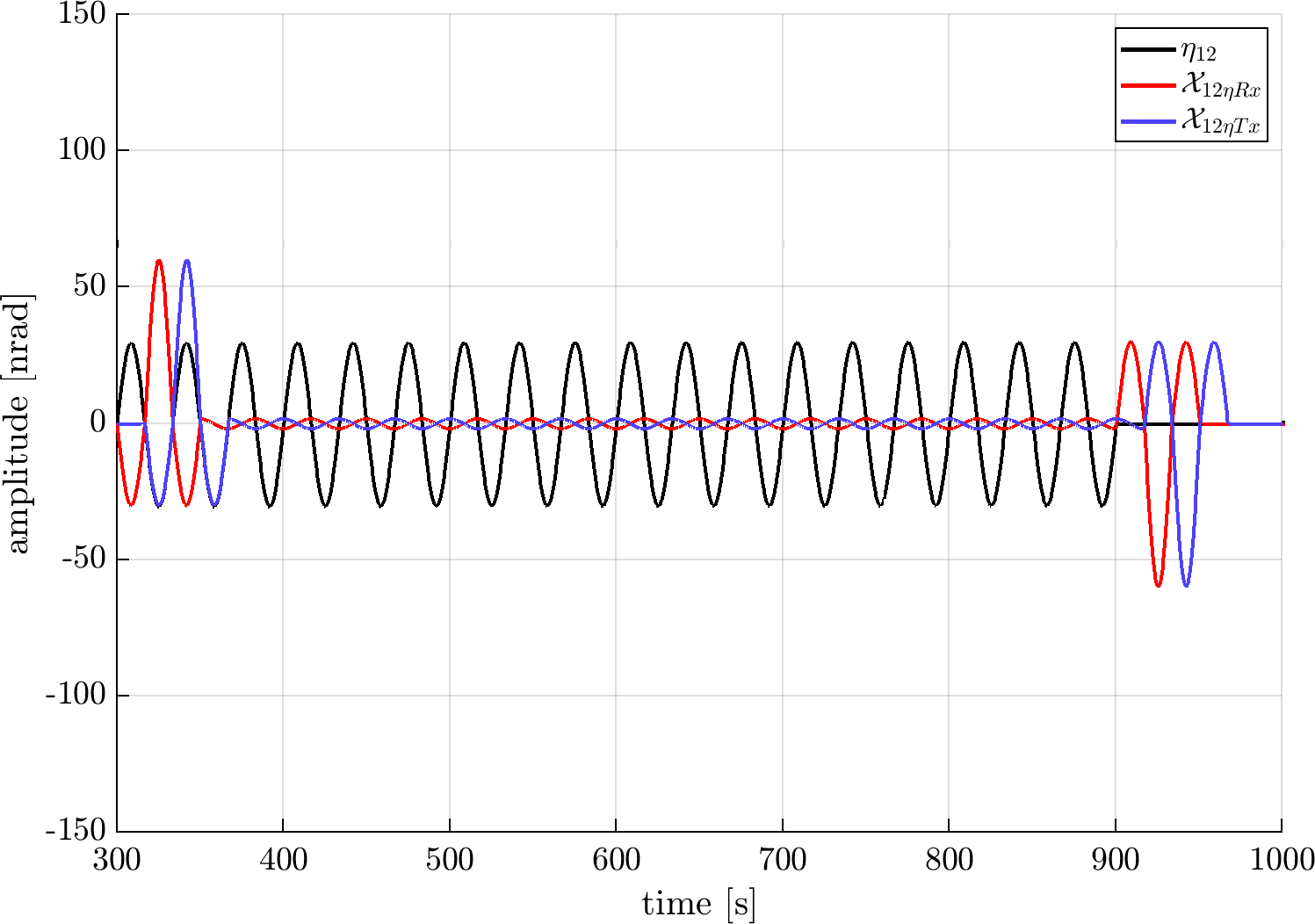}
	\end{minipage}
 \hfill
	\begin{minipage}[c]{0.49\textwidth}
	   \centering
	   \includegraphics[width=\textwidth]{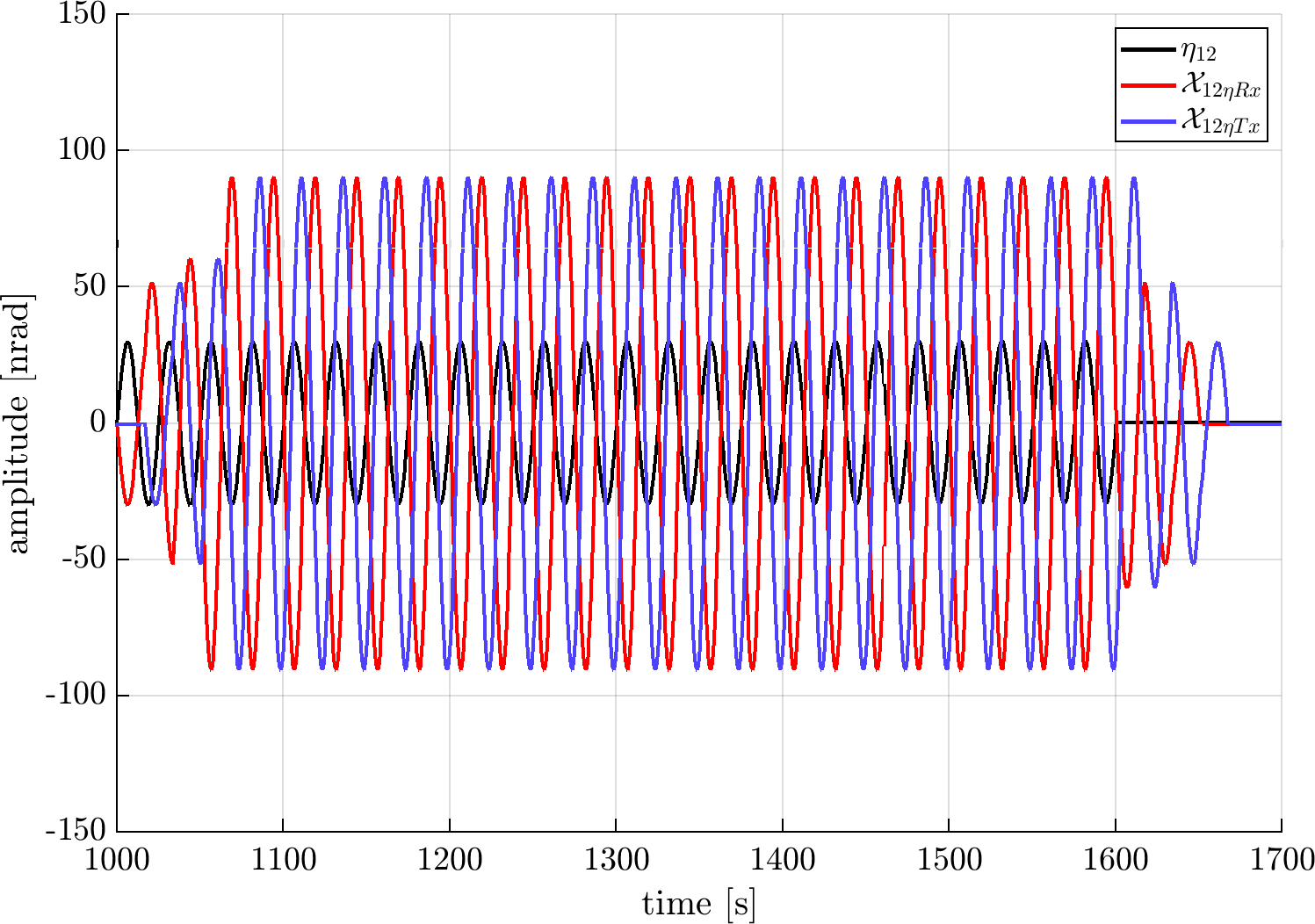}
	\end{minipage}
\caption{Illustration of bad and good examplary maneuver frequencies. Left (\SI{30}{\milli\Hz}; bad choice): destructive echo interference and high correlation between Rx and Tx contributions in the \ac{TDI} $X$ variable. Right (\SI{40}{\milli\Hz}; good choice): constructive interference and low correlation. The principle illustrated here for $\eta_{12}$ holds for all other angles as well. Note also that the Rx and Tx contributions in the left plot do not cancel each other entirely because the arm delays were not exactly equal in this simulation.}
\label{fig:f_comp}
\end{figure*}

Figure~\ref{fig:f_comp} illustrates simulations of two extreme cases. The black lines show sine waves injected into $\eta_{12}$, with frequencies \SI{30}{\milli\Hz} (left plot) and \SI{40}{\milli\Hz} (right plot). The red and blue lines show the \ac{TDI} angles $\mathcal{X}_{12 \eta Rx}$ and $\mathcal{X}_{12 \eta Tx}$, respectively, without sensing noise for the purpose of illustration. Note that the \ac{TDI} angles have not been multiplied by the coefficients and thus have the unit of \si{\radian}, cf. Eq.~\eqref{eq:X_12phirx_angle}. We make two observations here. Firstly, although both injection signals have the same amplitude, the \ac{TDI} angles almost vanish in the left plot and are large in the right plot. This is due to the destructive interference for \SI{30}{\milli\Hz}, and the constructive interference for \SI{40}{\milli\Hz}. This observation holds for both Rx and Tx, cf. Eqs.~\eqref{eq:approx_12etarx} and \eqref{eq:approx_12etatx}. The second observation is that the correlation between Rx and Tx \ac{TDI} angles is high for \SI{30}{\milli\Hz} and low for \SI{40}{\milli\Hz}.

We would like to quantify how these two observations affect the estimation uncertainty. In App.~\ref{c:std} we derived dependencies of $\sigma_\mathrm{LSQ}$ with the simplification made that merely two \ac{TTL} coefficients are estimated at a time, disregarding potential other correlations, see relation \eqref{eq:sigma_prop_all}. The relevant quantities that depend on the maneuver frequency are the strength of the \ac{TDI} angle, which can be expressed as $\sigma(\mathcal{X}_{ij\alpha\beta})$, cf. Eq.~\eqref{eq:sd}, and the correlation between Rx and Tx \ac{TDI} angles in $X$:
\begin{equation}
	c_{ij\alpha} := \mathrm{corr} \left(\mathcal{X}_{ij\alpha Rx},\mathcal{X}_{ij\alpha Tx} \right).
\end{equation}
Good choices for the frequency of a \ac{TTL} calibration maneuver can now be deduced from the relation
\begin{equation}
	\sigma_\mathrm{LSQ} \left(\hat C_{ij\alpha\beta} \right) \propto \frac{1}{\sigma \left(\mathcal{X}_{ij\alpha\beta} \right) \cdot \sqrt{1-c_{ij\alpha}^2}}, \label{eq:sigma_propto}
\end{equation}
which holds for any $i,j,\alpha$ for which $X$ is the crucial \ac{TDI} variable. This is the case for all angles on \ac{SC}\,$1$, i.e. when $i = 1$. For $i = 2$, an analog relation holds when $\mathcal{X}$ is replaced by $\mathcal{Y}$. For $i = 3$, $\mathcal{X}$ should be replaced by $\mathcal{Z}$.

An interpretation of relation \eqref{eq:sigma_propto} is that larger values of $\sigma(\mathcal{X}_{ij\alpha\beta})$ mean larger calibration signals and hence larger signal-to-noise ratio for the estimation. Secondly, high correlation complicates disentangling the two coefficients and hence increases individual uncertainties. Note that for the derivation of relation~\eqref{eq:sigma_propto} it was assumed that only two coefficients, namely $C_{ij\alpha Rx}$ and $C_{ij\alpha Tx}$, are estimated together, so it can merely serve as an approximation in the realistic scenario with $24$ coefficients.

\begin{figure*}[ht!]
	\begin{minipage}[c]{0.49\textwidth}
	   \centering
	   \includegraphics[width=\textwidth]{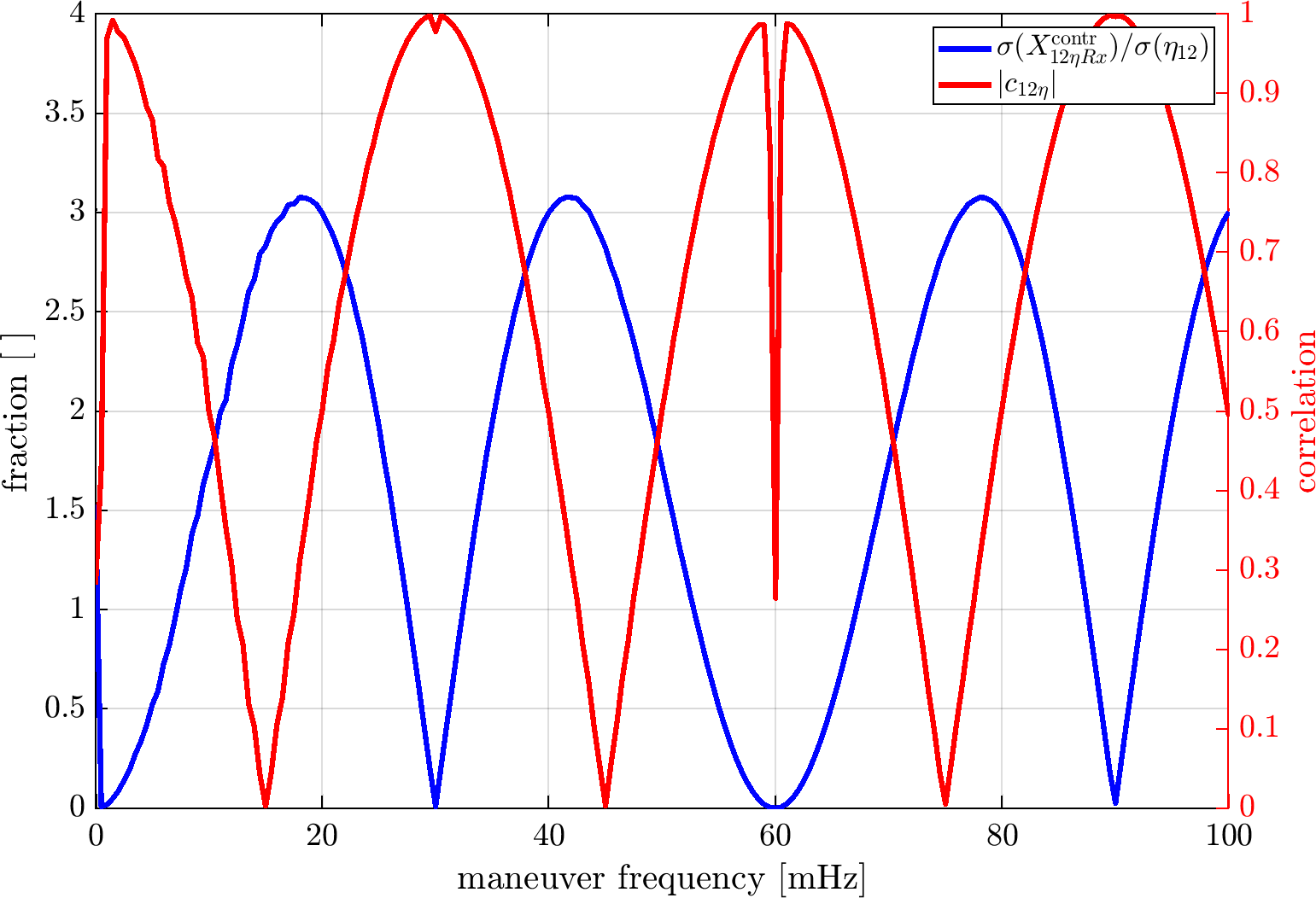}
	\end{minipage}
 \hfill
	\begin{minipage}[c]{0.49\textwidth}
	   \centering
	   \includegraphics[width=\textwidth]{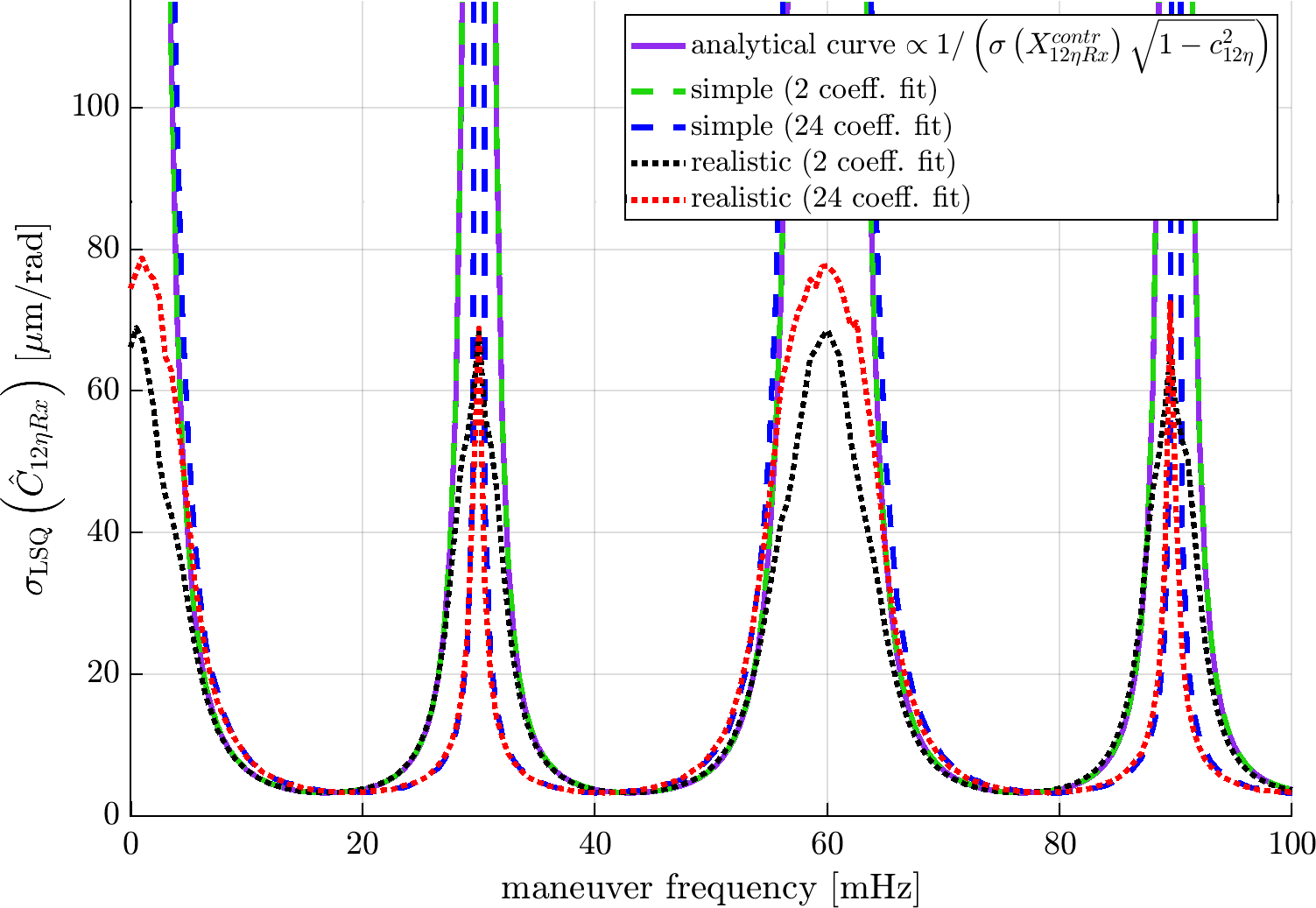}
	\end{minipage}
\caption{Left: relative standard deviation of the response of an injected signal in a \ac{TDI} angle (blue line), and absolute value $\vert c_{12\eta} \vert$ of the correlation between Rx and Tx \ac{TDI} angles (red line). Right: $\sigma_\mathrm{LSQ}$ depending on maneuver frequency (fixed amplitude) from $60$ simulations (black line) and from the analytical relation \eqref{eq:sigma_propto} (dashed red line).}
\label{fig:f_man}
\end{figure*}

We tested relation~\eqref{eq:sigma_propto} using simulations for $201$ different values of $f_\mathrm{man}$ between $0$ and \SI{100}{\milli\Hz}. Each simulation contains one $\eta_{12}$ maneuver with different $f_\mathrm{man}$ but fixed amplitude of \SI{100}{\nano\radian} and duration of \SI{600}{\s}. For each value of $f_\mathrm{man}$ we performed two simulations, a simple and a realistic case. In the realistic case, we used constant simulator settings as defined in Sec.~\ref{sec:simulation}. In the simple case, we set the \ac{DWS} noise and all \ac{TTL} coefficients to zero and all jitter levels to $1$\% of the normal level. For each simulation we computed $\sigma(\hat C_{12\eta Rx})$ in two ways, firstly by estimating only $C_{12\eta Rx}$ and $C_{12\eta Tx}$, secondly by estimating all $24$ coefficients.

For each simulation with simple settings, we computed $c_{12\eta}$ as well as $\sigma(\mathcal{X}_{12\eta Rx})$. We would prefer the result to be independent of the strength of the injection, which is a different topic addressed in Sec.~\ref{sec:inj}. Therefore we also computed $\sigma(\eta_{12})$ and examined the normalized signal strength $\sigma(\mathcal{X}_{12\eta Rx})/\sigma(\eta_{12})$. The left plot of Fig.~\ref{fig:f_man} shows this fraction in blue and $\vert c_{12\eta} \vert$ in red. From these values an analytical curve $\sigma_\mathrm{ana}$ satisfying relation~\eqref{eq:sigma_propto} was computed as
\begin{equation}
	\sigma_\mathrm{ana} = \frac{\sigma(n_V)}{\sigma(\mathcal{X}_{12\eta Rx}) \cdot \sqrt{3N \cdot (1 - c_{12\eta}^2)}}, \label{eq:sigma_ana}
\end{equation}
cf. App.~\ref{c:std}, with $\sigma(n_V)~=$~\SI{43.5}{\pico\m}. 

The right plot of Fig.~\ref{fig:f_man} shows the estimates of $\sigma_\mathrm{LSQ}(\hat C_{12\eta Rx})$ obtained in the different ways described above. The analytical curve from Eq.~\eqref{eq:sigma_ana} is plotted in purple. The values from estimating either two or all $24$ coefficients in the simple case are depicted as dashed green and blue lines, respectively. The values for the realistic case are shown as dotted black and red lines, respectively. The predicted values are nearly identical to the simple case estimating two coefficients. In all cases, the values strongly increase close to certain maneuver frequencies, e.g. \SI{30}{\milli\Hz}. For the realistic case, this increase is upper bounded due to the angular jitter providing information on the coefficients in addition to the maneuver. In both simple and realistic cases, the form of the spikes deviates from the analytical curve when estimating $24$ coefficients.

We find that the standard deviations of the estimated coefficients, $\sigma(\hat C_{ij\alpha\beta})$, is very close to its minimum for maneuver frequencies between \SI{40}{\milli\Hz} and \SI{45}{\milli\Hz}. We conclude that frequencies between \SI{40}{\milli\Hz} and \SI{45}{\milli\Hz} are nearly optimal choices for coefficient estimation, and we chose to use this frequency range for our maneuver simulations. Other good choices exist, e.g. $f_\mathrm{man}~\approx~$\SI{18}{\milli\Hz} or $f_\mathrm{man}~\approx$~\SI{78}{\milli\Hz}. On the other hand, \SI{30}{\milli\Hz} and multiples, more precisely the null frequencies of the \ac{TDI} transfer function, should be avoided. Note that the optimal frequency is independent of the amplitude of the injection signal.

\subsection{Signal Injection\label{sec:inj}}

In this section we investigate what injection signals are realistic, in particular what maneuver amplitudes are achievable. We start with injections via \ac{SC} rotation in Sec.~\ref{sec:inj_sc}, followed by injections via \ac{MOSA} rotation in Sec.~\ref{sec:inj_mosa}.

\subsubsection{Injection into SC Angles\label{sec:inj_sc}}

The \ac{LISA} satellites will likely utilize cold gas thrusters for attitude control. Our current best estimate is that these will be similar to the thrusters used in the \ac{LPF} mission, each of which could produce a maximum force of \SI{500}{\micro\N} \cite{armano_2019}, only part of which was intended for regular use, and part of it was allocated for potentially  necessary offsets. We conservatively assume that a force of $F_\mathrm{max} =$~\SI{10}{\micro\N} per thruster will be available for modulation. The thruster torque is computed as $\tau = r \times F$, where $r$ is the position vector of that thruster \ac{w.r.t.} the \ac{SC} \ac{CoM}. Since attitude thrusters are activated in pairs, both directed perpendicular to $r$, we may assume a maximum torque of about
\begin{equation}
	\vert \tau_\mathrm{max} \vert = 2 \cdot \vert r \vert \cdot \vert F_\mathrm{max} \vert \approx \SI{40}{\micro\N\m}
\end{equation}
per satellite axis if the thrusters are located $\vert r \vert~\approx$~\SI{2}{\m} away from the \ac{SC} \ac{CoM}, which we assume in lack of more detailed specifications. We model the control torque for a maneuver around one principal \ac{SC} axis by
\begin{equation}
	\tau_\mathrm{man}(t) = \left\vert \tau_\mathrm{max} \right\vert \cdot \sin \left(2\pi f_\mathrm{man} t \right). \label{eq:tau_man}
\end{equation}

We assume that the \ac{SC} will be constructed approximately symmetric, and the moments of inertia matrix $J$ will be approximately diagonal with entries $J_{x},J_{y},J_{z}$. Then we can approximate the derivative $\dot\omega$ of the angular velocity vector $\omega = (\omega_x,\omega_y,\omega_z)^T$ of the \ac{SC} by
\begin{equation}
	\dot{\omega} \approx J^{-1} \tau \approx \begin{pmatrix} \tau_x / J_x \\ \tau_y / J_y \\ \tau_z / J_z \end{pmatrix}, \label{eq:approx_anga}
\end{equation}
where $\tau$ is the three-dimensional total torque vector acting on the \ac{SC}. For now, in lack of more detailed specifications, we model the \ac{SC} as a cylinder with a mass of \SI{2.5}{\tonne} and a radius of \SI{2}{\m}, such that $J_x,J_y~\approx~$\SI{3333}{\kg\square\m} and $J_z~\approx~$\SI{5000}{\kg\square\m}.

A \ac{SC} Euler angle, e.g. around the $z$ axis, can be approximated by
\begin{align}
	\phi^\mathrm{SC}(t) &\approx \phi^\mathrm{SC}(t_0) + \int\limits_{t_0}^t \omega_z(s) ~ds \nonumber \\
	&= \phi^\mathrm{SC}(t_0) + (t-t_0) \cdot \omega_z(t_0) + &\int\limits_{t_0}^t \int\limits_{t_0}^s \dot\omega_z(\tilde s) ~d\tilde s ~ds
\end{align}
We neglect the constant and the linear term on the right hand side of the equation since they will be removed in data processing by the high-pass filter, cf. Sec.~\ref{sec:simulation}. Using approximation \eqref{eq:approx_anga} and inserting the modelled control torque as in Eq.~\eqref{eq:tau_man}, we obtain
\begin{align}
	 \phi^\mathrm{SC}(t) &\approx \frac{1}{J_{z}} \int\limits_{t_0}^t \int\limits_{t_0}^s \tau_\mathrm{man}(\tilde s) ~d\tilde s ~ds \nonumber \\
	 &= \frac{1}{J_{z}} \int\limits_{t_0}^t \int\limits_{t_0}^s \vert \tau_\mathrm{max} \vert \sin(2\pi f \tilde s) ~d\tilde s ~ds \nonumber \\
	 &= -\frac{\vert \tau_\mathrm{max} \vert}{J_z (2\pi f)^2} \sin(2\pi f t)
\end{align}
for a maneuver around the $z$ axis, and similarly for the $x$ and $y$ axes. We deduce that any \ac{SC} angle can be modulated by a sinusoid with an amplitude of
\begin{equation}
	A_\mathrm{max} \approx \frac{\SI{40}{\micro\N\m}}{\SI{5000}{\kg\square\m} \cdot (2\pi f)^2} \approx \SI{0.2}{\nano\radian} \cdot \left( \frac{\SI{1}{\Hz}}{f} \right)^2.
\end{equation}
For a maneuver frequency of $f_\mathrm{man} =$~\SI{43}{\milli\Hz}, for instance, we obtain $A_\mathrm{max}~\approx$~\SI{110}{\nano\radian} for a $\phi^\mathrm{SC}$ maneuver. Note that the achievable angular amplitude is proportional to $1/f^2$. If a maneuver frequency of \SI{20}{\milli\Hz} would be used, the maximal amplitude would be \SI{500}{\nano\radian} for rotations about the $z$ axis, and even larger for rotations about the $x$ or $y$ axes since $J_z > J_x,J_y$. We therefore conclude that it is rather conservative to work with a feasible maneuver amplitude of \SI{30}{\nano\radian} for maneuver frequencies near \SI{40}{\milli\Hz}.

In Sec.~\ref{sec:notation} we have defined $\eta_{ij}$ and $\phi_{ij}$ as the pitch and yaw angles of \ac{MOSA}\,$ij$ \ac{w.r.t.} incident beam. These are the physical angles that cause \ac{TTL}, so they are to be excited for the \ac{TTL} calibration. Equations~\eqref{eq:phi_ij} and \eqref{eq:eta_ij_2} show how these angles depend on the \ac{SC} angles. We would like to define a \ac{SC} rotation that results in a sinusoidal excitation of $\eta_{12}$, ideally without exciting $\eta_{13}$. This is achieved, according to Eq.~\eqref{eq:eta_ij_2}, if
\begin{equation}
	\eta_{1}^\mathrm{SC} = -\tan(\pi/6) \cdot \theta_{1}^\mathrm{SC} = -\frac{1}{\sqrt{3}} \cdot \theta_{1}^\mathrm{SC}. \label{eq:eta_12_relation}
\end{equation}
In the other case, for a pure $\eta_{13}$ excitation with $\eta_{12} = 0$, one must ensure
\begin{equation}
	\eta_{1}^\mathrm{SC} = \frac{1}{\sqrt{3}} \cdot \theta_{1}^\mathrm{SC}. \label{eq:eta_13_relation}
\end{equation}

\begin{figure}[ht!]
	\centering
	\includegraphics[width=.45\textwidth]{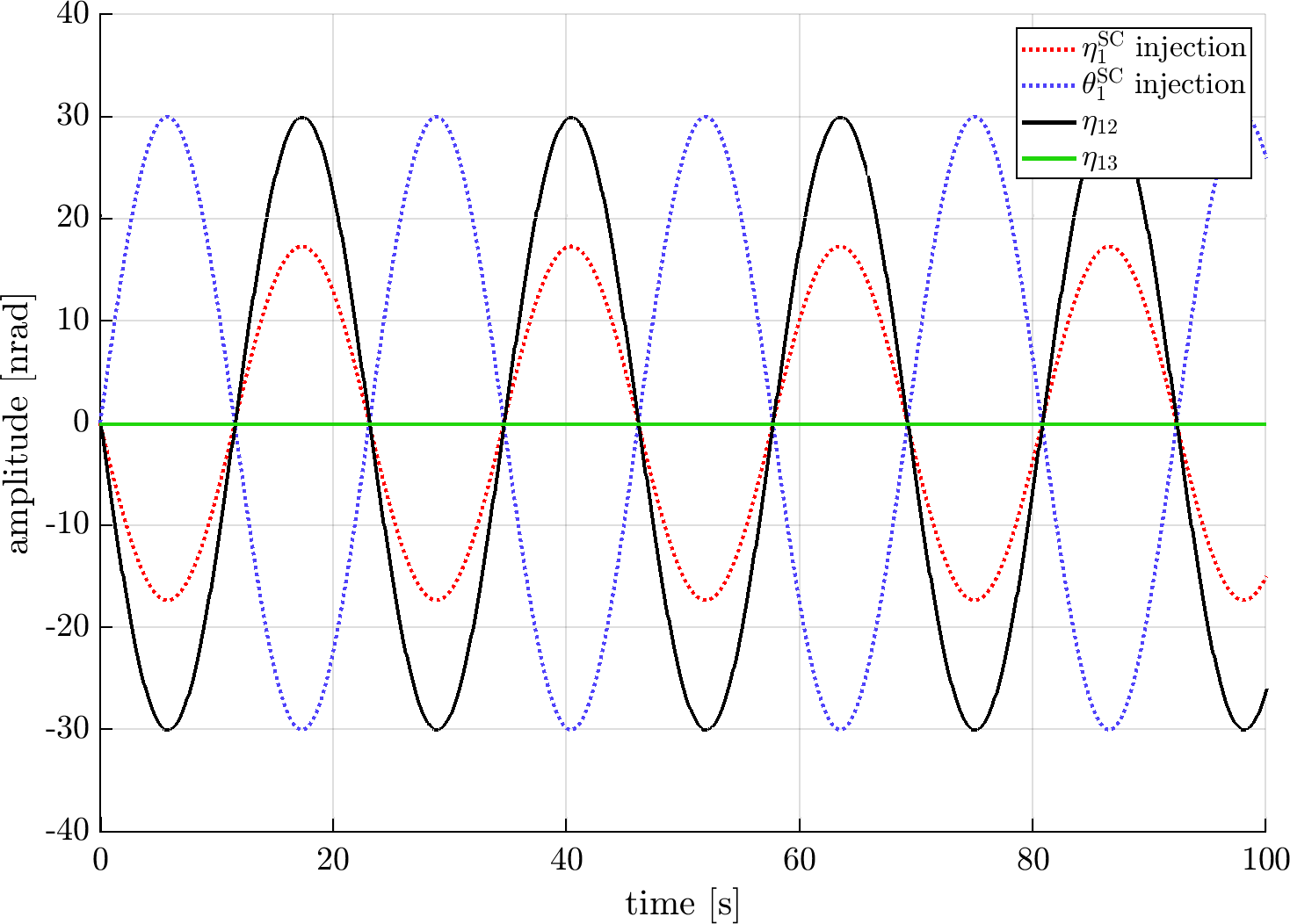}
	 \caption{Example: injections into the \ac{SC} angles $\eta_{1}^\mathrm{SC}$ and $\theta_{1}^\mathrm{SC}$ result in the depicted $\eta_{12}$ and $\eta_{13}$ angles, cf. Eq.~\eqref{eq:eta_ij_1}. The injections were chosen such that $\eta_{13} = 0$.}
 \label{fig:sc_inj_ex}
 \end{figure}

Figure~\ref{fig:sc_inj_ex} shows examplary sinusoidal injection signals for the \ac{SC} angles $\eta_{12}^\mathrm{SC}$ and $\theta_{12}^\mathrm{SC}$, as well as the $\eta_{12}$ and $\eta_{13}$ angles that result from these injections, cf. Eq.~\eqref{eq:eta_ij_2}. Since the injections were chosen such that Eq.~\eqref{eq:eta_12_relation} is fulfilled, merely $\eta_{12}$ is excited, while $\eta_{13} = 0$.

\subsubsection{Injection into MOSA Angles\label{sec:inj_mosa}}

The \acp{MOSA} will be controllable with use of the \ac{OATM}. The \ac{OATM} specifics are not finalized yet, however, one option is to use piezo electric actuators, providing stepwise actuation of each \ac{MOSA} in the yaw degree of freedom, i.e. around the $z$ axis. Based on internal discussions, we assume here a maximum tracking speed of \SI{5.5}{\nano\radian\per\s} and a motion step resolution of \SI{1}{\nano\radian} \cite{ESA-SCI-F-ESTEC-SOW-2019-028}.

\begin{figure}[ht!]
   \centering
   \includegraphics[width=.45\textwidth]{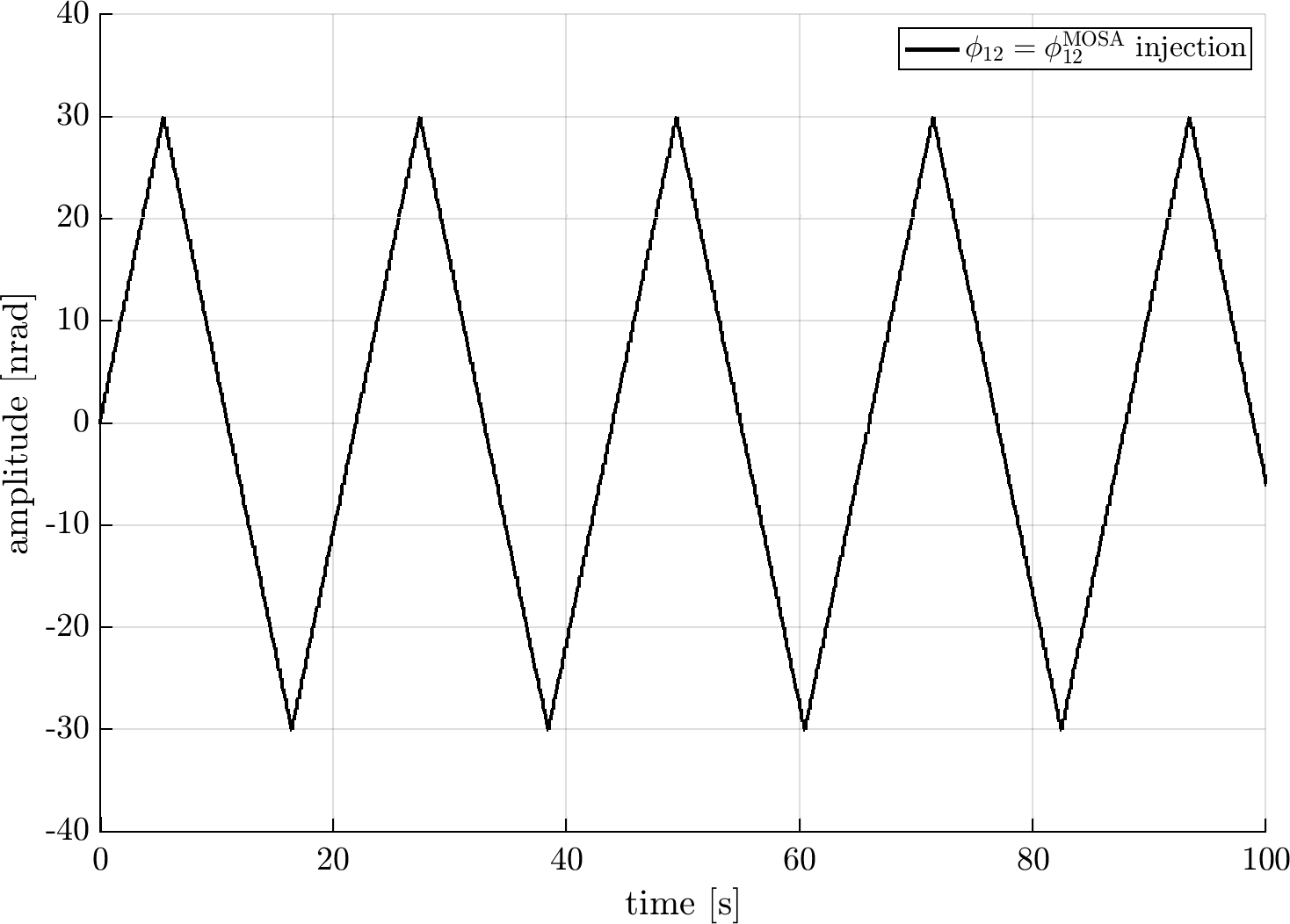}
	\caption{Example: injections directly into the \ac{MOSA} angle $\phi_{12}^\mathrm{MOSA}$ result in the depicted $\phi_{12}$ angle, which is identical to the injection, cf. Eq.~\eqref{eq:phi_ij}.}
\label{fig:mosa_inj_ex}
\end{figure}

Under these assumptions, a stair-like periodic signal with a triangular shape as shown examplarily in Fig.~\ref{fig:mosa_inj_ex} can be considered feasible. We aim at an angular excitation of \SI{30}{\nano\radian} since we used this amplitude here for the maneuvers via \ac{SC} rotation. Due to the maximum tracking speed of \SI{5.5}{\nano\radian\per\s}, the lowest achievable period of such an injection is about $21.8$ seconds, which corresponds to a maneuver frequency of about \SI{45.9}{\milli\Hz}.

Higher frequencies are not feasible if the amplitude shall not be less than \SI{30}{\nano\radian}. For simplicity, we consider motion steps of \SI{1}{\nano\radian}, i.e. $30$ steps are required to excite the yaw angle by \SI{30}{\nano\radian}. For instance, a possible injection signal comprising $25$ cycles within a maneuver duration of $600$ seconds would have a period of $24$ seconds. This would make a signal similar to a sine wave with a frequency of $41.\bar 6$\,\si{\milli\Hz}, i.e. similar to the injection signals that we assume for the \ac{SC} maneuvers.

Lower frequencies (longer periods) are achievable by slightly delaying each motion step. This way, by using different periods, different uncorrelated injection signals can be created, in analogy to using different frequencies for the sinusoidal \ac{SC} injections. This will allow for the performance of several \ac{MOSA} $\phi$ maneuvers simultaneously, cf. Sec.~\ref{sec:more_simultaneous} below.

\subsubsection{Imperfect Injections}
It is important to note that, in this document, it is not investigated in detail how the maneuvers proposed here can be implemented in the real mission. Here the \ac{DFACS} is assumed to be able to apply the necessary control torques to follow a prescribed commanded set point sequence. Consequently, imperfect injections are not considered here and are subject to further investigations. However, note also that slight deviations from the commanded sequence would not be crucial since the \ac{TTL} error is depending on the actual $\eta_{ij}$ and $\phi_{ij}$ angles, which are measured by the \ac{DWS} and used for the fit.

\subsection{Simultaneous Maneuver Performance\label{sec:simultaneous}}

\begin{table*}[!ht]
	\caption{Characterization of \ac{TTL} contributions in the \ac{TDI} variables. For each \ac{TTL} contribution, the variable in which Rx and Tx parts appear uncorrelated is marked in green. In the variables marked in red, Rx and Tx parts appear correlated, while both parts are zero in the variables marked in blue. From this pattern, uncorrelated \ac{TTL} pairs like $(\eta_{12},\eta_{32})$ can be deduced, for which maneuvers can be performed simultaneously without loss of estimation accuracy.}
	
\centering
\begin{tabularx}{.625\linewidth}{ c | c | c | c | c }

	~ & ~ & \multicolumn{3}{c}{appearance of Rx and Tx \ac{TTL} in \ac{TDI}} \\
	\# Rx/Tx $\eta$ ($\phi$) & angle & $X$ & $Y$ & $Z$ \\ \hline \hline

	1/13 (7/19) & $\eta_{12}$ ($\phi_{12}$) & \color{ForestGreen}{\bf uncorrelated} & \color{red}{\bf identical} & \color{blue}{\bf zero} \\
	2/14 (8/20) & $\eta_{23}$ ($\phi_{23}$) & \color{blue}{\bf zero} & \color{ForestGreen}{\bf uncorrelated} & \color{red}{\bf identical} \\
	3/15 (9/21) & $\eta_{31}$ ($\phi_{31}$) & \color{red}{\bf identical} & \color{blue}{\bf zero} & \color{ForestGreen}{\bf uncorrelated} \\
	4/16 (10/22) & $\eta_{13}$ ($\phi_{13}$) & \color{ForestGreen}{\bf uncorrelated} & \color{blue}{\bf zero} & \color{red}{\bf identical} \\
	5/17 (11/23) & $\eta_{32}$ ($\phi_{32}$) & \color{blue}{\bf zero} & \color{red}{\bf identical} & \color{ForestGreen}{\bf uncorrelated} \\
	6/18 (12/24) & $\eta_{21}$ ($\phi_{21}$) & \color{red}{\bf identical} & \color{ForestGreen}{\bf uncorrelated} & \color{blue}{\bf zero}

\end{tabularx}

\label{tab:pairs}
\end{table*}

\subsubsection{Uncorrelated Pairs}

When identical maneuvers are performed in more than one angle simultaneously, in general it might not be possible to disentangle the respective \ac{TTL} coefficients. In this section we show that there exist pairs of \ac{TTL} contributions that are naturally uncorrelated, allowing accurate coefficient estimation even if both angles are excited identically. Such pairs can be found by making the following observations regarding the equations given in App.~\ref{a:ttl_in_tdi}. 

For any given angle, there is always one \ac{TDI} variable in which both Rx and Tx contributions of this angle are zero. In one of the two remaining \ac{TDI} variables the Rx and Tx contributions of this angle are perfectly correlated since both appear with identical delays, compare e.g. Eqs.~\eqref{eq:y_eta12_rx} and \eqref{eq:y_eta12_tx}. Thus, for each angle, the main information for coefficient estimation is contained in only one of the three \ac{TDI} variables $X,Y,Z$.

For instance, for $\eta_{12}$, both Rx and Tx contributions appear in \ac{TDI} $X$ (uncorrelated), both are zero in $Z$, and both appear in $Y$ (correlated), see Eq.~\eqref{eq:Y_12eta_corr}. For $\eta_{32}$, the main information is contained in $Z$, but it contributes no \ac{TTL} to $X$. If $\eta_{12} = \eta_{32}$ for some period such as a maneuver, then each does not disturb the coefficient estimation for the other. In fact, all six $\eta$ angles can be divided this way into three pairs, and analogously for $\phi$. This fact is also shown in \cite{wanner_2024}.

Table~\ref{tab:pairs} shows this pattern for all $\eta$ and $\phi$ angles. This allows to identify pairs of angles for which \ac{TTL} calibration maneuvers can be performed simultaneously and with identical profile without loss of estimation accuracy: $(\eta_{12},\eta_{32})$, $(\eta_{23},\eta_{13})$, $(\eta_{31},\eta_{21})$, and $(\phi_{12},\phi_{32})$, $(\phi_{23},\phi_{13})$, $(\phi_{31},\phi_{21})$. Note that the same holds for all mixed pairs such as $(\eta_{12},\phi_{32})$, etc.

\subsubsection{More than one Pair Simultaneously\label{sec:more_simultaneous}}

The potential drawback of multiple simultaneous maneuvers is that the resulting \ac{TTL} contributions may be highly correlated, which would result in large estimation uncertainties, cf. App.~\ref{c:std}. At this point a useful observation is the fact that any two sines having different frequencies are uncorrelated, if both complete an integer number of cycles in the considered time span. Together with the naturally uncorrelated pairs that we found in the previous section, this allows to perform maneuvers for six angles at the same time by using three different frequencies. Hence, it is possible to cover all $12$ angles by performing two times six simultaneous maneuvers instead of $12$ maneuvers in a row. Note that, although not pursued further in this study, two signals with the same frequency but with a relative phase shift of $\pi/2$ would be another example of two uncorrelated signals.

\begin{figure}[ht!]
	\centering
	\includegraphics[width=.45\textwidth]{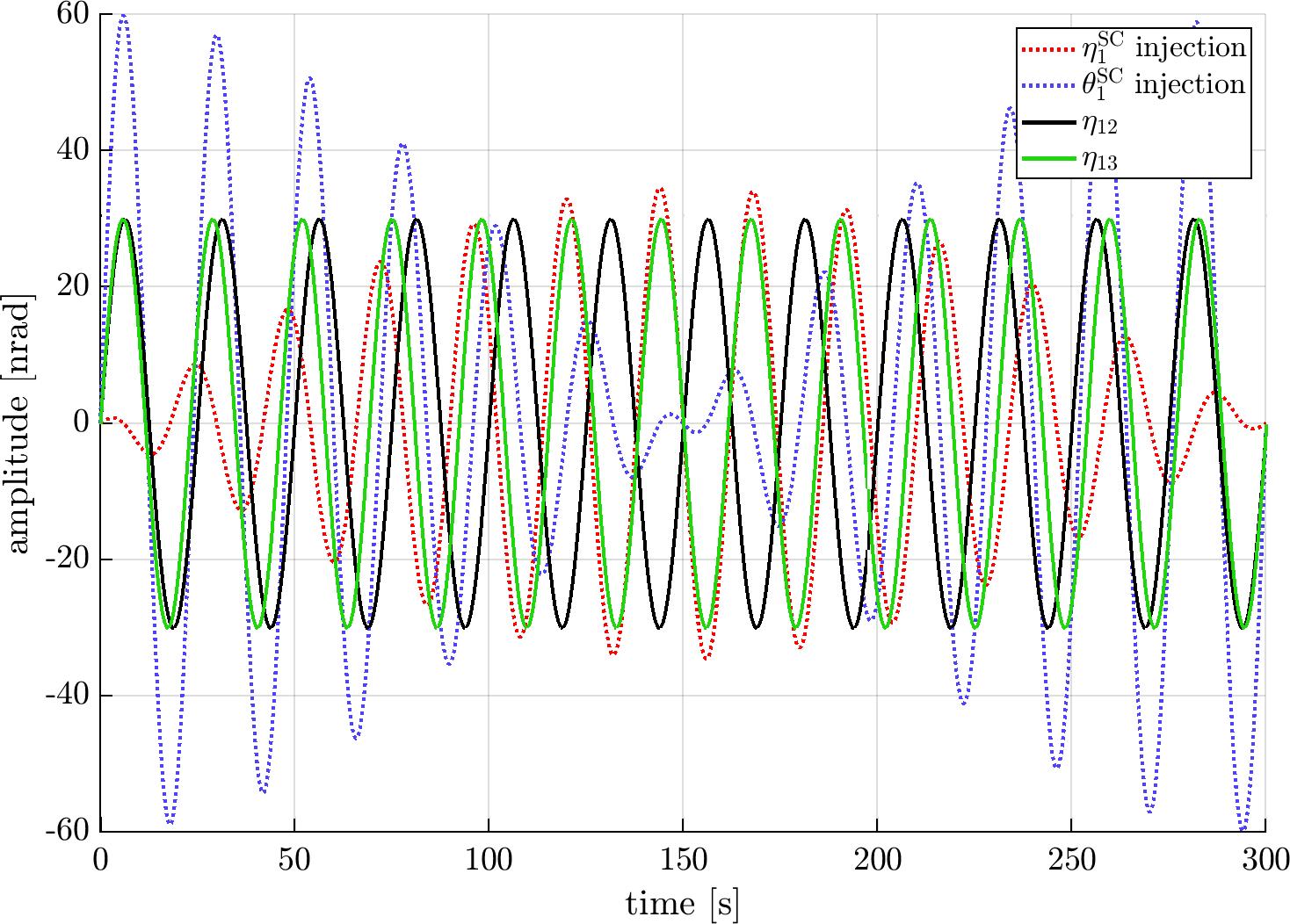}
	\caption{Example: simultaneous $\eta_{12}$ and $\eta_{13}$ maneuvers using two sinusoidal injections with different frequencies, $f_1~=$~\SI{40}{\milli\Hz} and $f_2~=~43.\bar 3$\,\si{\milli\Hz}.}
\label{fig:simultaneous_eta_man}
\end{figure}

One option would be to perform all six $\phi$ maneuvers at first, using a different maneuver frequency for each of the three pairs $(\phi_{12},\phi_{32})$, $(\phi_{23},\phi_{13})$, and $(\phi_{31},\phi_{21})$. For injections via \ac{SC} rotations this cannot be done, as will be explained in Sec.~\ref{sec:problem_phi}. For injections via \ac{MOSA} rotations it could be achieved by defining three signals as described in Sec.~\ref{sec:inj_mosa}, using three different periods. Subsequently all six $\eta$ maneuvers could be performed in the following way. Using Eq.~\eqref{eq:eta_13_relation}, we construct injections into $\theta_1^\mathrm{SC}$ and $\eta_1^\mathrm{SC}$ as
\begin{align}
	\theta_1^\mathrm{SC,12inj}(t) &= \sin(2 \pi f_1 t) \\
	\eta_1^\mathrm{SC,12inj}(t) &= -\frac{1}{\sqrt{3}} \sin(2 \pi f_1 t),
\end{align}
where the amplitude is set to $1$ for this illustration of principle. Using Eq.~\eqref{eq:eta_ij_2}, we find that the total injected \ac{MOSA} $\eta$ angles relative to the incident beam are
\begin{align}
	\eta_{12}^\mathrm{12inj} &= \sin(2 \pi f_1 t) \\
	\eta_{13}^\mathrm{12inj} &= 0.
\end{align}
Because relation \eqref{eq:eta_12_relation} is satisfied, such \ac{SC} rotations do not excite $\eta_{13}$. Additionally, another rotation can be constructed which excites $\eta_{13}$, but not $\eta_{12}$, by
\begin{align}
	\theta_1^\mathrm{SC,13inj}(t) &= \sin(2 \pi f_2 t) \\
	\eta_1^\mathrm{SC,13inj}(t) &= \frac{1}{\sqrt{3}} \sin(2 \pi f_2 t),
\end{align}
using a second frequency $f_2$, resulting in the following \ac{MOSA} injection angles:
\begin{align}
	\eta_{12}^\mathrm{13inj} &= 0 \\
	\eta_{13}^\mathrm{13inj} &= \sin(2 \pi f_2 t).
\end{align}
The two rotations can be added up to obtain simultaneous uncorrelated signals in the two \ac{MOSA} $\eta$ angles:
\begin{align}
	\eta_{12} &= \sin(2 \pi f_1 t) \\
	\eta_{13} &= \sin(2 \pi f_2 t).
\end{align}
An example of such an injection is shown in Fig.~\ref{fig:simultaneous_eta_man}. Clearly, this can be done analogously for \ac{SC}\,$2$ and \ac{SC}\,$3$. It can be seen that this type of injection in fact requires exciting $\theta_1^\mathrm{SC}$ to amplitudes larger than \SI{30}{\nano\radian}, which we excluded above. Therefore we consider another option in the following.

\begin{figure}[ht!]
	\centering
	\includegraphics[width=.45\textwidth]{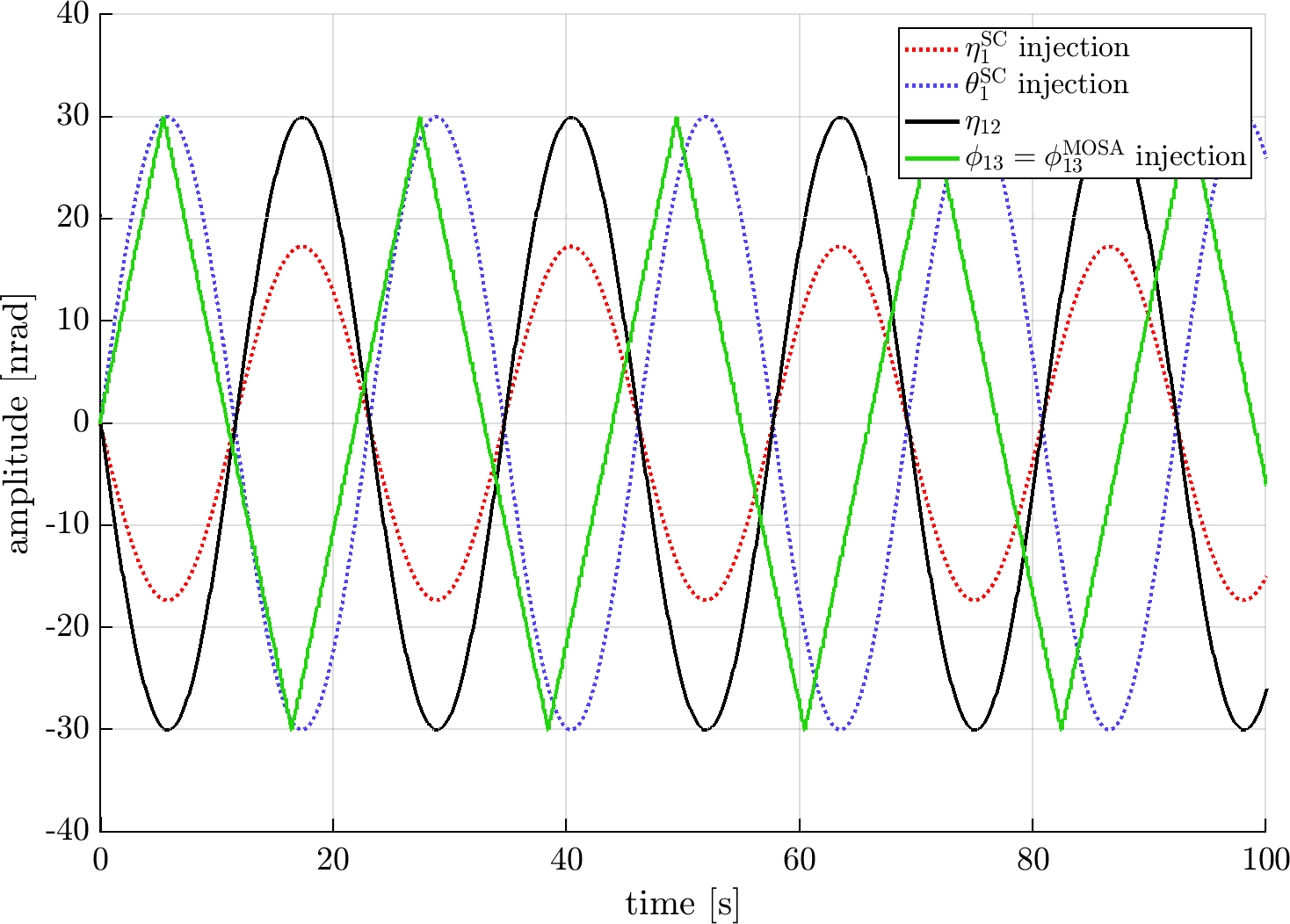}
	\caption{Example: sinusoidal $\eta_{12}$ \ac{SC} maneuver with $f_1~=~43.\bar 3$\,\si{\milli\Hz}, and simultaneous $\phi_{13}$ \ac{MOSA} maneuver with triangular shape and $f_2~=$~\SI{45}{\milli\Hz}, respectively.}
\label{fig:simultaneous_optC}
\end{figure}

Another way to implement two times six maneuvers to cover all $12$ angles is to include mixed uncorrelated pairs. I.e., it is possible to rotate one of the \acp{MOSA} in $\eta$ via \ac{SC} injection, while the other \ac{MOSA} on the same \ac{SC} is rotated in $\phi$ via \ac{MOSA} injection. In the scenario which we will call \textbf{case A}, we chose to excite $(\eta_{12},\eta_{32})$ with a frequency $f_1$, $(\eta_{21},\phi_{31})$ with $f_2$, and $(\phi_{13},\phi_{23})$ with $f_3$. Subsequently the pairs $(\phi_{12},\phi_{32})$ $(\phi_{21},\eta_{31})$, and $(\eta_{13},\eta_{23})$ were excited using the same three frequencies $f_1,f_2,f_3$, respectively. Note that this is one of many possible configurations. Figure~\ref{fig:simultaneous_optC} examplarily shows simultaneous injections for \ac{SC}\,$1$.

\subsection{Issue with SC Phi Injections\label{sec:problem_phi}}

The procedure to excite one \ac{MOSA}'s $\eta$ angle via \ac{SC} rotation, while the other \ac{MOSA}'s $\eta$ angle is unaffected, as described in Sec.~\ref{sec:inj_sc} and depicted in Fig.\ref{fig:sc_inj_ex}, is not feasible for the $\phi$ angles. In this section we discuss the impact of this fact on the \ac{TTL} calibration via injection into \ac{SC} angles.

\subsubsection{Problem Statement}

Equation \eqref{eq:phi_ij} shows that any rotation of \ac{SC}\,$i$ in the yaw angle $\phi_i^\mathrm{SC}$ necessarily causes identical rotations of $\phi_{ij}$ and $\phi_{ik}$, $j \neq k$. Exciting $\phi_i^\mathrm{SC}$ thereby stimulates four \ac{TTL} contributions, i.e. the Rx and Tx contributions from the two local \acp{MOSA}' $\phi$ angles. Taking $\phi_1^\mathrm{SC}$ as an example, this means that the coefficients $C_{12 \phi Rx}$, $C_{12 \phi Tx}$, $C_{13 \phi Rx}$, and $C_{13 \phi Tx}$ are involved. The \ac{TTL} contributions associated with these coefficients can be obtained from Eqs.~\eqref{eq:x_eta12_rx}, \eqref{eq:y_eta12_rx}, \eqref{eq:x_eta13_rx}, \eqref{eq:z_eta13_rx}, \eqref{eq:x_eta12_tx}, \eqref{eq:y_eta12_tx}, \eqref{eq:x_eta13_tx}, and \eqref{eq:z_eta13_tx} by replacing all $\eta$'s with $\phi$'s. Since we consider calibration signals injected into $\phi_1^\mathrm{SC}$, we assume for the moment $\phi_{12}~\approx~\phi_1^\mathrm{SC}~\approx~\phi_{13}$, neglecting the differential \ac{MOSA} jitter in $\phi_{12}^\mathrm{MOSA}$ and $\phi_{13}^\mathrm{MOSA}$, only considering common \ac{MOSA} motion caused by \ac{SC} rotation.

The terms in the \ac{TDI} $X$ variable depending on $C_{12 \phi Rx}$ and $C_{13 \phi Rx}$ differ only due to different armlengths, and similarly for $C_{12 \phi Tx}$ and $C_{13 \phi Tx}$. If one assumes almost equal armlengths, one can approximate the sum of \ac{TTL} contributions in \ac{TDI} $X$ due to $\phi_1^\mathrm{SC}$ as
\begin{align}
	X_{\phi_1^\mathrm{SC}}^\mathrm{TTL} &\approx [ (C_{13 \phi Rx} - C_{12 \phi Rx}) \cdot (1 - \mathcal{D}^2 - \mathcal{D}^4 + \mathcal{D}^6) \\
	&+ (C_{13 \phi Tx} - C_{12 \phi Tx}) \cdot (\mathcal{D}^2 - \mathcal{D}^4 - \mathcal{D}^6 + \mathcal{D}^8) ] \phi_1^\mathrm{SC}, \nonumber
\end{align}
where $\mathcal{D}$ is a one-arm delay, as in approximations \eqref{eq:approx_12etarx} and \eqref{eq:approx_12etatx}. Then $X$ allows only to estimate the differences $(C_{13 \phi Rx} - C_{12 \phi Rx})$ and $(C_{13 \phi Tx} - C_{12 \phi Tx})$.

In Sec.~\ref{sec:f_man} we have seen that the Rx and Tx \ac{TTL} contributions of $\phi_{12}$ in $Y$ are perfectly correlated, cf. Eq.~\eqref{eq:Y_12eta_corr}. The same is true for the contributions of $\phi_{13}$ in $Z$. This implies that, from the $Y$ and $Z$ variables, one can only estimate the combinations $(C_{12 \phi Rx} + C_{12 \phi Tx})$ and $(C_{13 \phi Rx} + C_{13 \phi Tx})$, but not the individual coefficients.

In total, we can estimate the right-hand side $\tilde C$ of
\begin{equation} \label{eq:c_comb}
	\begin{pmatrix} -1 & 1 & 0 & 0 \\ 0 & 0 & -1 & 1 \\ 1 & 0 & 1 & 0 \\ 0 & 1 & 0 & 1 \end{pmatrix} \cdot \begin{pmatrix} C_{12 \phi Rx} \\ C_{13 \phi Rx} \\ C_{12 \phi Tx} \\ C_{13 \phi Tx} \end{pmatrix} = \tilde C.
\end{equation}
Since the matrix on the left-hand side has rank 3, it is not possible to determine the 4 individual coefficients from the mere knowledge of $\tilde C$.

\subsubsection{Correlation with Unequal Armlengths}

Above we have seen that \ac{SC} $\phi$ maneuvers do not allow the estimation of the individual $\phi$ \ac{TTL} coefficients, assuming that the three \ac{LISA} arms have equal length. Here we examine if this conclusion still holds when we consider unequal arm lengths. For the duration of a maneuver, assume that $\phi_1^\mathrm{SC}$ is dominated by the sinusoidal injection signal, i.e.
\begin{equation}
	\phi_1^\mathrm{SC} = \sin(\omega t) = \sin(2 \pi f_\mathrm{man} t).
\end{equation}
The crucial quantity is the correlation between Eqs.~\eqref{eq:x_eta12_rx} and \eqref{eq:x_eta13_rx} since the perfect correlations in $Y$ and $Z$ hold independent of the armlengths. For simplicity we assume static armlengths for a short period such as the maneuver time, so that the delay operators are commutative. Then one can show that
\begin{align*}
	\left\vert \mathrm{corr}(X_{12 \phi Rx}^\mathrm{TTL}, X_{13 \phi Rx}^\mathrm{TTL}) \right\vert &\ge \left\vert \mathrm{corr} \left(D_{121} \phi_1^\mathrm{SC}, D_{131} \phi_1^\mathrm{SC} \right) \right\vert \\
	&= \left\vert \mathrm{corr} \left( \sin(\omega t),\sin(\omega (t-\delta_t)) \right) \right\vert \\
	&= \left\vert \cos\left(\omega \delta_t \right) \right\vert,
\end{align*}
where $\delta_t$ denotes the differential time between the delays $D_{131}$ and $D_{121}$. The correlation is $1$ for $\omega \delta_t = 0$ and $0$ for $\omega \delta_t = \pi/2$. If, for instance, $f_\mathrm{man}~=$~\SI{40}{\milli\Hz}, the correlation is $0$ for $\delta_t~=$~\SI{6.25}{\s}, which will never occur. For $\delta_t~\approx$~\SI{0.167}{\s}, which corresponds to a $1$\,\% armlength difference, we have
\begin{equation}
	\left\vert \mathrm{corr} \left(X_{12 \phi Rx}^\mathrm{TTL}, X_{13 \phi Rx}^\mathrm{TTL} \right) \right\vert > 0.999.
\end{equation}
That is, for the purpose of \ac{TTL} maneuver injections into the $\phi_1^\mathrm{SC}$ angle, the terms $X_{12 \phi Rx}^\mathrm{TTL}$ and $X_{13 \phi Rx}^\mathrm{TTL}$, as well as the terms $X_{12 \phi Tx}^\mathrm{TTL}$ and $X_{13 \phi Tx}^\mathrm{TTL}$, can be considered almost perfectly correlated. The same holds for \ac{SC}\,$2$ and \ac{SC}\,$3$ as well. If the armlength difference was $5$\,\% instead of $1$\,\%, the respective absolute correlation would still be larger than $0.98$. This shows that \ac{SC} maneuvers in $\phi$ are impractical for the estimation of \ac{TTL} coefficients.

\subsubsection{Workaround}

A potential way to circumvent the problem discussed above is to utilize \ac{MOSA} maneuvers via the \ac{OATM}. In Sec.~\ref{sec:inj_mosa} above, it is described what \ac{MOSA} maneuvers are feasible and how different uncorrelated injection signals can be obtained. The results of a full simulation, comprising \ac{SC} maneuvers for the $\eta$ angles and \ac{MOSA} maneuvers for the $\phi$ angles, are presented in Sec.~\ref{sec:fullsim}.

\subsubsection{Subtraction without Knowledge of Individual Coefficients}

Knowledge of the coefficient combinations defined by the left hand side of Eq.~\ref{eq:c_comb} is sufficient in order to subtract \ac{TTL} from the \ac{TDI} variables. This was confirmed with simulations and the results are presented in Sec.~\ref{sec:sc_man}.

\section{Simulation Results\label{sec:results}}

\subsection{Full Simulation with Maneuvers (case A)}
\label{sec:fullsim}

\begin{figure*}[ht!]
   \centering
   \includegraphics[width=\textwidth]{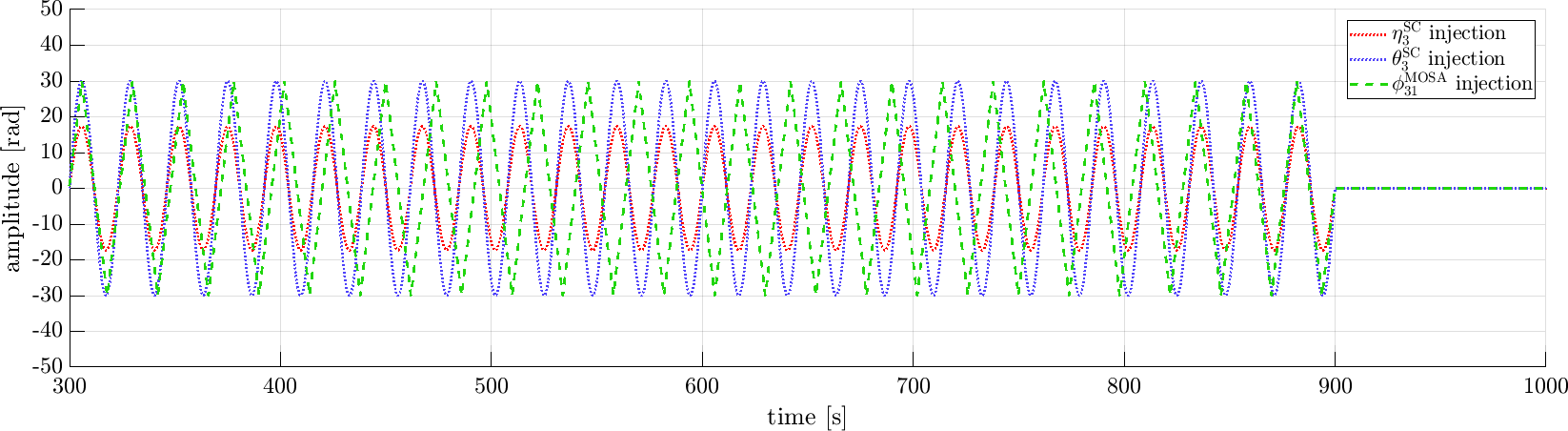}
   \includegraphics[width=\textwidth]{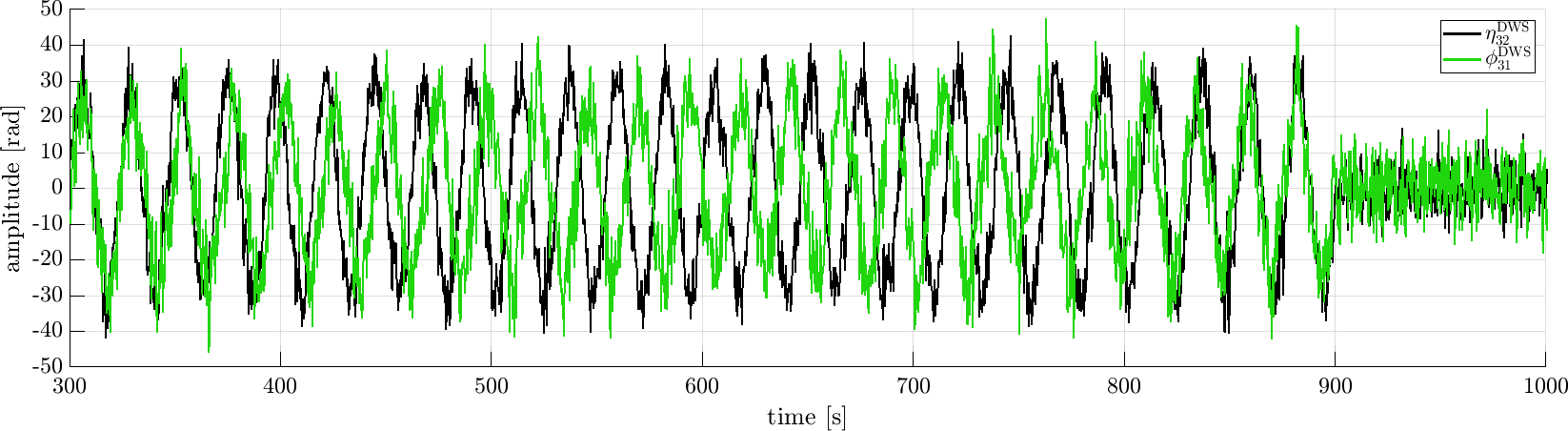}
   \includegraphics[width=\textwidth]{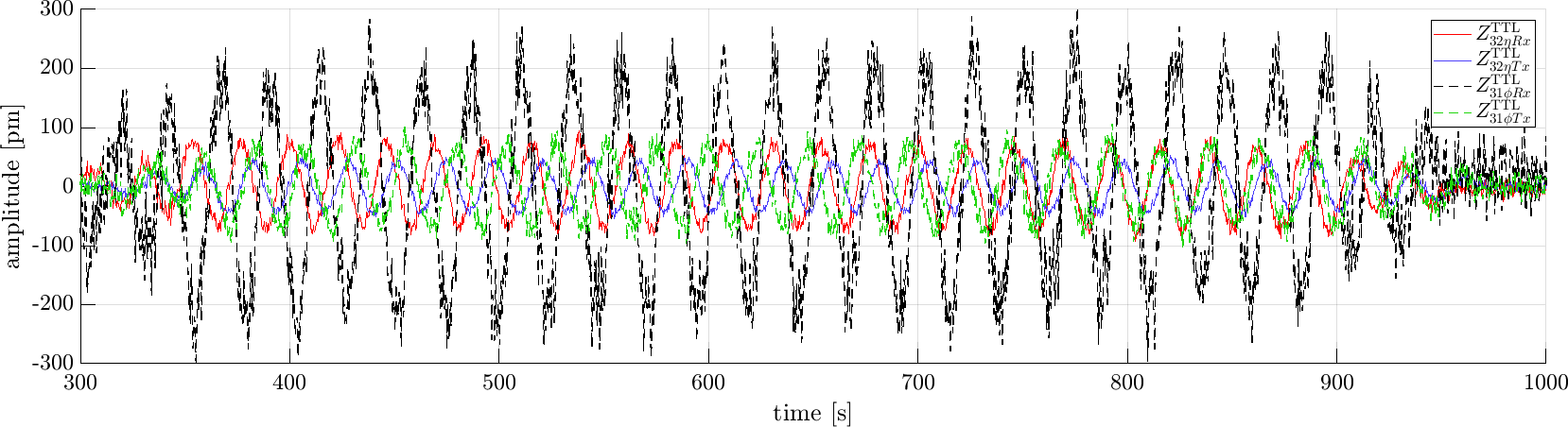}
	\caption{$\eta_{32}$ and $\phi_{31}$ maneuvers from the full simulation. Top: angle injections; Middle: simulated \ac{DWS} angles; Bottom: resulting \ac{TTL} contributions in \ac{TDI} $Z$.}
\label{fig:phi1_mans}
\end{figure*}

We performed a full simulation, called \textbf{case A}, comprising maneuvers for all angles. The simulation was performed in Matlab with the LISASim simulator \cite{lisasim}. Noise and jitter levels were chosen according to Sec.~\ref{sec:simulation}. The full \ac{MOSA} jitter was applied: \SI{1}{\nano\radian\per\sqrt\Hz} for $\eta$ and \SI{5}{\nano\radian\per\sqrt\Hz} for $\phi$. The true \ac{TTL} coupling coefficients were uniformly distributed random values between \SI{-3}{\milli\m\per\radian} and \SI{3}{\milli\m\per\radian}. The estimated coefficients and their uncertainties were obtained by a \ac{LSQ} fit in the time domain, as described in Sec.~\ref{sec:pe}. In total \SI{2000}{\s} of data were simulated, while the first and last \SI{300}{\s} were cut off to remove potential filter effects, so the integration time for estimation was $T~=$~\SI{1400}{\s}.

We defined two sets of maneuvers, where we used mixed uncorrelated pairs and employed three different frequencies. This approach was described in Sec.~\ref{sec:more_simultaneous}. The modulations in $\eta$ were sinusoidal injections via \ac{SC} angles, cf. Sec.~\ref{sec:inj_sc}. For $\phi$ we used triangular-shaped signals injected into the \ac{MOSA} yaw angles, cf. Sec.~\ref{sec:inj_mosa}. In the first set, $(\eta_{12},\eta_{32})$ were excited at a frequency $f_1~=~43.\bar 3$\,\si{\milli\Hz}, $(\eta_{21},\phi_{31})$ at $f_2~=~41.\bar 6$\,\si{\milli\Hz}, and $(\phi_{13},\phi_{23})$ at $f_3~\approx~44.7$\,\si{\milli\Hz}. The second set comprised maneuvers in $(\phi_{12},\phi_{32})$ at $f_3$, $(\phi_{21},\eta_{31})$ at $f_2$, and $(\eta_{13},\eta_{23})$ at $f_1$.

The first set of maneuvers started at $t_0~=$~\SI{300}{\s}, the second set started at $t_0~=$~\SI{1000}{\s}, each set comprising six simultaneous maneuvers with a duration of $T_\mathrm{man}~=$~\SI{600}{\s}. \SI{100}{\s} after each set of maneuvers was used additionally for the estimation to capture the entire effect of the maneuvers in the \ac{TDI} variables, including delayed terms. Thus, the total integration time for estimation was $T~=$~\SI{1400}{\s}. All maneuvers had an injection amplitude of $A_\mathrm{man}~=$~\SI{30}{\nano\radian}. Table~\ref{tab:man_param} lists all maneuver parameters including the maneuver frequencies.

\begin{table}[!ht]
	\caption{Maneuver parameters for the full simulation}
	
	\begin{tabularx}{\linewidth}{ c | c | c | c | c | c | c }

		set & angle & $t_0$ [\si{\s}] & $T_\mathrm{man}$ [\si{\s}] & $A_\mathrm{man}$ [\si{\nano\radian}] & $f_\mathrm{man}$ [\si{\milli\Hz}] & P [\si{\s}] \\ \hline \hline

		$1$ & $\eta_{12}$ & $300$ & $600$ & $30$ & $43.\bar 3$ & $23.08$ \\
		$1$ & $\eta_{32}$ & $300$ & $600$ & $30$ & $43.\bar 3$ & $23.08$ \\
		$1$ & $\eta_{21}$ & $300$ & $600$ & $30$ & $41.\bar 6$ & $24$ \\
		$1$ & $\phi_{31}$ & $300$ & $600$ & $30$ & $41.\bar 6$ & $24$ \\
		$1$ & $\phi_{13}$ & $300$ & $604$ & $30$ & $44.7$ & $22.36$ \\
		$1$ & $\phi_{23}$ & $300$ & $604$ & $30$ & $44.7$ & $22.36$ \\

		$2$ & $\phi_{12}$ & $1000$ & $604$ & $30$ & $44.7$ & $22.36$ \\
		$2$ & $\phi_{32}$ & $1000$ & $604$ & $30$ & $44.7$ & $22.36$ \\
		$2$ & $\phi_{21}$ & $1000$ & $600$ & $30$ & $41.\bar 6$ & $24$ \\
		$2$ & $\eta_{31}$ & $1000$ & $600$ & $30$ & $41.\bar 6$ & $24$ \\
		$2$ & $\eta_{13}$ & $1000$ & $600$ & $30$ & $43.\bar 3$ & $23.08$ \\
		$2$ & $\eta_{23}$ & $1000$ & $600$ & $30$ & $43.\bar 3$ & $23.08$ \\

	\end{tabularx}
	
	\label{tab:man_param}
\end{table}

Figure~\ref{fig:phi1_mans} illustrates the $\eta_{32}$ and $\phi_{31}$ maneuvers, with maneuver parameters as specified in Tab.~\ref{tab:man_param}. The top plot shows the injections into $\eta_3^\mathrm{SC}$ and $\theta_3^\mathrm{SC}$ to excite $\eta_{32}$ and into $\phi_{31}^\mathrm{MOSA}$ to excite $\phi_{31}$. The \ac{DWS} measurements of the excited angles are displayed in the middle plot, where it can be seen that the correlation between $\eta_{32}$ and $\phi_{31}$ is minimal due to the two different frequencies. The bottom plot shows the four \ac{TTL} contributions which result from these angular injections, i.e. $Z_{32 \eta Rx}^\mathrm{TTL}$, $Z_{32 \eta Tx}^\mathrm{TTL}$, $Z_{31 \phi Rx}^\mathrm{TTL}$, and $Z_{31 \phi Tx}^\mathrm{TTL}$. The amplitudes of the \ac{TTL} terms are not equal as they depend on the coupling coefficients, which were $\{794.2,-469.4,2745.0,934.4\}$\,\si{\micro\m\per\radian}, respectively.

In this \textbf{case A} simulation, with full \ac{MOSA} jitter and a complete set of maneuvers, we obtained uncertainties $\sigma_\mathrm{LSQ}(\hat C_{ij\alpha\beta})$ between \SI{10.2}{\micro\m\per\radian} and \SI{13.1}{\micro\m\per\radian}. The detailed results will be discussed in the next section and can be found in Tab.~\ref{tab:fullsim_coeffs}. We performed two more simulations excluding maneuvers, called \textbf{case B} and \textbf{case C}, in order to assess the potential of maneuvers by comparing the cases. These simulations are described in the following section, and the detailed estimation results of all three cases are stated and interpreted.

\subsection{Simulations without Maneuvers (cases B and C)\label{sec:man_noman}}

For comparison, we performed two additional simulations with the same integration time of \SI{1400}{\s}, but without using maneuvers, so that the information on the coefficients came exclusively from the angular \ac{SC} and \ac{MOSA} jitters. In one of these simulations, called \textbf{case B}, the full \ac{MOSA} jitter was applied: \SI{1}{\nano\radian\per\sqrt\Hz} for $\eta$ and \SI{5}{\nano\radian\per\sqrt\Hz} for $\phi$. For the other simulation, called \textbf{case C}, the \ac{MOSA} jitter was set to the reduced levels of \SI{0.1}{\nano\radian\per\sqrt\Hz} for $\eta$ and \SI{0.5}{\nano\radian\per\sqrt\Hz} for $\phi$. As mentioned in Sec.~\ref{sec:simulation}, the full jitter levels were chosen according to the performance model \cite{lisa_perf_model} (version of 2021), while the reduced levels describe an alternative scenario that is worth considering. In both cases the \ac{SC} angular jitter levels of \SI{5}{\nano\radian\per\sqrt\Hz} were used, see Tab.~\ref{tab:LISASim_spec}.

\begin{table*}[!ht]
\caption{For all cases: true \ac{TTL} coefficients $C_{ij\alpha\beta}$. For \textbf{case A} (including maneuvers): estimated coefficients $\hat C_{ij\alpha\beta}$, estimation errors $\Delta_{ij\alpha\beta} = \hat C_{ij\alpha\beta} - C_{ij\alpha\beta}$, as well as \ac{LSQ} uncertainties $\sigma_\mathrm{LSQ}(C_{ij\alpha\beta})$. For \textbf{case B} (no maneuvers, full jitter) and \textbf{case C} (no maneuvers, reduced \ac{MOSA} jitter): \ac{LSQ} uncertainties $\sigma_\mathrm{LSQ}(C_{ij\alpha\beta})$. All cases are using the same simulator settings and the same integration time $T~=$~\SI{1400}{\s}. All values are in \si{\micro\m\per\radian}.}

\centering
\begin{tabularx}{.825\linewidth}{ c | c | c | c | c | c | c | c }

coefficient \# & indices $ij\,\alpha\,\beta$ & $C_{ij\alpha\beta}$ & $\hat C_{ij\alpha\beta}$ & $\Delta_{ij\alpha\beta}$ & $\sigma_\mathrm{LSQ}(\hat C_{ij\alpha\beta})$ & $\sigma_\mathrm{LSQ}(\hat C_{ij\alpha\beta})$ & $\sigma_\mathrm{LSQ}(\hat C_{ij\alpha\beta})$ \\
~ & ~ & ~ & (\textbf{case A}) & (\textbf{case A}) & (\textbf{case A}) & (\textbf{case B}) & (\textbf{case C}) \\ \hline\hline

1 & $12\,\eta\,Rx$ & 1888.3 & 1855.9 & -32.4 & 10.8 & 40.5 & 46.1 \\
2 & $23\,\eta\,Rx$ & 2434.8 & 2421.7 & -13.1 & 10.8 & 42.1 & 47.8 \\
3 & $31\,\eta\,Rx$ & -2238.1 & -2253.2 & -15.1 & 10.3 & 43.0 & 46.6 \\
4 & $13\,\eta\,Rx$ & 2480.3 & 2475.9 & -4.3 & 10.8 & 41.5 & 45.1 \\
5 & $32\,\eta\,Rx$ & 794.2 & 811.0 & 16.9 & 11.0 & 42.8 & 48.3 \\
6 & $12\,\eta\,Rx$ & -2414.8 & -2404.1 & 10.7 & 10.4 & 43.0 & 47.4 \\ \hline

7 & $12\,\phi\,Rx$ & -1329.0 & -1352.4 & -23.4 & 13.0 & 30.1 & 218.0 \\
8 & $23\,\phi\,Rx$ & 281.3 & 265.3 & -16.0 & 12.8 & 30.9 & 223.0 \\
9 & $31\,\phi\,Rx$ & 2745.0 & 2748.4 & 3.4 & 12.1 & 29.4 & 206.9 \\
10 & $13\,\phi\,Rx$ & 2789.3 & 2804.3 & 14.9 & 12.9 & 30.2 & 214.1 \\
11 & $32\,\phi\,Rx$ & -2054.3 & -2060.9 & -6.6 & 12.7 & 29.9 & 205.5 \\
12 & $21\,\phi\,Rx$ & 2823.6 & 2824.6 & 1.0 & 12.1 & 30.4 & 223.1 \\ \hline

13 & $12\,\eta\,Tx$ & 2743.0 & 2747.4 & 4.4 & 10.9 & 40.6 & 46.1 \\
14 & $23\,\eta\,Tx$ & -87.7 & -81.4 & 6.4 & 10.7 & 42.0 & 47.9 \\
15 & $31\,\eta\,Tx$ & 1801.7 & 1787.9 & -13.8 & 10.3 & 43.1 & 45.7 \\
16 & $13\,\eta\,Tx$ & -2148.7 & -2126.3 & 22.4 & 10.9 & 41.4 & 45.3 \\
17 & $32\,\eta\,Tx$ & -469.4 & -453.5 & 15.9 & 10.8 & 42.8 & 47.8 \\
18 & $21\,\eta\,Tx$ & 2494.4 & 2490.6 & -3.8 & 10.4 & 43.1 & 47.4 \\ \hline

19 & $12\,\phi\,Tx$ & 1753.2 & 1748.9 & -4.4 & 12.9 & 30.1 & 216.4 \\
20 & $23\,\phi\,Tx$ & 2757.0 & 2785.7 & 28.8 & 12.9 & 30.7 & 223.0 \\
21 & $31\,\phi\,Tx$ & 934.4 & 928.9 & -5.6 & 12.1 & 29.3 & 206.6 \\
22 & $13\,\phi\,Tx$ & -2785.7 & -2799.2 & -13.5 & 12.9 & 30.3 & 216.0 \\
23 & $32\,\phi\,Tx$ & 2094.8 & 2087.1 & -7.7 & 12.8 & 30.0 & 205.6 \\
24 & $21\,\phi\,Tx$ & 2604.0 & 2606.8 & 2.9 & 12.0 & 30.5 & 223.1

\end{tabularx}
\label{tab:fullsim_coeffs}
\end{table*}

Detailed results for the three cases are listed in Tab.~\ref{tab:fullsim_coeffs}:
\begin{itemize}
	\item the true coefficients $C_{ij\alpha\beta}$
	\item \textbf{case A}: estimated coefficients $\hat C_{ij\alpha\beta}$
	\item \textbf{case A}: estimation errors $\Delta_{ij\alpha\beta}~=~\hat C_{ij\alpha\beta}~-~C_{ij\alpha\beta}$
	\item \textbf{cases A,B,C}: \ac{LSQ} uncertainties $\sigma_\mathrm{LSQ}\left( \hat C_{ij\alpha\beta} \right)$
\end{itemize}
In the full simulation inclunding maneuvers, \textbf{case A}, we obtained uncertainties $\sigma_\mathrm{LSQ}(\hat C_{ij\alpha\beta})$ between \SI{10.2}{\micro\m\per\radian} and \SI{11.0}{\micro\m\per\radian} for the $\eta$ coefficients, and between \SI{12.0}{\micro\m\per\radian} and \SI{13.1}{\micro\m\per\radian} for the $\phi$ coefficients. In \textbf{case B}, with the full jitter, we obtained uncertainties of about \SI{42}{\micro\m\per\radian} and \SI{30}{\micro\m\per\radian} for the $\eta$ and $\phi$ coefficients, respectively. In \textbf{case C}, with the reduced \ac{MOSA} jitter, we found larger uncertainties of about \SI{47}{\micro\m\per\radian} and \SI{215}{\micro\m\per\radian} for the $\eta$ and $\phi$ coefficients, respectively.

These results allow the following general interpretations. In the absence of maneuvers, the coefficient uncertainties are strongly depending on the jitter levels. In the case of $\eta$, both \ac{SC} and \ac{MOSA} jitter contribute information. Since the \ac{SC} $\eta$ jitter is larger than the \ac{MOSA} $\eta$ jitter in both cases, the reduction of \ac{MOSA} jitter has merely a small effect. However, for $\phi$, the information on the individual coefficients is mainly contained in the \ac{MOSA} jitter. This is because \ac{SC} $\phi$ jitter causes common angular motion of both \acp{MOSA}, as we recall from Sec.~\ref{sec:problem_phi}. This explains why reducing the \ac{MOSA} jitter has a much more significant impact on the estimation of $\phi$ coefficients. On the other hand, when maneuvers are performed, the coefficient uncertainties are mainly driven by the injection signals.

We would like to address the question of what can be gained by the utilization of maneuvers. So far, it has become clear that the amount of improvement depends on the level and spectral shapes of naturally occurring \ac{SC} and \ac{MOSA} jitter. We observe from Tab.~\ref{tab:fullsim_coeffs} that all coefficient uncertainties $\sigma_\mathrm{LSQ}(\hat C_{ij\alpha\beta})$ in \textbf{case A} are below \SI{15}{\micro\m\per\radian}. I.e., maneuvers with an amplitude of \SI{30}{\nano\radian} and a total integration time of \SI{1400}{\s} allow to determine all individual coefficients with an uncertainty of at most \SI{15}{\micro\m\per\radian}. Recall that each uncertainty is inversely proportional to the square root of the integration time $T$, and inversely proportional to the maneuver amplitude $A_\mathrm{man}$, cf. App.~\ref{c:std}. Hence, when using maneuvers, the achievable uncertainties can be related to $A_\mathrm{man}$ and $T$ by the following approximate inequality:
\begin{equation}
	\sigma_\mathrm{LSQ}(\hat C_{ij\alpha\beta}) \lessapprox \SI{15}{\micro\m\per\radian} \cdot \frac{\SI{30}{\nano\radian}}{A_\mathrm{man}} \cdot \sqrt{\frac{\SI{1400}{\s}}{T}}, \label{eq:sigma_ineq}
\end{equation}
for all $i,j,\alpha,\beta$.

From the relations given in App.~\ref{c:std}, we can also extrapolate how much integration time would be required to obtain comparable uncertainties without using maneuvers. In the scenario with full jitter, about \SI{3.5}{\hour} would be needed, limited by the uncertainties of the $\eta$ coefficients. In the alternative scenario with reduced \ac{MOSA} jitter, about \SI{90}{\hour} would be needed, in this case due to the large uncertainties of the $\phi$ coefficients. Note that this is a comparison of scenarios with maneuvers versus without maneuvers. The uncertainty of \SI{15}{\micro\m\per\radian} is not required, instead \SI{100}{\micro\m\per\radian} were used as preliminary requirement, as in \cite{paczkowski_2022}.

In conclusion, maneuvers may help to reduce the time required to estimate the coefficients, however, the amount of improvement depends strongly on the jitter levels and spectral shapes. If the \ac{MOSA} $\phi$ jitter level is \SI{5}{\nano\radian\per\sqrt\Hz}, maneuvers may not be necessary. On the other hand, with the reduced level of \SI{0.5}{\nano\radian\per\sqrt\Hz}, utilizing maneuvers might be a beneficial option, if one wants to determine all coefficients, e.g. for adjusting the \acp{BAM}.

\subsection{SC Maneuvers for Estimating Combined Coefficients}
\label{sec:sc_man}

As discussed in Sec.~\ref{sec:problem_phi}, only combined $\phi$ coefficients can be determined well from \ac{SC} maneuvers, not their individual values. Because of the high correlation between the respective \ac{TTL} contributions, they are indistinguishable in the \ac{TDI} variables. On the other hand, this means that knowledge of the combined coefficients might be sufficient if the goal is merely to subtract \ac{TTL} from the \ac{TDI} variables. We have confirmed this with a simulation comprising \ac{SC} maneuvers for all $\eta$ and $\phi$ angles, i.e. without \ac{MOSA} maneuvers.

In this simulation, six simultaneous $\eta$ maneuvers were performed via injection into the \ac{SC} angles, utilizing uncorrelated pairs and three different frequencies as developed in Sec.~\ref{sec:simultaneous}. Furthermore, $\phi$ maneuvers were performed on all three \ac{SC}, simultaneous but separate from the $\eta$ maneuvers, also using three different frequencies. For $\phi$, uncorrelated pairs could not be used since an injection into a \ac{SC} $\phi$ angle necessarily causes common angular motion of both \acp{MOSA} on that \ac{SC}, cf. Eq.~\eqref{eq:phi_ij}. All maneuvers were sinusoidal injections with an amplitude of \SI{30}{\nano\radian} and a duration of $600$ seconds. For the sake of a better illustration of the principle, we chose to use the reduced \ac{MOSA} jitter levels of \SI{0.1}{\nano\radian\per\sqrt\Hz} and \SI{0.5}{\nano\radian\per\sqrt\Hz} for $\eta$ and $\phi$, respectively. All other settings were kept unchanged.

\begin{figure*}[ht!]
	\begin{minipage}[c]{0.49\textwidth}
	   \centering
	   \includegraphics[width=\textwidth]{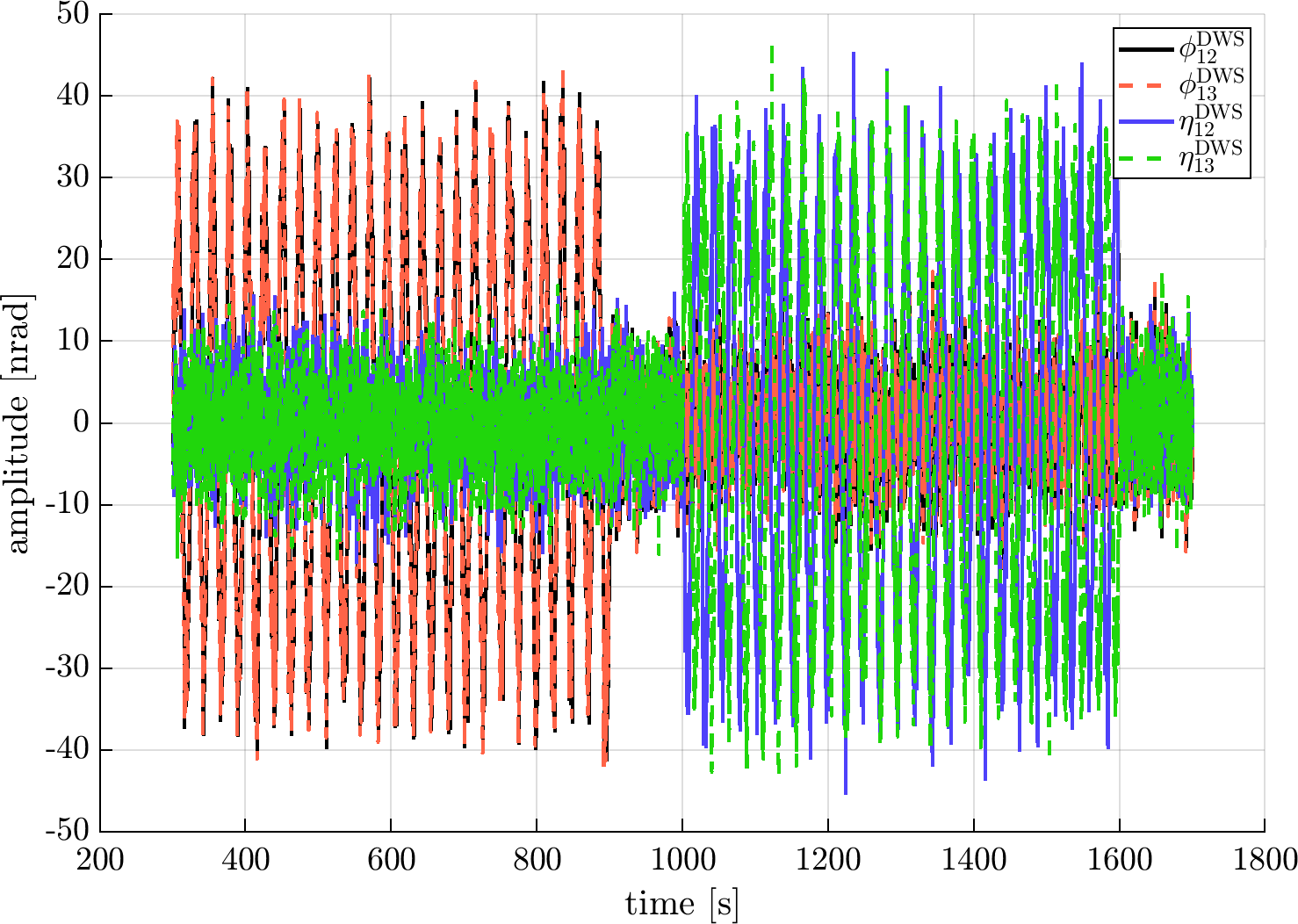}
	   \includegraphics[width=\textwidth]{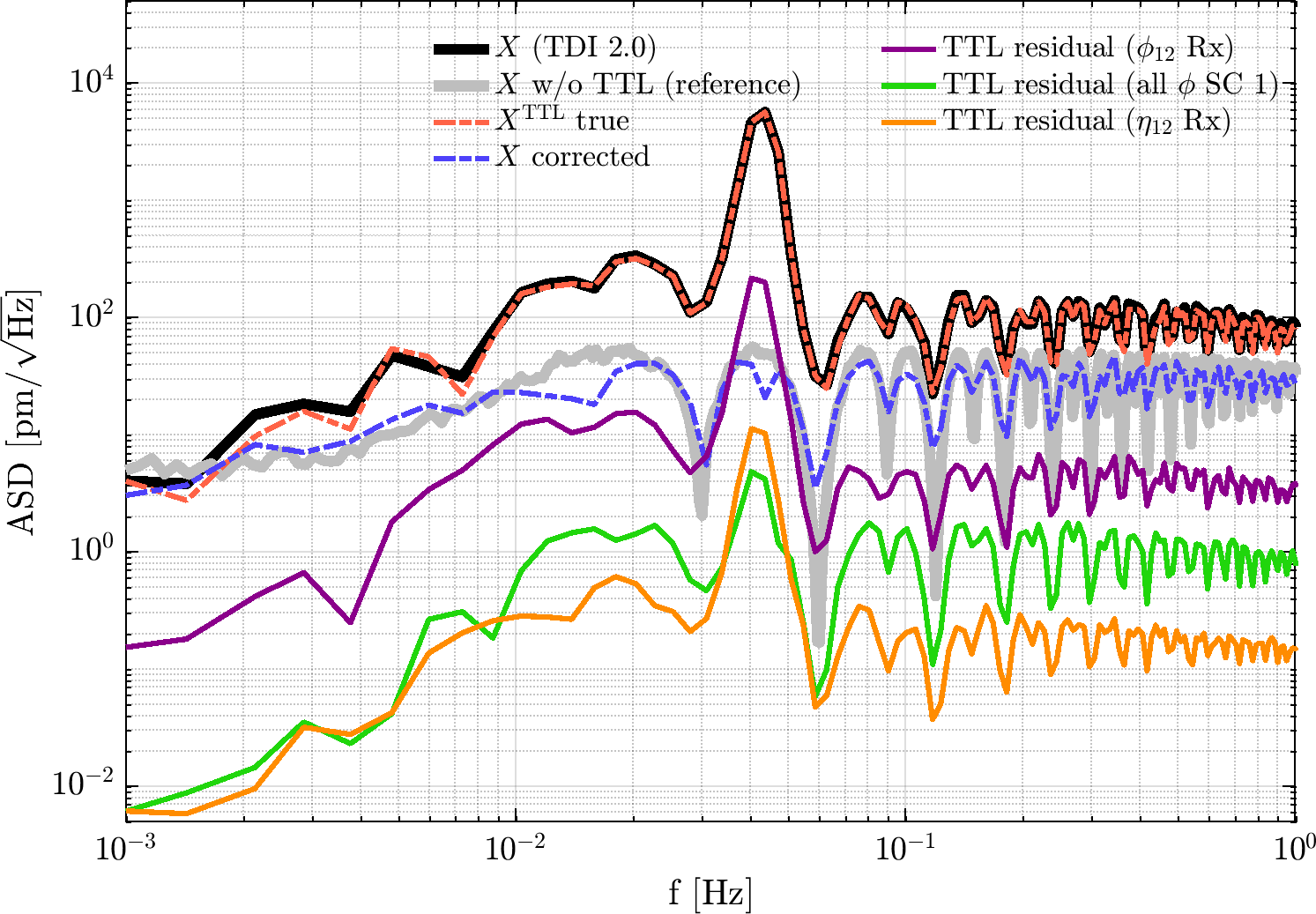}
	\end{minipage}
 \hfill
	\begin{minipage}[c]{0.49\textwidth}
	   \centering
	   \includegraphics[width=\textwidth]{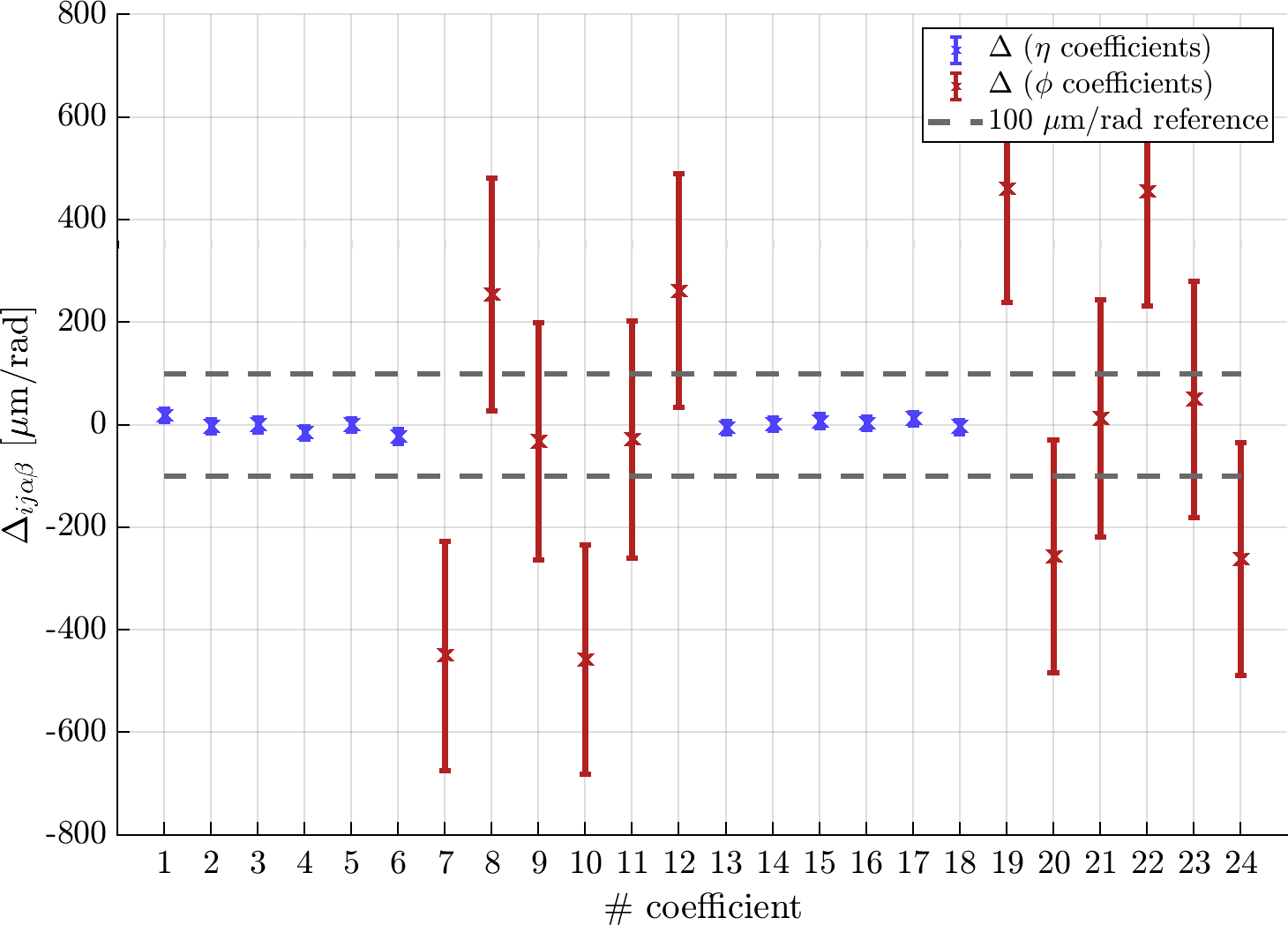}
	   \includegraphics[width=\textwidth]{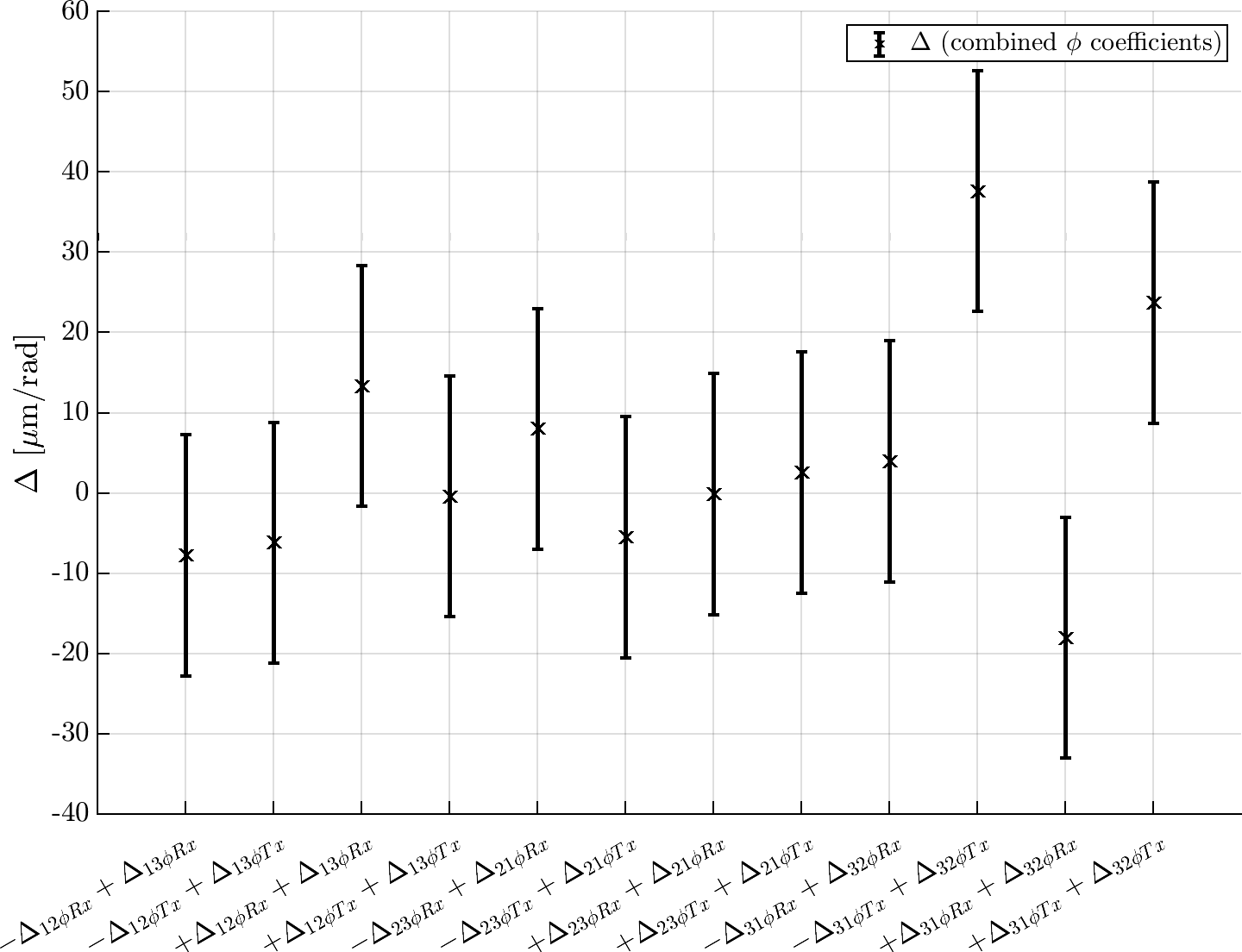}
	\end{minipage}
\caption{Simulation results for \ac{SC} maneuvers. Top left: simulated DWS angles measured on \ac{SC}\,$1$. Bottom left: \acp{ASD} of $X$, $X^\mathrm{TTL}$ and different residuals, see explanation in the text. Top right: estimation errors $\Delta_{ij\alpha\beta}$. Bottom right: deviations of estimated combined $\phi$ coefficients.}
\label{fig:sc_man}
\end{figure*}

Figure \ref{fig:sc_man} shows some results of this simulation. The top left plot shows the simulated \ac{DWS} measurements of both $\phi$ and $\eta$ angles on \ac{SC} 1. The $\phi$ maneuvers take place from second $300$ to $900$, the $\eta$ maneuvers from second $1000$ to $1600$. For each \ac{SC}, the excitations of the two \ac{MOSA} $\eta$ angles have low correlation due to different frequencies, while the excitations of the two \ac{MOSA} $\phi$ angles are highly correlated as explained in Sec.~\ref{sec:problem_phi}. For this reason, the uncertainties of $\eta$ \ac{TTL} coefficients are low, i.e. for coefficients \# 1-6 (Rx) and 13-18 (Tx), and the uncertainties of the $\phi$ coefficients (\# 7-12 and 19-24) are well above \SI{0.1}{\milli\m\per\radian}, the temporary requirement used e.g. in \cite{paczkowski_2022}. As illustrated in the top right plot, the actual deviations $\Delta_{ij\alpha\beta} = \hat C_{ij\alpha\beta} - C_{ij\alpha\beta}$ reflect these uncertainties, which are shown as error bars. This shows that $\phi$ maneuvers via \ac{SC} rotation do not allow precise coefficient estimation. The \ac{RMS} deviation of the $\eta$ coefficients (\# 1-6 and 13-18) was \SI{10.6}{\micro\m\per\radian}, for the individual $\phi$ coefficients (\# 7-12 and 19-24) it was \SI{303.4}{\micro\m\per\radian}.

The deviations of the combined $\phi$ coefficients, however, were significantly lower than for the individual values. Estimated values for all $12$ such combinations are plotted in the bottom right of Fig.~\ref{fig:sc_man}. For example, the deviations for \ac{SC}\,$1$ were:
\begin{align}
	-\Delta_{12 \phi Rx} + \Delta_{13 \phi Rx} &= \SI{-7.7}{\micro\m\per\radian} \\
	-\Delta_{12 \phi Tx} + \Delta_{13 \phi Tx} &= \SI{-6.1}{\micro\m\per\radian} \\
	\Delta_{12 \phi Rx} + \Delta_{12 \phi Tx} &= \SI{13.4}{\micro\m\per\radian} \\
	\Delta_{13 \phi Rx} + \Delta_{13 \phi Tx} &= \SI{-0.3}{\micro\m\per\radian}
\end{align}
Similar results were obtained for \ac{SC}\,$2$ and \ac{SC}\,$3$. The \ac{RMS} deviation of all $12$ combined $\phi$ coefficients was \SI{15.0}{\micro\m\per\radian}, and therefore well below the temporary requirement of \SI{100}{\micro\m\per\radian}.

The bottom left plot of Fig.~\ref{fig:sc_man} shows several \acp{ASD}, all of which stem from the \ac{SC} maneuver simulation described in this subsection \ref{sec:sc_man}, except for the thick light gray line. The light gray line serves as a reference for $X$ without \ac{TTL} and was obtained from a separate simulation with the coefficients set to zero. The \ac{TDI} $X$ variable from the \ac{SC} maneuver simulation is shown as a thick black line. It is dominated by the dashed red line, which shows the true total \ac{TTL} computed with the true angles and the true coupling factors. The corrected $X$ variable (dashed blue line), i.e. $X - X^\mathrm{TTL}(\hat C)$ computed with estimated coefficients and \ac{DWS} angles, roughly coincides with the reference for $X$ without \ac{TTL}, which shows that the subtraction of the estimated total \ac{TTL} works. However, let us take a closer look at the residuals of the individual \ac{TTL} contributions, i.e.
\begin{equation}
	C_{ij\alpha\beta} \cdot \mathcal{X}_{ij\alpha\beta} - \hat C_{ij\alpha\beta} \cdot \mathcal{X}_{ij\alpha\beta}^\mathrm{DWS}.
\end{equation}
It can bee seen that the residual for $X_{12 \phi Rx}^\mathrm{TTL}$ alone (purple line) is much larger than for the sum of all $\phi$ \ac{TTL} contributions for \ac{SC}\,$1$ (green line). This is due to the large uncertainty of the individual estimated $\phi$ coefficients, but low uncertainty for the combined coefficients. For comparison, the dotted yellow line shows that the residual for $X_{12 \eta Rx}^\mathrm{TTL}$ is much smaller than for $X_{12 \phi Rx}^\mathrm{TTL}$ due to the low uncertainty of the $\eta$ coefficients. In summary, \ac{SC} $\phi$ maneuvers can be used to subtract the combined \ac{TTL} originating from $\phi$ angles, but do not allow precise estimation of the individual $\phi$ coefficients.

\subsection{Comparison of Estimation Uncertainties}
\label{sec:sim_results}

We have compared our results to two other sources in the literature. Both sources assume the same linear \ac{TTL} model for \ac{LISA} and make statements about the uncertainty of the $24$ coefficients. In both cases, the angular jitter is modeled with \ac{ASD} shapes similar to ours, cf. Fig.~\ref{fig:lisasim_spec}, merely with different jitter levels. The results are summarized in Tab.~\ref{tab:comparison}, where our values are given in terms of $\sigma_\mathrm{stat} = 1.5 \cdot \sigma_\mathrm{LSQ}$, cf. Sec.~\ref{sec:pe} and App.~\ref{b:lsq}.

The first reference is \cite{paczkowski_2022}, where the \ac{TTL} coefficients are determined with an MCMC estimation in the frequency domain. For the main simulations, \cite{paczkowski_2022} used \SI{2}{\nano\radian\per\sqrt\Hz} as the \ac{MOSA} $\phi$ jitter level. However, a result is also given for \SI{5}{\nano\radian\per\sqrt\Hz}. We have performed two simulations in order to compare the results in both cases, either in terms of $\sigma(\hat C_{ij\alpha\beta})$ or in terms of \ac{RMS} values, which are taken over the $24$ deviations $\hat C_{ij\alpha\beta} - C_{ij\alpha\beta}$. Here we find that we predict slightly lower uncertainties. Our computed standard deviations $\sigma_\mathrm{stat}$ are about a factor of $2$ smaller, cf. Tab.~\ref{tab:comparison}. A similar observation is made for the \ac{RMS} values. Although the different estimation methods yield slightly different results, the overall picture of the feasibility of post-processing subtraction is still in agreement with our findings.

The second source we compared our results to is \cite{george_2022}. There, the authors obtain lower bounds for $\sigma(\hat C)$ using Fisher information theory. We can compare to what is referred to as case A in \cite{george_2022}, but not to the case B, where different jitter \ac{ASD} shapes are used. Note also that this reference uses a slightly lower \ac{DWS} noise level of $50/300$\,\si{\nano\radian\per\sqrt\Hz}. We have performed a simulation with the same settings given in \cite{george_2022} for case A. The lower bounds they obtain for the $\sigma(\hat C_{ij\alpha\beta})$ match well with our values for $\sigma_\mathrm{stat}$, cf. Tab.~\ref{tab:comparison}. This strengthens the confidence in our results and moreover suggests that the uncertainties we obtain are very close to optimal.

\begin{table*}[!ht]
\caption{Comparing coefficient uncertainties to results from two other sources in the literature, \cite{paczkowski_2022} and \cite{george_2022}. Based on simulated data with one day of integration time, without using maneuvers.}

\centering
\begin{tabularx}{.765\linewidth}{ c | c | c | c | c | c | c }

~ & \multicolumn{3}{c|}{jitter levels~[\si{\nano\radian\per\sqrt\Hz}]} & \multicolumn{2}{c|}{uncertainty [\si{\micro\m\per\radian}]} & \si{\nano\radian\per\sqrt\Hz} \\
Reference & \ac{SC} (per axis) & \ac{MOSA} $\eta$ & \ac{MOSA} $\phi$ & $\sigma$ & \ac{RMS} & \ac{DWS} noise level \\ \hline

source \cite{paczkowski_2022} & $5$ & $1$ & $2$ & $23.7 \le \sigma \le 38.6$ & $\approx 35$ & $70/335$ \\
this study & $5$ & $1$ & $2$ & $\sigma_\mathrm{stat} \approx (8.7,13.1)$ & $12.6$ & $70/335$ \\ \hline
source \cite{paczkowski_2022} & $5$ & $1$ & $5$ & N/A & $\approx 25$ & $70/335$ \\
this study & $5$ & $1$ & $5$ & $\sigma_\mathrm{stat} \approx (8.7,6.3)$ & $11.3$ & $70/335$ \\ \hline
-
source \cite{george_2022} & $10$ & $10$ & $10$ & $\sigma \ge (2.98,3.23)$ & N/A & $50/300$ \\
this study & $10$ & $10$ & $10$ & $\sigma_\mathrm{stat} \approx (2.94,3.20)$ & $2.6$ & $50/300$ \\

\end{tabularx}
\label{tab:comparison}
\end{table*}

\section{Summary and Outlook\label{sec:summary}}

In Sec.~\ref{sec:design}, we discussed the feasibility and optimal design of \ac{TTL} calibration maneuvers. We found that both \ac{SC} and \ac{MOSA} maneuvers are likely possible with amplitudes of about \SI{30}{\nano\radian}. The optimal frequency for periodic injection signals is $43$ (40-45)\,\si{\milli\Hz}, or, alternatively, about $18$ or $78$ mHz. Uncorrelated pairs can be used to perform maneuvers partly simultaneously. Using these pairs and three different frequencies allows to perform two sets of six simultaneous maneuvers each, covering all $12$ angles which cause \ac{TTL}. Since Rx and Tx coefficients can also be estimated at the same time, this is sufficient to recover all $24$ \ac{TTL} coefficients. While $\eta$ maneuvers can be performed via injection into \ac{SC} angles, $\phi$ maneuvers should be done via \ac{MOSA} injections if one wants to disentangle the individual $\phi$ coefficients. Alternatively, three \ac{SC} $\phi$ maneuvers allow to estimate combined $\phi$ coefficients, which is sufficient for \ac{TTL} subtraction, as we confirmed in Sec.~\ref{sec:sc_man}.

In Sec.~\ref{sec:results}, we have shown that with a total maneuver duration of $20$~minutes, all \ac{TTL} coupling coefficients could be recovered with uncertainties below \SI{15}{\micro\m\per\radian}, which is more than sufficient for \ac{TTL} subtraction. When comparing the estimation using maneuvers to a fit-of-noise approach, the improvement factor in terms of integration time required to achieve this uncertainty level was about $9$ in the case of full angular jitter. In the alternative scenario with the reduced \ac{MOSA} $\phi$ jitter level of \SI{0.5}{\nano\radian\per\sqrt\Hz}, maneuvers would even reduce the required integration time by a factor of about $230$. This shows that maneuvers can be used to determine the \ac{TTL} coefficients significantly quicker. However, whether the potential improvement justifies performing maneuvers will likely depend on the \ac{MOSA} $\phi$ jitter level and frequency shape in a closed-loop control system. Finally, we found that the estimation uncertainties obtained from our simulations are compatible with results from \cite{paczkowski_2022} and \cite{george_2022}.

This document provides a first-level assessment of the feasibility of \ac{TTL} calibration maneuvers in \ac{LISA}, and of the potential improvement for coefficient estimation. For concrete planning, the actual realization of the maneuvers should be studied in more detail.

% APPENDIX

\appendix

\section{TTL Contributions in the TDI Variables\label{a:ttl_in_tdi}}

We are working with the \ac{TDI} 2.0 variables \cite{tinto_2020,otto_phd,bayle_phd}. The second generation Michelson combination $X$ is shown in equation $(4.33)$ in \cite{bayle_phd}. From this, one can derive the formula for all \ac{TTL} in $X$ due to $\eta$ angles:
\begin{widetext}
\begin{align}
	X^\mathrm{TTL}_\eta = - &\left( 1 - \mathcal{D}_{131} - \mathcal{D}_{13121} + \mathcal{D}_{1213131} \right) \left( (C_{12 \eta Rx} + C_{12 \eta Tx}\mathcal{D}_{121})\eta_{12} + (C_{21 \eta Rx} + C_{21 \eta Tx}) \mathcal{D}_{12} \eta_{21} \right) \nonumber \\
	+ &\left( 1 - \mathcal{D}_{121} - \mathcal{D}_{12131} + \mathcal{D}_{1312121} \right) \left( (C_{13 \eta Rx} + C_{13 \eta Tx}\mathcal{D}_{131})\eta_{13} + (C_{31 \eta Rx} + C_{31 \eta Tx}) \mathcal{D}_{13} \eta_{31} \right). \label{eq:X_eta_ttl}
\end{align}
\end{widetext}
Note that the signs in Eq.~\eqref{eq:X_eta_ttl} are conventional and may be swapped in other publications. Replacing all $\eta$'s with $\phi$'s yields the \ac{TTL} in $X$ due to the $\phi$ angles, $X^\mathrm{TTL}_\phi$, and the entire \ac{TTL} in $X$ is given by the sum $X^\mathrm{TTL} = X^\mathrm{TTL}_\eta + X^\mathrm{TTL}_\phi$. The combinations $Y$ and $Z$ are obtained by circular permutation of the indices. Here $\phi_{ij}$ and $\eta_{ij}$ denote the pitch and yaw angles of \ac{MOSA}\,$ij$ \ac{w.r.t.} incident beam originating from \ac{SC}\,$j$, cf. Sec.\ref{sec:angles}. We contracted the indices according to the convention given in Sec.~\ref{sec:tdi}. For reference, we write out the individual equations for all \ac{TTL} contributions depending on $\eta$ angles in the following. The \ac{TTL} contributions of the $\phi$ angles are built identically, i.e. they can be obtained by replacing each $\eta$ with a $\phi$.

For $\eta_{12}$ \textbf{Rx} (coefficient \textbf{\#1}, using the numbering convention given in Tab.~\ref{tab:C_order}) we have:
\begin{align}
X_{12 \eta Rx}^\mathrm{TTL} = &C_{12 \eta Rx} \cdot (-1 + \mathcal{D}_{131} + \mathcal{D}_{13121} - \mathcal{D}_{1213131}) \eta_{12} \label{eq:x_eta12_rx} \\
Y_{12 \eta Rx}^\mathrm{TTL} = &C_{12 \eta Rx} \cdot (1 - \mathcal{D}_{232} - \mathcal{D}_{23212} + \mathcal{D}_{2123232})\mathcal{D}_{21} \eta_{12} \label{eq:y_eta12_rx} \\
Z_{12 \eta Rx}^\mathrm{TTL} = &0 \label{eq:z_eta12_rx}
\end{align}
For $\eta_{12}$ \textbf{Tx} (\textbf{\#13}) we have:
\begin{align}
X_{12 \eta Tx}^\mathrm{TTL} &= C_{12 \eta Tx} \cdot (-1 + \mathcal{D}_{131} + \mathcal{D}_{13121} - \mathcal{D}_{1213131}) \mathcal{D}_{121} \eta_{12} \label{eq:x_eta12_tx} \\
Y_{12 \eta Tx}^\mathrm{TTL} &= C_{12 \eta Tx} \cdot (1 - \mathcal{D}_{232} - \mathcal{D}_{23212} + \mathcal{D}_{2123232}) \mathcal{D}_{21} \eta_{12} \label{eq:y_eta12_tx} \\
Z_{12 \eta Tx}^\mathrm{TTL} &= 0 \label{eq:z_eta12_tx}
\end{align}

For $\eta_{23}$ \textbf{Rx} (\textbf{\#2}) we have:
\begin{align}
X_{23 \eta Rx}^\mathrm{TTL} &= 0 \label{eq:x_eta23_rx} \\
Y_{23 \eta Rx}^\mathrm{TTL} &= C_{23 \eta Rx} \cdot (-1 + \mathcal{D}_{212} + \mathcal{D}_{21232} - \mathcal{D}_{2321212}) \eta_{23} \label{eq:y_eta23_rx} \\
Z_{23 \eta Rx}^\mathrm{TTL} &= C_{23 \eta Rx} \cdot (1 - \mathcal{D}_{313} - \mathcal{D}_{31323} + \mathcal{D}_{3231313})\mathcal{D}_{32} \eta_{23} \label{eq:z_eta23_rx}
\end{align}
For $\eta_{23}$ \textbf{Tx} (\textbf{\#14}) we have:
\begin{align}
X_{23 \eta Tx}^\mathrm{TTL} &= 0 \label{eq:x_eta23_tx} \\
Y_{23 \eta Tx}^\mathrm{TTL} &= C_{23 \eta Tx} \cdot (-1 + \mathcal{D}_{212} + \mathcal{D}_{21232} - \mathcal{D}_{2321212}) \mathcal{D}_{232} \eta_{23} \label{eq:y_eta23_tx} \\
Z_{23 \eta Tx}^\mathrm{TTL} &= C_{23 \eta Tx} \cdot (1 - \mathcal{D}_{313} - \mathcal{D}_{31323} + \mathcal{D}_{3231313}) \mathcal{D}_{32} \eta_{23} \label{eq:z_eta23_tx}
\end{align}

For $\eta_{31}$ \textbf{Rx} (\textbf{\#3}) we have:
\begin{align}
X_{31 \eta Rx}^\mathrm{TTL} &= C_{31 \eta Rx} \cdot (1 - \mathcal{D}_{121} - \mathcal{D}_{12131} + \mathcal{D}_{1312121})\mathcal{D}_{13} \eta_{31} \label{eq:x_eta31_rx} \\
Y_{31 \eta Rx}^\mathrm{TTL} &= 0 \label{eq:y_eta31_rx} \\
Z_{31 \eta Rx}^\mathrm{TTL} &= C_{31 \eta Rx} \cdot (-1 + \mathcal{D}_{323} + \mathcal{D}_{32313} - \mathcal{D}_{3132323}) \eta_{31} \label{eq:z_eta31_rx}
\end{align}
For $\eta_{31}$ \textbf{Tx} (\textbf{\#15}) we have:
\begin{align}
X_{31 \eta Tx}^\mathrm{TTL} &= C_{31 \eta Tx} \cdot (1 - \mathcal{D}_{121} - \mathcal{D}_{12131} + \mathcal{D}_{1312121}) \mathcal{D}_{13} \eta_{31} \label{eq:x_eta31_tx} \\
Y_{31 \eta Tx}^\mathrm{TTL} &= 0 \label{eq:y_eta31_tx} \\
Z_{31 \eta Tx}^\mathrm{TTL} &= C_{31 \eta Tx} \cdot (-1 + \mathcal{D}_{323} + \mathcal{D}_{32313} - \mathcal{D}_{3132323}) \mathcal{D}_{313} \eta_{31} \label{eq:z_eta31_tx}
\end{align}

For $\eta_{13}$ \textbf{Rx} (\textbf{\#4}) we have:
\begin{align}
X_{13 \eta Rx}^\mathrm{TTL} &= C_{13 \eta Rx} \cdot (1 - \mathcal{D}_{121} - \mathcal{D}_{12131} + \mathcal{D}_{1312121}) \eta_{13} \label{eq:x_eta13_rx} \\
Y_{13 \eta Rx}^\mathrm{TTL} &= 0 \label{eq:y_eta13_rx} \\
Z_{13 \eta Rx}^\mathrm{TTL} &= C_{13 \eta Rx} \cdot (-1 + \mathcal{D}_{323} + \mathcal{D}_{32313} - \mathcal{D}_{3132323})\mathcal{D}_{31} \eta_{13} \label{eq:z_eta13_rx}
\end{align}
For $\eta_{13}$ \textbf{Tx} (\textbf{\#16}) we have:
\begin{align}
X_{13 \eta Tx}^\mathrm{TTL} &= C_{13 \eta Tx} \cdot (1 - \mathcal{D}_{121} - \mathcal{D}_{12131} + \mathcal{D}_{1312121}) \mathcal{D}_{131} \eta_{13} \label{eq:x_eta13_tx} \\
Y_{13 \eta Tx}^\mathrm{TTL} &= 0 \label{eq:y_eta13_tx} \\
Z_{13 \eta Tx}^\mathrm{TTL} &= C_{13 \eta Tx} \cdot (-1 + \mathcal{D}_{323} + \mathcal{D}_{32313} - \mathcal{D}_{3132323}) \mathcal{D}_{31} \eta_{13} \label{eq:z_eta13_tx}
\end{align}

For $\eta_{32}$ \textbf{Rx} (\textbf{\#5}) we have:
\begin{align}
X_{32 \eta Rx}^\mathrm{TTL} &= 0 \label{eq:x_eta32_rx} \\
Y_{32 \eta Rx}^\mathrm{TTL} &= C_{32 \eta Rx} \cdot (-1 + \mathcal{D}_{212} + \mathcal{D}_{21232} - \mathcal{D}_{2321212})\mathcal{D}_{23} \eta_{32} \label{eq:y_eta32_rx} \\
Z_{32 \eta Rx}^\mathrm{TTL} &= C_{32 \eta Rx} \cdot (1 - \mathcal{D}_{313} - \mathcal{D}_{31323} + \mathcal{D}_{3231313}) \eta_{32} \label{eq:z_eta32_rx}
\end{align}
For $\eta_{32}$ \textbf{Tx} (\textbf{\#17}) we have:
\begin{align}
X_{32 \eta Tx}^\mathrm{TTL} &= 0 \label{eq:x_eta32_tx} \\
Y_{32 \eta Tx}^\mathrm{TTL} &= C_{32 \eta Tx} \cdot (-1 + \mathcal{D}_{212} + \mathcal{D}_{21232} - \mathcal{D}_{2321212}) \mathcal{D}_{23} \eta_{32} \label{eq:y_eta32_tx} \\
Z_{32 \eta Tx}^\mathrm{TTL} &= C_{32 \eta Tx} \cdot (1 - \mathcal{D}_{313} - \mathcal{D}_{31323} + \mathcal{D}_{3231313}) \mathcal{D}_{323} \eta_{32} \label{eq:z_eta32_tx}
\end{align}

For $\eta_{21}$ \textbf{Rx} (\textbf{\#6}) we have:
\begin{align}
X_{21 \eta Rx}^\mathrm{TTL} &= C_{21 \eta Rx} \cdot (-1 + \mathcal{D}_{131} + \mathcal{D}_{13121} - \mathcal{D}_{1213131})\mathcal{D}_{12} \eta_{21} \label{eq:x_eta21_rx} \\
Y_{21 \eta Rx}^\mathrm{TTL} &= C_{21 \eta Rx} \cdot (1 - \mathcal{D}_{232} - \mathcal{D}_{23212} + \mathcal{D}_{2123232}) \eta_{21} \label{eq:y_eta21_rx} \\
Z_{21 \eta Rx}^\mathrm{TTL} &= 0 \label{eq:z_eta21_rx}
\end{align}
For $\eta_{21}$ \textbf{Tx} (\textbf{\#18}) we have:
\begin{align}
X_{21 \eta Tx}^\mathrm{TTL} &= C_{21 \eta Tx} \cdot (-1 + \mathcal{D}_{131} + \mathcal{D}_{13121} - \mathcal{D}_{1213131}) \mathcal{D}_{12} \eta_{21} \label{eq:x_eta21_tx} \\
Y_{21 \eta Tx}^\mathrm{TTL} &= C_{21 \eta Tx} \cdot (1 - \mathcal{D}_{232} - \mathcal{D}_{23212} + \mathcal{D}_{2123232}) \mathcal{D}_{212} \eta_{21} \label{eq:y_eta21_tx} \\
Z_{21 \eta Tx}^\mathrm{TTL} &= 0 \label{eq:z_eta21_tx}
\end{align}

\section{Statistical estimator uncertainty\label{b:lsq}}

In order to gain a better understanding of realistic uncertainties, we performed a test by simulating $500$ sets of \ac{LISA} data, each of length \SI{2000}{\s} and with identical noise settings as specified in Tab.~\ref{tab:LISASim_spec}. In each simulation, different random coefficients were used, uniformly distributed within the interval $[-3,3]$\,\si{\milli\m\per\radian}. For each coefficient $C_{ij\alpha\beta}$ we computed the standard deviation of all deviations $\Delta_{ij\alpha\beta} = \hat C_{ij\alpha\beta} - C_{ij\alpha\beta}$ over $500$ runs. These statistical standard deviations were within $[61.5,68]$\,\si{\micro\m\per\radian} for the $\eta$ coefficients and within $[45.5,49]$\,\si{\micro\m\per\radian} for $\phi$, while the standard deviations obtained by formula \eqref{eq:std_lsq} were within $[42.8,47.8]$\,\si{\micro\m\per\radian} for $\eta$ and within $[31.4,35]$\,\si{\micro\m\per\radian} for $\phi$. For each coefficient, the ratio of statistical standard deviations over $\sigma_\mathrm{LSQ}$ was between $1.4$ and $1.5$. We conclude that a more realistic uncertainty is given by the value $\sigma_\mathrm{stat}$, which we define as
\begin{equation}
    \sigma_\mathrm{stat} = 1.5 \cdot \sigma_\mathrm{LSQ}.
\end{equation}

\section{Dependencies of LSQ estimator uncertainty\label{c:std}}

Suppose we estimate merely two of the \ac{TTL} coefficients and denote these by $C_1,C_2$. We can use the same notation as described in Sec.~\ref{sec:notation}, just considering two instead of all $24$ \ac{TTL} contributions. That is, we define $\mathcal{A} = [\mathcal{A}_1,\mathcal{A}_2] \in \mathbb{R}^{3N \times 2}$ as in Eq.\eqref{eq:alpha}, but taking only the two relevant columns. The sum of these two \ac{TTL} contributions in $V = (X^T,Y^T,Z^T)^T$ is given by $\mathcal{A} \cdot C$, with $C = [C_1,C_2]^T$, just as in Eq.\eqref{eq:V_ttl}. Assume further without loss of generality that each column of $\mathcal{A}$ has zero mean. Recall that the formal error of the \ac{LSQ} estimator is given in Eq.~\eqref{eq:std_lsq}. If we define $M := (\mathcal{A}^T\mathcal{A})^{-1}$, with $M_{11}$ denoting the top left matrix entry, we have
\begin{equation}
    \sigma_\mathrm{LSQ}(\hat C_1) = \sigma(n_V) \cdot \sqrt{M_{11}} \propto \sqrt{M_{11}}
\end{equation}
where $\hat C_1$ is the LSQ estimator of $C_1$. Since $\mathcal{A}^T\mathcal{A}$ is a $2\times 2$ matrix, its inverse $M$ is given by an explicit formula. In particular,
\begin{equation}
    M_{11} = \frac{1}{\det(\mathcal{A}^T\mathcal{A})} \mathcal{A}_2^T\mathcal{A}_2 \label{eq:M_11}
\end{equation}
with
\begin{equation}
    \det(\mathcal{A}^T\mathcal{A}) = (\mathcal{A}_1^T\mathcal{A}_1)(\mathcal{A}_2^T\mathcal{A}_2) - (\mathcal{A}_1^T\mathcal{A}_2)^2.
\end{equation}

Equation~\eqref{eq:M_11} can be written in terms of the standard deviation $\sigma(\mathcal{A}_1)$ and the correlation $\mathrm{corr}(\mathcal{A}_1,\mathcal{A}_2)$. In our matrix notation these can be expressed as
\begin{equation}
    \sigma(\mathcal{A}_i)^2 = \frac{1}{3N} \mathcal{A}_i^T\mathcal{A}_i, \label{eq:sd}
\end{equation}
$i=1,2$, and
\begin{align}
    \mathrm{corr}(\mathcal{A}_1,\mathcal{A}_2) &= \frac{\mathrm{cov}(\mathcal{A}_1,\mathcal{A}_2)}{\sigma(\mathcal{A}_1)\sigma(\mathcal{A}_2)} \\
    &= \frac{\mathcal{A}_1^T\mathcal{A}_2}{\sqrt{(\mathcal{A}_1^T\mathcal{A}_1)(\mathcal{A}_2^T\mathcal{A}_2)}},
\end{align}
provided that each $\mathcal{A}_i$ has zero mean. Now, rearranging the terms of Eq.~\eqref{eq:M_11} yields
\begin{align}
    M_{11} &= \frac{\mathcal{A}_2^T\mathcal{A}_2}{(\mathcal{A}_1^T\mathcal{A}_1)(\mathcal{A}_2^T\mathcal{A}_2) - (\mathcal{A}_1^T\mathcal{A}_2)^2} \\
    &= \frac{\mathcal{A}_2^T\mathcal{A}_2}{(\mathcal{A}_1^T\mathcal{A}_1)(\mathcal{A}_2^T\mathcal{A}_2) ( 1 - \mathrm{corr}(\mathcal{A}_1,\mathcal{A}_2)^2 )} \\
    &= \frac{1}{3N \cdot \sigma(\mathcal{A}_1)^2 \cdot (1-\mathrm{corr}(\mathcal{A}_1,\mathcal{A}_2)^2)},
\end{align}
provided $\mathrm{corr}(\mathcal{A}_1,\mathcal{A}_2)^2 \neq 1$ since otherwise the two columns of $\mathcal{A}$ are linearly dependent and $M$ does not exist to start with. Hence it follows that
\begin{equation}
    \sigma_\mathrm{LSQ}(\hat C_1) \propto \frac{1}{\sqrt{N}} \frac{1}{\sigma(\mathcal{A}_1)} \frac{1}{\sqrt{1 - \mathrm{corr}(\mathcal{A}_1,\mathcal{A}_2)^2}}. \label{eq:sigma_prop_all}
\end{equation}
In particular, if $\mathcal{A}_1$ is dominated by a sinusoidal injection angle with amplitude $A_\mathrm{man}$, then $\mathcal{A}_1$ is approximately proportional to $A_\mathrm{man}$ as well, and we clearly have
\begin{equation}
    \sigma_\mathrm{LSQ}(\hat C_1) \propto \frac{1}{A_\mathrm{man}} \label{eq:prop_A}
\end{equation}
for large enough $A_\mathrm{man}$. Moreover, for the maneuver duration $T_\mathrm{man}$ we then have
\begin{equation}
    \sigma_\mathrm{LSQ}(\hat C_1) \propto \frac{1}{\sqrt{T_\mathrm{man}}} \label{eq:prop_T}
\end{equation}
since $N \propto T_\mathrm{man}$ for a fixed sampling rate, where $T_\mathrm{man}$ is the maneuver time that is used entirely for the estimation. Note that this dependency requires the noise $n_V$ in the \ac{TDI} variables to be white, which the \ac{LSQ} estimator does assume. In reality the noise spectrum is frequency dependent and it is clear that the estimation uncertainty cannot be lowered arbitrarily by increasing the amount of sampling points $N$.

\begin{figure*}[ht!]
	\begin{minipage}[c]{0.49\textwidth}
	   \centering
	   \includegraphics[width=\textwidth]{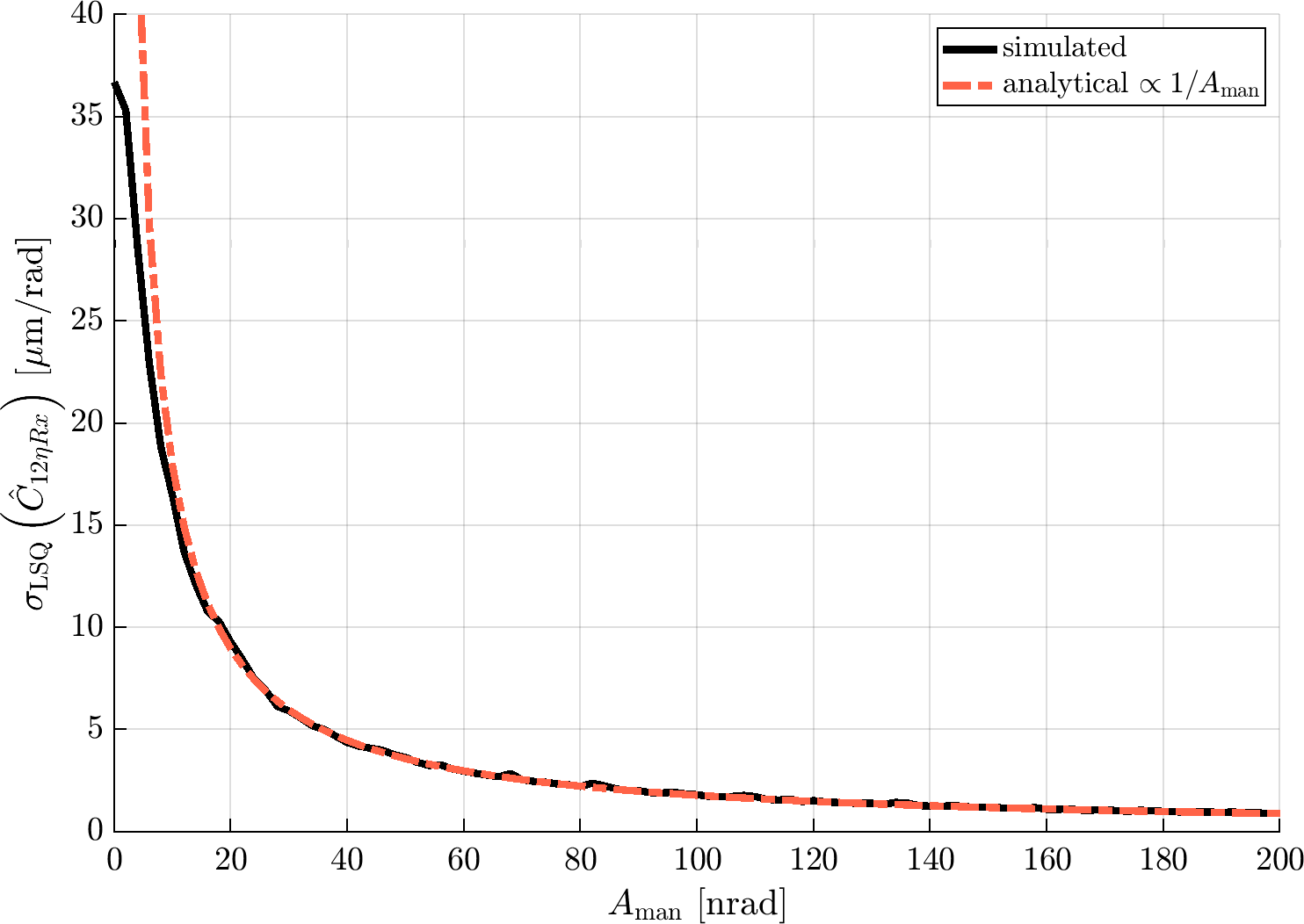}
	\end{minipage}
 \hfill
	\begin{minipage}[c]{0.49\textwidth}
	   \centering
	   \includegraphics[width=\textwidth]{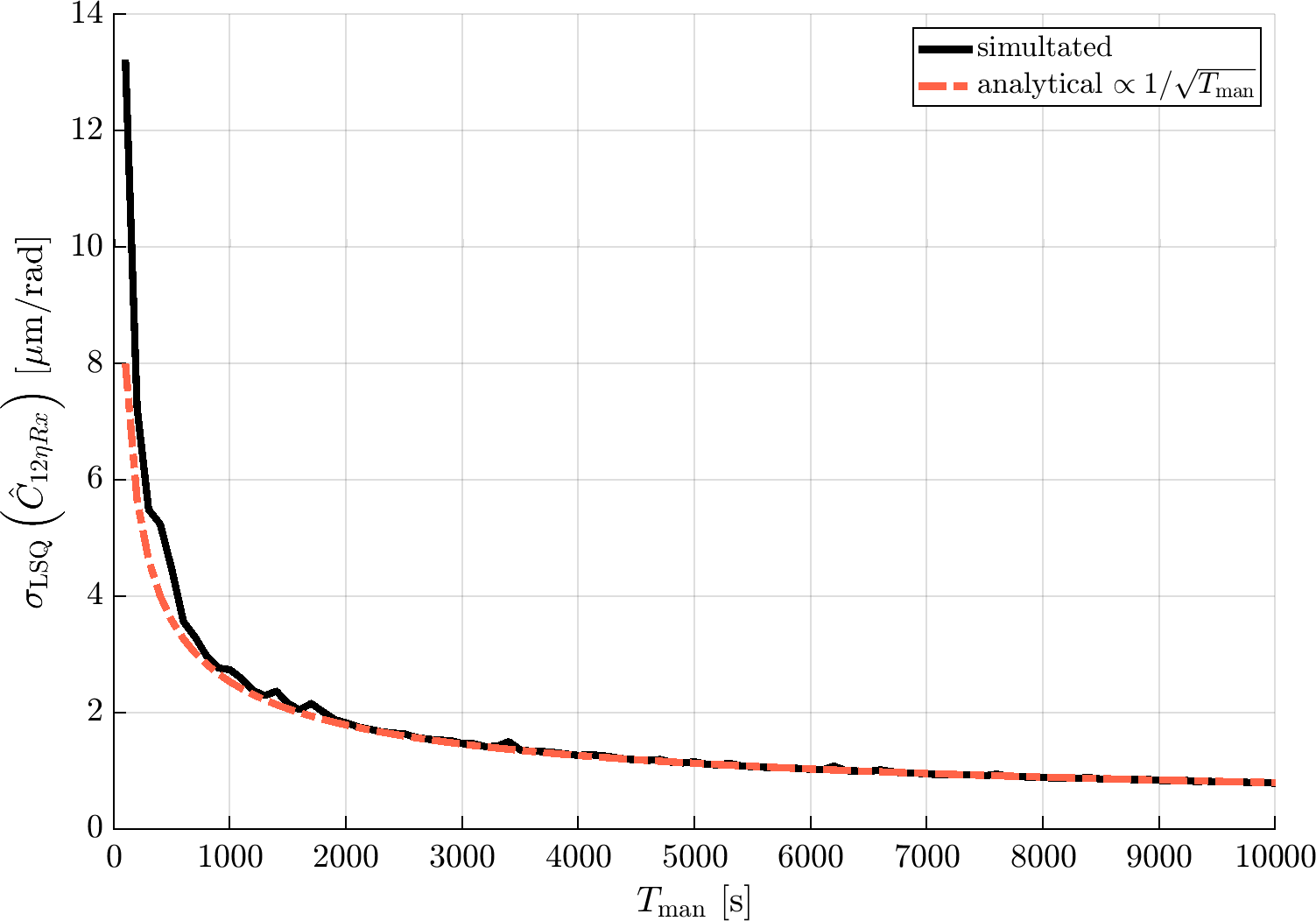}
	\end{minipage}
\caption{Dependency of estimator uncertainties to $1/A_\mathrm{man}$ (left) and $1/\sqrt{T_\mathrm{man}}$ (right), confirmed with simulations.}
\label{fig:simulations_prop}
\end{figure*}

Since the two \ac{TTL} contributions were chosen arbitrarily in the beginning, the result holds for any two \ac{TTL} contributions, if only these two are estimated and there are no correlations with other contributions. When all $24$ \ac{TTL} coefficients are estimated at the same time, the relations \ref{eq:prop_A} and \ref{eq:prop_T} still hold approximately. We have confirmed this with simulations. The results for the coefficient $C_{12\eta Rx}$ are shown in Fig.~\ref{fig:simulations_prop}. For the left plot, $100$ simulations were performed, each including an $\eta_{12}$ maneuver with an amplitude between $0$ and \SI{200}{\nano\radian}, while all other settings were fixed. The maneuver duration was fixed to $T_\mathrm{man}~=$~\SI{2000}{\s}, the maneuver frequency was $f_\mathrm{man}~=$~\SI{43}{\milli\Hz}. Shown in blue are $\sigma_\mathrm{LSQ}(\hat C_{12 \eta Rx})$, plotted over $A_\mathrm{man}$. The red dashed line shows the predicted values, which were obtained by extrapolating an example value according to the relation \ref{eq:prop_A}. For significant amplitudes $>$~\SI{20}{\nano\radian}, when the \ac{TTL} contribution of $\eta_{12}$ is dominated by the maneuver signal, the standard deviations from the simulations are compatible with the predicted values, confirming proportionality $\sigma \propto 1/A_\mathrm{man}$.

The right plot of Fig.~\ref{fig:simulations_prop} shows again $100$ different values of $\sigma(\hat C_{12 \eta Rx})$ in blue. A long $\eta_{12}$ maneuver with a duration of \SI{10000}{\s} was simulated. The maneuver amplitude was \SI{100}{\nano\radian}, the maneuver frequency was $f_\mathrm{man} =$~\SI{43}{\milli\Hz}. For each of the $100$ plotted values, a different data length was used for the estimation and computation of the standard deviation. The red dashed line shows the predicted standard deviations, according to relation \ref{eq:prop_T}, plotted over $T_\mathrm{man}$. In particular, the plot confirms the proportionality $\sigma \propto 1/\sqrt{T_\mathrm{man}}$.

Recall that, for \ac{SC} maneuvers, the maneuver amplitude may be limited by the maximally achievable torque, which we did not assume. Should that be the case, we have furthermore
\begin{equation}
	A_\mathrm{man} \propto \frac{1}{f_\mathrm{man}^2},
\end{equation}
and hence
\begin{equation}
	\sigma \propto \frac{f_\mathrm{man}^2}{\sqrt{T_\mathrm{man}}}.
\end{equation}

% ACRONYMS

\begin{acronym}
		\acro{AIVT}{Assembly, Integration, Verification, Test}
		\acro{ASD}{Amplitude Spectral Density}
		\acrodefplural{ASD}{Amplitude Spectral Densities}
		\acro{BAM}{Beam Alignment Mechanism}
		\acrodefplural{BAM}{Beam Alignment Mechanisms}
		\acro{CoM}{center-of-mass}
		\acro{CPSD}{Cross-Power Spectral Density}
		\acrodefplural{CPSD}{Cross-Power Spectral Densities}
		\acro{DFACS}{Drag Free Attitude Control System}
		\acro{DWS}{Differential Wavefront Sensing}
		\acro{ESA}{European Space Agency}
		\acro{FEE}{Front End Electronics}
		\acro{GFO}{GRACE Follow-On}
		\acro{GRS}{Gravity Reference Sensor}
		\acro{GW}{Gravitational Wave}
		\acro{IFO}{Interferometer}
		\acro{ISI}{Inter-Satellite Interferometer}
		\acro{INReP}{Initial Noise Reduction Pipeline}
		\acro{LCA}{LISA Core Assembly}
		\acro{LDPG}{LISA Data Processing Group}
		\acro{LIG}{LISA Instrument Group}
		\acro{LISA}{Laser Interferometer Space Antenna}
		\acro{LPF}{LISA Pathfinder}
		\acro{LSQ}{least squares}
		\acro{MCMC}{Markov Chain Monte Carlo}
		\acro{MOSA}{Moving Optical Sub-Assembly}
		\acrodefplural{MOSA}{Moving Optical Sub-Assemblies}
		\acro{OATM}{Optical Assembly Tracking Mechanism}
		\acro{OB}{Optical Bench}
		\acro{OBC}{On-Board Computer}
		\acro{OMS}{Optical Metrology System}
		\acro{PSD}{Power Spectral Density}
		\acrodefplural{PSD}{Power Spectral Densities}
		\acro{QPR}{Quadrant Photo-Receiver}
		\acro{QPD}{Quadrant Photodiode}
		\acro{RFI}{Reference Interferometer}
		\acro{RIN}{Relative Intensity Noise}
		\acro{RMS}{Root Mean Square}
		\acro{SC}{spacecraft}
		\acrodefplural{SD}{standard deviations}
		\acro{SGWB}{Stochastic Gravitational Wave Background}
		\acro{SNR}{signal-to-noise ratio}
		\acro{TDI}{Time Delay Interferometry}
		\acro{TLS}{total least squares}
		\acro{TM}{Test Mass}
		\acrodefplural{TM}{Test Masses}
		\acro{TMI}{Test Mass Interferometer}
		\acro{TS}{time series}
		\acro{TTL}{Tilt-To-Length}
		\acro{w.r.t.}{with respect to}
\end{acronym}

% ACKNOWLEDGEMENTS
\begin{acknowledgments}
We gratefully acknowledge support by the Deutsches Zentrum für Luft- und Raumfahrt (DLR) with funding of the Bundesministerium für Wirtschaft und Klimaschutz with a decision of the Deutsche Bundestag (DLR Project Reference No. FKZ 50 OQ 1801). Additionally, we acknowledge funding by Deutsche Forschungsgemeinschaft (DFG) via its Cluster of Excellence QuantumFrontiers (EXC 2123, Project ID 390837967).
\end{acknowledgments}

% REFERENCES
\bibliography{literature.bib}

%apsrev4-2.bst 2019-01-14 (MD) hand-edited version of apsrev4-1.bst
%Control: key (0)
%Control: author (8) initials jnrlst
%Control: editor formatted (1) identically to author
%Control: production of article title (0) allowed
%Control: page (0) single
%Control: year (1) truncated
%Control: production of eprint (0) enabled
\begin{thebibliography}{21}%
\makeatletter
\providecommand \@ifxundefined [1]{%
 \@ifx{#1\undefined}
}%
\providecommand \@ifnum [1]{%
 \ifnum #1\expandafter \@firstoftwo
 \else \expandafter \@secondoftwo
 \fi
}%
\providecommand \@ifx [1]{%
 \ifx #1\expandafter \@firstoftwo
 \else \expandafter \@secondoftwo
 \fi
}%
\providecommand \natexlab [1]{#1}%
\providecommand \enquote  [1]{``#1''}%
\providecommand \bibnamefont  [1]{#1}%
\providecommand \bibfnamefont [1]{#1}%
\providecommand \citenamefont [1]{#1}%
\providecommand \href@noop [0]{\@secondoftwo}%
\providecommand \href [0]{\begingroup \@sanitize@url \@href}%
\providecommand \@href[1]{\@@startlink{#1}\@@href}%
\providecommand \@@href[1]{\endgroup#1\@@endlink}%
\providecommand \@sanitize@url [0]{\catcode `\\12\catcode `\$12\catcode `\&12\catcode `\#12\catcode `\^12\catcode `\_12\catcode `\%12\relax}%
\providecommand \@@startlink[1]{}%
\providecommand \@@endlink[0]{}%
\providecommand \url  [0]{\begingroup\@sanitize@url \@url }%
\providecommand \@url [1]{\endgroup\@href {#1}{\urlprefix }}%
\providecommand \urlprefix  [0]{URL }%
\providecommand \Eprint [0]{\href }%
\providecommand \doibase [0]{https://doi.org/}%
\providecommand \selectlanguage [0]{\@gobble}%
\providecommand \bibinfo  [0]{\@secondoftwo}%
\providecommand \bibfield  [0]{\@secondoftwo}%
\providecommand \translation [1]{[#1]}%
\providecommand \BibitemOpen [0]{}%
\providecommand \bibitemStop [0]{}%
\providecommand \bibitemNoStop [0]{.\EOS\space}%
\providecommand \EOS [0]{\spacefactor3000\relax}%
\providecommand \BibitemShut  [1]{\csname bibitem#1\endcsname}%
\let\auto@bib@innerbib\@empty
%</preamble>
\bibitem [{\citenamefont {Colpi}\ \emph {et~al.}(2024)\citenamefont {Colpi} \emph {et~al.}}]{lisa_redbook}%
  \BibitemOpen
  \bibfield  {author} {\bibinfo {author} {\bibfnamefont {M.}~\bibnamefont {Colpi}} \emph {et~al.},\ }\href@noop {} {\bibinfo {title} {{LISA Definition Study Report}}} (\bibinfo {year} {2024}),\ \Eprint {https://arxiv.org/abs/2402.07571} {arXiv:2402.07571 [astro-ph.CO]} \BibitemShut {NoStop}%
\bibitem [{\citenamefont {Tinto}\ and\ \citenamefont {Dhurandhar}(2020)}]{tinto_2020}%
  \BibitemOpen
  \bibfield  {author} {\bibinfo {author} {\bibfnamefont {M.}~\bibnamefont {Tinto}}\ and\ \bibinfo {author} {\bibfnamefont {S.~V.}\ \bibnamefont {Dhurandhar}},\ }\bibfield  {title} {\bibinfo {title} {{Time-delay interferometry}},\ }\bibfield  {journal} {\bibinfo  {journal} {Living Reviews in Relativity}\ }\textbf {\bibinfo {volume} {24}},\ \href {https://doi.org/10.1007/s41114-020-00029-6} {10.1007/s41114-020-00029-6} (\bibinfo {year} {2020})\BibitemShut {NoStop}%
\bibitem [{\citenamefont {Armano}\ \emph {et~al.}(2023)\citenamefont {Armano} \emph {et~al.}}]{armano_2023}%
  \BibitemOpen
  \bibfield  {author} {\bibinfo {author} {\bibfnamefont {M.}~\bibnamefont {Armano}} \emph {et~al.},\ }\bibfield  {title} {\bibinfo {title} {{Tilt-to-length coupling in LISA Pathfinder: A data analysis}},\ }\href {https://doi.org/10.1103/PhysRevD.108.102003} {\bibfield  {journal} {\bibinfo  {journal} {Phys. Rev. D}\ }\textbf {\bibinfo {volume} {108}},\ \bibinfo {pages} {102003} (\bibinfo {year} {2023})}\BibitemShut {NoStop}%
\bibitem [{\citenamefont {Heinzel}\ \emph {et~al.}(2020)\citenamefont {Heinzel}, \citenamefont {Hewitson}, \citenamefont {Born}, \citenamefont {Karnesis}, \citenamefont {Wissel}, \citenamefont {Kaune}, \citenamefont {Wanner}, \citenamefont {Danzmann}, \citenamefont {Paczkowski}, \citenamefont {Wittchen}, \citenamefont {Hartig}, \citenamefont {Audley},\ and\ \citenamefont {Reiche}}]{audley_2020}%
  \BibitemOpen
  \bibfield  {author} {\bibinfo {author} {\bibfnamefont {G.}~\bibnamefont {Heinzel}}, \bibinfo {author} {\bibfnamefont {M.}~\bibnamefont {Hewitson}}, \bibinfo {author} {\bibfnamefont {M.}~\bibnamefont {Born}}, \bibinfo {author} {\bibfnamefont {N.}~\bibnamefont {Karnesis}}, \bibinfo {author} {\bibfnamefont {L.}~\bibnamefont {Wissel}}, \bibinfo {author} {\bibfnamefont {B.}~\bibnamefont {Kaune}}, \bibinfo {author} {\bibfnamefont {G.}~\bibnamefont {Wanner}}, \bibinfo {author} {\bibfnamefont {K.}~\bibnamefont {Danzmann}}, \bibinfo {author} {\bibfnamefont {S.}~\bibnamefont {Paczkowski}}, \bibinfo {author} {\bibfnamefont {A.}~\bibnamefont {Wittchen}}, \bibinfo {author} {\bibfnamefont {M.-S.}\ \bibnamefont {Hartig}}, \bibinfo {author} {\bibfnamefont {H.}~\bibnamefont {Audley}},\ and\ \bibinfo {author} {\bibfnamefont {J.}~\bibnamefont {Reiche}},\ }\href {https://doi.org/10.2314/KXP:1692401564} {\emph {\bibinfo {title} {{LISA Pathfinder mission extension report for the German contribution}}}},\ \bibinfo {type} {Tech. Rep.}\ (\bibinfo  {institution} {[Max-Planck-Institut für Gravitationsphysik, Albert-Einstein-Institut, Teilinstitut Hannover]},\ \bibinfo {year} {2020})\BibitemShut {NoStop}%
\bibitem [{\citenamefont {Wanner}\ \emph {et~al.}(2024)\citenamefont {Wanner} \emph {et~al.}}]{wanner_2024}%
  \BibitemOpen
  \bibfield  {author} {\bibinfo {author} {\bibfnamefont {G.}~\bibnamefont {Wanner}} \emph {et~al.},\ }\bibfield  {title} {\bibinfo {title} {{In-Depth Modeling of Tilt-To-Length Coupling in LISA’s Interferometers and TDI Michelson Observables}},\ }\href@noop {} {\bibfield  {journal} {\bibinfo  {journal} {Physical Review D}\ } (\bibinfo {year} {2024})}\BibitemShut {NoStop}%
\bibitem [{\citenamefont {Paczkowski}\ \emph {et~al.}(2022)\citenamefont {Paczkowski}, \citenamefont {Giusteri}, \citenamefont {Hewitson}, \citenamefont {Karnesis}, \citenamefont {Fitzsimons}, \citenamefont {Wanner},\ and\ \citenamefont {Heinzel}}]{paczkowski_2022}%
  \BibitemOpen
  \bibfield  {author} {\bibinfo {author} {\bibfnamefont {S.}~\bibnamefont {Paczkowski}}, \bibinfo {author} {\bibfnamefont {R.}~\bibnamefont {Giusteri}}, \bibinfo {author} {\bibfnamefont {M.}~\bibnamefont {Hewitson}}, \bibinfo {author} {\bibfnamefont {N.}~\bibnamefont {Karnesis}}, \bibinfo {author} {\bibfnamefont {E.~D.}\ \bibnamefont {Fitzsimons}}, \bibinfo {author} {\bibfnamefont {G.}~\bibnamefont {Wanner}},\ and\ \bibinfo {author} {\bibfnamefont {G.}~\bibnamefont {Heinzel}},\ }\bibfield  {title} {\bibinfo {title} {{Postprocessing subtraction of tilt-to-length noise in LISA}},\ }\href {https://doi.org/10.1103/PhysRevD.106.042005} {\bibfield  {journal} {\bibinfo  {journal} {Phys. Rev. D}\ }\textbf {\bibinfo {volume} {106}},\ \bibinfo {pages} {042005} (\bibinfo {year} {2022})}\BibitemShut {NoStop}%
\bibitem [{\citenamefont {Paczkowski}\ \emph {et~al.}(2025)\citenamefont {Paczkowski} \emph {et~al.}}]{paczkowski_2025}%
  \BibitemOpen
  \bibfield  {author} {\bibinfo {author} {\bibfnamefont {S.}~\bibnamefont {Paczkowski}} \emph {et~al.},\ }\bibfield  {title} {\bibinfo {title} {{Update on TTL coefficient estimation using noise minimisation}},\ }\href@noop {} {\bibfield  {journal} {\bibinfo  {journal} {in preparation}\ } (\bibinfo {year} {2025})}\BibitemShut {NoStop}%
\bibitem [{\citenamefont {George}\ \emph {et~al.}(2023)\citenamefont {George}, \citenamefont {Sanjuan}, \citenamefont {Fulda},\ and\ \citenamefont {Mueller}}]{george_2022}%
  \BibitemOpen
  \bibfield  {author} {\bibinfo {author} {\bibfnamefont {D.}~\bibnamefont {George}}, \bibinfo {author} {\bibfnamefont {J.}~\bibnamefont {Sanjuan}}, \bibinfo {author} {\bibfnamefont {P.}~\bibnamefont {Fulda}},\ and\ \bibinfo {author} {\bibfnamefont {G.}~\bibnamefont {Mueller}},\ }\bibfield  {title} {\bibinfo {title} {{Calculating the precision of tilt-to-length coupling estimation and noise subtraction in LISA using Fisher information}},\ }\href {https://doi.org/10.1103/PhysRevD.107.022005} {\bibfield  {journal} {\bibinfo  {journal} {Phys. Rev. D}\ }\textbf {\bibinfo {volume} {107}},\ \bibinfo {pages} {022005} (\bibinfo {year} {2023})}\BibitemShut {NoStop}%
\bibitem [{\citenamefont {Houba}\ \emph {et~al.}(2022{\natexlab{a}})\citenamefont {Houba}, \citenamefont {Delchambre}, \citenamefont {Ziegler},\ and\ \citenamefont {Fichter}}]{houba_2022a}%
  \BibitemOpen
  \bibfield  {author} {\bibinfo {author} {\bibfnamefont {N.}~\bibnamefont {Houba}}, \bibinfo {author} {\bibfnamefont {S.}~\bibnamefont {Delchambre}}, \bibinfo {author} {\bibfnamefont {T.}~\bibnamefont {Ziegler}},\ and\ \bibinfo {author} {\bibfnamefont {W.}~\bibnamefont {Fichter}},\ }\bibfield  {title} {\bibinfo {title} {{Optimal Estimation of Tilt-to-Length Noise for Spaceborne Gravitational-Wave Observatories}},\ }\href {https://doi.org/10.2514/1.G006064} {\bibfield  {journal} {\bibinfo  {journal} {Journal of Guidance, Control, and Dynamics}\ }\textbf {\bibinfo {volume} {45}},\ \bibinfo {pages} {1078} (\bibinfo {year} {2022}{\natexlab{a}})}\BibitemShut {NoStop}%
\bibitem [{\citenamefont {Houba}\ \emph {et~al.}(2022{\natexlab{b}})\citenamefont {Houba}, \citenamefont {Delchambre}, \citenamefont {Ziegler}, \citenamefont {Hechenblaikner},\ and\ \citenamefont {Fichter}}]{houba_2022b}%
  \BibitemOpen
  \bibfield  {author} {\bibinfo {author} {\bibfnamefont {N.}~\bibnamefont {Houba}}, \bibinfo {author} {\bibfnamefont {S.}~\bibnamefont {Delchambre}}, \bibinfo {author} {\bibfnamefont {T.}~\bibnamefont {Ziegler}}, \bibinfo {author} {\bibfnamefont {G.}~\bibnamefont {Hechenblaikner}},\ and\ \bibinfo {author} {\bibfnamefont {W.}~\bibnamefont {Fichter}},\ }\bibfield  {title} {\bibinfo {title} {{LISA spacecraft maneuver design to estimate tilt-to-length noise during gravitational wave events}},\ }\href {https://doi.org/10.1103/PhysRevD.106.022004} {\bibfield  {journal} {\bibinfo  {journal} {Phys. Rev. D}\ }\textbf {\bibinfo {volume} {106}},\ \bibinfo {pages} {022004} (\bibinfo {year} {2022}{\natexlab{b}})}\BibitemShut {NoStop}%
\bibitem [{\citenamefont {Morrison}\ \emph {et~al.}(1994)\citenamefont {Morrison}, \citenamefont {Meers}, \citenamefont {Robertson},\ and\ \citenamefont {Ward}}]{morrison_1994}%
  \BibitemOpen
  \bibfield  {author} {\bibinfo {author} {\bibfnamefont {E.}~\bibnamefont {Morrison}}, \bibinfo {author} {\bibfnamefont {B.~J.}\ \bibnamefont {Meers}}, \bibinfo {author} {\bibfnamefont {D.~I.}\ \bibnamefont {Robertson}},\ and\ \bibinfo {author} {\bibfnamefont {H.}~\bibnamefont {Ward}},\ }\bibfield  {title} {\bibinfo {title} {{Automatic alignment of optical interferometers}},\ }\href {https://doi.org/10.1364/AO.33.005041} {\bibfield  {journal} {\bibinfo  {journal} {Appl. Opt.}\ }\textbf {\bibinfo {volume} {33}},\ \bibinfo {pages} {5041} (\bibinfo {year} {1994})}\BibitemShut {NoStop}%
\bibitem [{\citenamefont {Bayle}(2019)}]{bayle_phd}%
  \BibitemOpen
  \bibfield  {author} {\bibinfo {author} {\bibfnamefont {J.-B.}\ \bibnamefont {Bayle}},\ }\emph {\bibinfo {title} {{Simulation and Data Analysis for LISA}}},\ \href {https://hal.archives-ouvertes.fr/tel-03120731/document} {Ph.D. thesis},\ \bibinfo  {school} {Université de Paris} (\bibinfo {year} {2019})\BibitemShut {NoStop}%
\bibitem [{\citenamefont {Otto}(2015)}]{otto_phd}%
  \BibitemOpen
  \bibfield  {author} {\bibinfo {author} {\bibfnamefont {M.}~\bibnamefont {Otto}},\ }\emph {\bibinfo {title} {{Time-Delay Interferometry Simulations for the Laser Interferometer Space Antenna}}},\ \href {https://doi.org/10.15488/8545} {Ph.D. thesis},\ \bibinfo  {school} {Leibniz Universit\"at Hannover}, \bibinfo {address} {Germany} (\bibinfo {year} {2015})\BibitemShut {NoStop}%
\bibitem [{\citenamefont {Hewitson}\ \emph {et~al.}(2021{\natexlab{a}})\citenamefont {Hewitson} \emph {et~al.}}]{lisasim}%
  \BibitemOpen
  \bibfield  {author} {\bibinfo {author} {\bibfnamefont {M.}~\bibnamefont {Hewitson}} \emph {et~al.},\ }\bibfield  {title} {\bibinfo {title} {{LISASim: An open-loop LISA simulator in MATLAB}}} (\bibinfo {year} {2021}{\natexlab{a}}),\ \bibinfo {note} {{LISA-LCST-INST-DD-003}}\BibitemShut {NoStop}%
\bibitem [{\citenamefont {Paczkowski}\ \emph {et~al.}(2023)\citenamefont {Paczkowski}, \citenamefont {Hartig},\ and\ \citenamefont {Giusteri}}]{LISA-LCST-INST-TN-017}%
  \BibitemOpen
  \bibfield  {author} {\bibinfo {author} {\bibfnamefont {S.}~\bibnamefont {Paczkowski}}, \bibinfo {author} {\bibfnamefont {M.}~\bibnamefont {Hartig}},\ and\ \bibinfo {author} {\bibfnamefont {R.}~\bibnamefont {Giusteri}},\ }\bibfield  {title} {\bibinfo {title} {{Post-processing subtraction of Tilt-To-Length noise in LISA - Phase B1 investigations}}} (\bibinfo {year} {2023}),\ \bibinfo {note} {{LISA-LCST-INST-TN-017 i1.0}}\BibitemShut {NoStop}%
\bibitem [{\citenamefont {Chwalla}\ \emph {et~al.}(2020)\citenamefont {Chwalla} \emph {et~al.}}]{chwalla_2020}%
  \BibitemOpen
  \bibfield  {author} {\bibinfo {author} {\bibfnamefont {M.}~\bibnamefont {Chwalla}} \emph {et~al.},\ }\bibfield  {title} {\bibinfo {title} {{Optical Suppression of Tilt-to-Length Coupling in the LISA Long-Arm Interferometer}},\ }\href {https://doi.org/10.1103/PhysRevApplied.14.014030} {\bibfield  {journal} {\bibinfo  {journal} {Phys. Rev. Applied}\ }\textbf {\bibinfo {volume} {14}},\ \bibinfo {pages} {014030} (\bibinfo {year} {2020})}\BibitemShut {NoStop}%
\bibitem [{\citenamefont {Hewitson}\ \emph {et~al.}(2021{\natexlab{b}})\citenamefont {Hewitson} \emph {et~al.}}]{lisa_perf_model}%
  \BibitemOpen
  \bibfield  {author} {\bibinfo {author} {\bibfnamefont {M.}~\bibnamefont {Hewitson}} \emph {et~al.},\ }\bibfield  {title} {\bibinfo {title} {{LISA Performance Model}}} (\bibinfo {year} {2021}{\natexlab{b}}),\ \bibinfo {note} {{LISA-LCST-INST-TN-003 i2.1}}\BibitemShut {NoStop}%
\bibitem [{\citenamefont {Hartig}\ \emph {et~al.}(2024)\citenamefont {Hartig}, \citenamefont {Marmor}, \citenamefont {George}, \citenamefont {Paczkowski},\ and\ \citenamefont {Sanjuan}}]{hartig_2024}%
  \BibitemOpen
  \bibfield  {author} {\bibinfo {author} {\bibfnamefont {M.-S.}\ \bibnamefont {Hartig}}, \bibinfo {author} {\bibfnamefont {J.}~\bibnamefont {Marmor}}, \bibinfo {author} {\bibfnamefont {D.}~\bibnamefont {George}}, \bibinfo {author} {\bibfnamefont {S.}~\bibnamefont {Paczkowski}},\ and\ \bibinfo {author} {\bibfnamefont {J.}~\bibnamefont {Sanjuan}},\ }\bibfield  {title} {\bibinfo {title} {Tilt-to-length coupling in lisa -- uncertainty and biases},\ }\href {https://arxiv.org/abs/2410.16475} {\bibfield  {journal} {\bibinfo  {journal} {arXiv}\ } (\bibinfo {year} {2024})},\ \Eprint {https://arxiv.org/abs/2410.16475} {2410.16475} \BibitemShut {NoStop}%
\bibitem [{\citenamefont {Crassidis}\ and\ \citenamefont {Junkins}(2004)}]{crassidis_2004}%
  \BibitemOpen
  \bibfield  {author} {\bibinfo {author} {\bibfnamefont {J.~L.}\ \bibnamefont {Crassidis}}\ and\ \bibinfo {author} {\bibfnamefont {J.~L.}\ \bibnamefont {Junkins}},\ }\href@noop {} {\emph {\bibinfo {title} {Optimal Estimation of Dynamic Systems}}}\ (\bibinfo  {publisher} {CRC Press LLC},\ \bibinfo {year} {2004})\BibitemShut {NoStop}%
\bibitem [{\citenamefont {Armano}\ \emph {et~al.}(2019)\citenamefont {Armano} \emph {et~al.}}]{armano_2019}%
  \BibitemOpen
  \bibfield  {author} {\bibinfo {author} {\bibfnamefont {M.}~\bibnamefont {Armano}} \emph {et~al.} (\bibinfo {collaboration} {LISA Pathfinder Collaboration}),\ }\bibfield  {title} {\bibinfo {title} {{LISA Pathfinder micronewton cold gas thrusters: In-flight characterization}},\ }\href {https://doi.org/10.1103/PhysRevD.99.122003} {\bibfield  {journal} {\bibinfo  {journal} {Phys. Rev. D}\ }\textbf {\bibinfo {volume} {99}},\ \bibinfo {pages} {122003} (\bibinfo {year} {2019})}\BibitemShut {NoStop}%
\bibitem [{\citenamefont {TEB}(2021)}]{ESA-SCI-F-ESTEC-SOW-2019-028}%
  \BibitemOpen
  \bibfield  {author} {\bibinfo {author} {\bibnamefont {TEB}},\ }\bibfield  {title} {\bibinfo {title} {{LISA - Optical Assembly Tracking Mechanism Development}}} (\bibinfo {year} {2021}),\ \bibinfo {note} {{ESA-SCI-F-ESTEC-SOW-2019-028, Programme Reference: C215-137FT}}\BibitemShut {NoStop}%
\end{thebibliography}%

\end{document}